\documentclass[aps,prd,floats,floatfix,twocolumn,showpacs,eqsecnum,
superscriptaddress,nofootinbib]{revtex4-2}

\usepackage[utf8]{inputenc}
\usepackage{lmodern}
\usepackage[T1]{fontenc}
\usepackage[centertags]{amsmath}
\usepackage{amssymb}
\usepackage{latexsym}
\usepackage{enumerate}
\usepackage{graphicx}
\usepackage{mathrsfs}
\usepackage{hyperref}
\hypersetup{
  colorlinks,
  citecolor={cyan!50!black},
  linkcolor={orange!60!black},
  urlcolor=violet,
}
\usepackage{stmaryrd}
\usepackage{import}
\usepackage{tensor}
\usepackage{mathtools}  
\usepackage{commath}  
\usepackage{physics}
\usepackage[usenames,dvipsnames]{xcolor}
\usepackage{bm}
\usepackage{multirow}
\usepackage{breakurl}
\usepackage{float}
\usepackage{verbatim}
\usepackage{mathrsfs}
\usepackage{booktabs}
\usepackage{orcidlink}
\usepackage{microtype}
\usepackage[capitalise]{cleveref}
\usepackage[caption=false]{subfig}
\usepackage[normalem]{ulem}
\crefname{section}{Sec.}{Sec.}
\Crefname{section}{Section}{Sections}
\newcommand{\ccite}[1]{Ref.~\cite{#1}}
\newcommand{\ccites}[1]{Refs.~\cite{#1}}

\newcommand{\defeq}{\coloneqq}

\newcommand{\dgF}{\mathcal{F}}

\newcommand{\dgFA}{\mathcal{A}}

\newcommand{\calP}{\mathcal{P}}
\newcommand{\calR}{\mathcal{R}}
\newcommand{\calE}{\mathcal{E}}
\newcommand{\calO}{\mathcal{O}}

\newcommand{\Lz}{\mathcal{L}_z}

\newcommand{\spectre}{\texttt{SpECTRE}}
\newcommand{\pybhpt}{\texttt{pybhpt}}
\newcommand{\matplotlib}{\texttt{matplotlib}}
\newcommand{\paraview}{\texttt{ParaView}}

\allowdisplaybreaks[1]

\definecolor{darkgreen}{rgb}{0.2,0.7,0.2}

\begin{document}
\newcommand{\caltech}{\affiliation{Theoretical Astrophysics, Walter Burke
Institute for Theoretical Physics, California Institute of Technology, Pasadena,
California 91125, USA}}
\newcommand{\uzh}{\affiliation{Department of Astrophysics, University of Zurich, Winterthurerstrasse 190, 8057 Zurich, Switzerland}}
\newcommand{\aei}{\affiliation{Max Planck Institute for Gravitational Physics (Albert Einstein Institute), Am M\" uhlenberg 1, Potsdam 14476, Germany}}
\newcommand{\maryland}{\affiliation{Department of Physics, University of Maryland, College Park, Maryland 20742, USA}}
\newcommand{\geneseo}{\affiliation{Department of Physics and Astronomy, State University of New York at Geneseo, New York 14454, USA}}
\newcommand{\soton}{\affiliation{ School of Mathematical Sciences and STAG Research Centre, University of Southampton, Southampton, SO17 1BJ, United Kingdom}}
\newcommand{\ucd}{\affiliation{ School of Mathematics and Statistics, University College Dublin, Belfield, Dublin 4, Ireland}}
\newcommand{\cornell}{\affiliation{Cornell Center for Astrophysics and Planetary
Science, Cornell University, Ithaca, New York 14853, USA}}
\author{Nami Nishimura\,\orcidlink{0009-0002-4398-9130}} \aei \maryland
\author{Nils L.\ Vu\,\orcidlink{0000-0002-5767-3949}} \uzh
\author{Adrià Gispert-Fernández\,\orcidlink{0009-0006-2658-339X}} \ucd
\author{Cillian Kelly} \soton
\author{Lawrence E.\ Kidder\,\orcidlink{0000-0001-5392-7342}} \cornell
\author{Benjamin Leather\,\orcidlink{0000-0001-6186-7271}} \soton
\author{Thomas Osburn\,\orcidlink{0000-0003-2747-3994}} \geneseo \ucd
\author{Jonathan E.\ Thompson\,\orcidlink{0000-0002-0419-5517}} \soton
\author{Samuel D.\ Upton\,\orcidlink{0000-0003-2965-7674}} \ucd \soton
\author{Barry Wardell\,\orcidlink{0000-0001-6176-9006}} \ucd
\title{Kerr metric perturbations in Lorenz gauge via numerical relativity methods}

\date{\today}

\begin{abstract}
  Gravitational self-force calculations are needed to accurately describe the motion of a small body in a spinning (Kerr) black hole spacetime, which is vital to predict gravitational wave signals from asymmetric-mass compact binaries.
  Accurate calculations for the upcoming LISA mission require a minimum of second perturbative order in the mass ratio, which has thus far only been achieved for nonspinning (Schwarzschild) black holes due to nonlinear complexities of the Einstein equations.
  In this work, we calculate Kerr metric perturbations with a new computational approach based on numerical relativity methods, extending recent scalar self-force results to full general relativity. We formulate the perturbative Lorenz-gauge Einstein equations in terms of azimuthal $m$-modes in the frequency domain and solve the resulting elliptic partial differential equations using a spectral discontinuous Galerkin discretization, adaptive mesh refinement, and preconditioned iterative solvers.
  With our new method, we achieve exponentially convergent first-order gravitational self-force calculations for circular equatorial orbits in Kerr spacetime up to spin $|a|/M=0.998$ (the Thorne limit).
  Our framework is designed to carry over to second order by delivering accurate and well-behaved first-order data for nonlinear source construction, and as a solver for the second-order metric perturbations themselves.
\end{abstract}

\maketitle

\section{Introduction}
Since the first detection of gravitational waves (GWs) by LIGO in 2015, substantial effort has gone into modeling GWs from compact binaries. 
Next-generation detectors, including the Laser Interferometer Space Antenna (LISA)~\cite{LISA,LISA:2024hlh,LISAWaveformWG:2023arg}, Einstein Telescope~\cite{ET:2025xjr} and Cosmic Explorer~\cite{Evans:2021gyd} are currently under development. 
With their improved sensitivity, and in LISA's case access to lower frequencies, these detectors will observe extreme- and intermediate-mass-ratio inspirals, which require more accurate waveforms.
Compact binary modeling in the LIGO-Virgo-KAGRA era has been centered on comparable-mass systems, since current detectors are most sensitive to stellar-mass binary components~\cite{Abbott_2019,Abbott_2021,Abbott_2022}.
This has justifiably driven considerable investment in numerical relativity~\cite{Scheel:2025jct, Healy:2022wdn, Jani:2016wkt, Ferguson:2023vta, BAMcatalog}, which works well for comparable masses but struggles with increasingly asymmetric masses due to a large separation of scales~\cite{LISAWaveformWG:2023arg}. 

Asymmetric-mass binaries are primarily modeled using the black hole perturbation theory and gravitational self-force framework~\cite{Mino_1997,Quinn_1997,Poisson:2011nh,Barack:2018yvs,Pound_2021}, in which the Einstein field equations are expanded in powers of the mass ratio $\epsilon=\mu/M \ll 1$, where $M$ and $\mu$ are the masses of the primary and the secondary bodies, respectively.
Meeting LISA's accuracy requirements is expected to require this expansion to be carried out to second order in $\epsilon$~\cite{Pound_2005,Burke_2024}.
Modeling second-order effects is substantially more difficult than modeling first-order ones, since second-order calculations involve inherently nonlinear terms. While the foundations of second-order self-force theory are well understood~\cite{Rosenthal:2006iy,Pound:2009sm,Detweiler:2011tt,Pound:2012nt,Gralla:2012db, Pound:2012dk,Pound:2017psq,Upton:2021oxf}, concrete calculations have so far been carried out only for quasicircular orbits around a nonspinning primary.
These calculations are based on a decomposition of the metric perturbation into tensor spherical harmonics, solved mode-by-mode in Lorenz gauge~\cite{Barack:2005nr,Barack:2007tm,Akcay:2010dx,Akcay_2013,Osburn:2014hoa,Hopper:2015jxa}. This approach has yielded the binding energy~\cite{Pound_2020, Bonetto:2021exn}, the energy flux~\cite{Warburton:2021kwk}, and the waveform~\cite{Wardell_2023} for a nonspinning secondary orbiting a Schwarzschild black hole. The spin of the secondary was also added to those models in Ref.~\cite{Mathews:2021rod}, and they were further extended to slowly spinning primaries with a linear-in-spin approximation~\cite{Mathews:2025txc}.
However, linear-in-spin corrections are not sufficient for astrophysical extreme-mass-ratio inspirals, since the massive black holes hosting them are expected to spin not just moderately but rapidly~\cite{Reynolds:2020jwt}, making second-order calculations on a Kerr background essential.

Extending these calculations to Kerr introduces a fundamental obstacle: the reduced symmetry of Kerr spacetime means that, unlike in Schwarzschild, the field equations no longer directly separate in a basis of tensor spherical harmonics. Instead of solving the metric perturbation equations directly, one direction of effort has focused on a single, fully separable scalar equation: the Teukolsky equation~\cite{PhysRevLett.29.1114,1973ApJ...185..635T}, written for either of the two gauge-invariant Weyl scalars $\psi_0$ or $\psi_4$. These scalars carry the complete radiative flux content at first order, and the metric perturbation can be reconstructed from them, a technique that underlies many first-order Kerr calculations~\cite{Shah:2012gu,Pound:2013faa,vandeMeent:2015lxa,Bini:2016dvs,Merlin:2016boc, Kavanagh:2016idg,vandeMeent:2017bcc, Bini:2018ylh, Toomani:2021jlo, Dolan_2022, Dolan_2024, Wardell_2024, Bourg:2024vre, Nasipak:2025tby}. Recent work has carried out second-order Teukolsky calculations on a Schwarzschild background~\cite{Leather:2026zhl}, and a formalism for second-order Teukolsky calculations has been developed for Kerr spacetime~\cite{Spiers:2023cip}. Its practical implementation in Kerr spacetime, however, faces a key difficulty. Although the Teukolsky operator itself still separates, the second-order source is quadratic in the first-order metric perturbation and has a mixed radial and polar dependence. Near the secondary body, the first-order field modes converge slowly, and each mode of the quadratic source couples to infinitely many modes of the first-order field~\cite{Miller_2016}. 
Handling this coupling requires dedicated techniques~\cite{Spiers_2024}. 

These difficulties motivate a promising alternative: giving up on a full $(\ell,m)$ decomposition altogether. Since the Lorenz-gauge Einstein equations are not directly separable in such a decomposition, we instead decompose the metric perturbation into azimuthal $m$-modes---a natural choice given the axisymmetry of Kerr spacetime. Moreover, this choice is also advantageous for constructing the second-order source. Since the $(r,\theta)$ dependence is resolved directly on a two-dimensional mesh rather than through an angular mode sum, the discretization near the secondary body is approximately isotropic, which avoids the slow convergence of the angular mode sum there; the coupling of the quadratic source remains, but reduces to a single sum over $m$-modes rather than a double sum over $(\ell, m)$. Such $m$-mode methods have been developed in both the time domain~\cite{Barack_2007a,Barack_2007b,Dolan_2011a,Dolan_2011b,Dolan_2013,Thornburg_2017} and the frequency domain~\cite{Macedo_2024, Osburn_2022, Vu:2026ypc,Osburn:2026uct, Cownden:2026msq}. Here, we adopt the frequency-domain approach because it is well suited to the multiscale expansion used in second-order self-force calculations, and because time-domain evolutions in the Lorenz gauge have encountered unchecked growth in time~\cite{Barack:2007tm,Barack:2010tm,Dolan_2013} (although this issue might be avoided by extending the recent $\ell m$-mode techniques of Ref.~\cite{Thornburg:2026eeb} to the $m$-mode setting). After entering the frequency domain, we are left with coupled two-dimensional elliptic partial differential equations (PDEs) in $(r,\theta)$, rather than the radial ordinary differential equations of an $\ell m$-mode approach. Solving these is more computationally demanding, but we argue that this cost is outweighed by the advantages above.

Our approach builds on previous work on frequency-domain $m$-mode self-force calculations. The method was first introduced for a scalar-field toy model and solved in \textit{Mathematica}~\cite{Osburn_2022}, and was later extended to the gravitational case in Schwarzschild spacetime~\cite{Osburn:2026uct}. In parallel, the scalar model was solved in Ref.~\cite{Vu:2026ypc} by leveraging computational methods from the adjacent field of numerical relativity (NR). Every NR simulation of merging black holes begins by solving the elliptic Einstein constraint equations for the initial slice of spacetime, known as the initial-data problem. To this end, the NR community has developed sophisticated methods for solving elliptic PDEs numerically~\cite{Vu_2022,Lorene,Ansorg:2004ds,Pfeiffer2003-mt,Papenfort:2021hod,Rashti:2021ihv,Uryu2012-bo,Assumpcao2021-es}.
In the present work, we extend this implementation to gravitational perturbations of Kerr spacetime, adapting the elliptic solver of the NR code \spectre{}~\cite{spectrecode,Vu_2022} to the $m$-mode self-force problem. The solver combines a discontinuous Galerkin (DG) discretization with adaptive mesh refinement (AMR) and an iterative Krylov-type linear solver with multigrid-Schwarz preconditioning.

This paper is organized as follows. Section~\ref{sec:formulation} develops the elliptic $m$-mode formulation of the Lorenz-gauge field equations, including the hyperboloidal $vtu$ slicing and the compactification toward future null infinity. Section~\ref{sec:regularization} describes the effective-source regularization, and Section~\ref{sec:solver} details our implementation in the \spectre{} elliptic solver. In Section~\ref{sec:results}, we validate our solutions against independent computations of the $m$-mode metric perturbation, the redshift invariant and the energy flux, and we conclude in Section~\ref{sec:conclusion}.

\section{Formulation}\label{sec:formulation}
Throughout, we use geometrized units, $G=c=1$, and the metric signature $(-+++)$.

\subsection{Circular geodesics in Kerr spacetime} 
We consider a point particle of mass $\mu$ on a circular, equatorial geodesic around a Kerr black hole of mass $M$ and spin parameter $a$, neglecting self-force effects on the orbit. Positive (negative) $a$ corresponds to a prograde (retrograde) orbit.
In Boyer--Lindquist coordinates $x^\alpha = (t, r, \theta, \phi)$, the Kerr line element is
\begin{align}
ds^2 &= g_{\mu\nu} \, dx^\mu dx^\nu \notag
\\&= -\left(1-\frac{2Mr}{\Sigma}\right)dt^2-\frac{4Ma \, r\sin^2{\theta}}{\Sigma}d\phi \, dt + \frac{\Sigma}{\Delta} dr^2 \notag
\\&\;\;\;\; +\Sigma \, d\theta^2+\left( r^2+a^2+\frac{2Mr \, a^2\sin^2{\theta}}{\Sigma} \right) \sin^2{\theta} \, d\phi^2
\text{,}
\end{align}
where $\Sigma \equiv r^2 + a^2 \cos^2{\theta}$ and $\Delta \equiv r^2-2Mr+a^2 = (r-r_+)(r-r_-)$, with outer ($+$) and inner ($-$) horizons radii $r_\pm = M \pm \sqrt{M^2-a^2}$.
The particle's worldline $x_p^\alpha(\tau)$, parametrized by proper time $\tau$, has $r_p=r_0$ and $\theta_p=\pi/2$. Its four-velocity $u^\alpha = dx_p^\alpha/d\tau$ therefore has only $t$ and $\phi$ components, given by
\begin{align}\label{eq:ut}
u^t = \frac{a + \sqrt{r_0^3/M}}{\sqrt{r_0^3/M-3r_0^2+2a\sqrt{r_0^3/M}}} \, , \qquad u^\phi = \Omega\, u^t \, ,
\end{align}
where
\begin{align}
\Omega = \frac{d\phi_p}{dt} = \frac{1}{a+\sqrt{r_0^3/M}}
\end{align}
is the orbital angular frequency, so that $\phi_p=\Omega t$.

\subsection{Lorenz-gauge field equations}
Next, we expand the exact metric in powers of the mass ratio $\epsilon$ about the Kerr background $g_{\mu\nu}$,
\begin{align}
g^{\text{ exact}}_{\mu\nu}=g_{\mu\nu}+\epsilon \, h_{\mu\nu}+O(\epsilon^{2}).
\end{align}
The metric perturbation $h_{\mu\nu}$ satisfies the linearized Einstein equation
\begin{align}
G^{(1)}_{\mu\nu}[h]=8\pi T_{\mu\nu} \, ,
\end{align}
where $T_{\mu\nu}$ is the stress-energy tensor of the small body, and
\begin{align}
G^{(1)}_{\mu\nu}[h]=-\frac{1}{2}\left[\square\bar{h}_{\mu\nu}+2R^{\alpha\;\beta}_{\:\mu\:\nu}\bar{h}_{\alpha\beta}+g_{\mu\nu}\nabla_{\sigma}Z^{\sigma}-2\nabla_{(\mu}Z_{\nu)}\right]
\end{align}
is the linearized Einstein operator with $Z_{\mu}\equiv\nabla^{\nu}\bar{h}_{\mu\nu}$. 
Here, the covariant derivative $\nabla_{\mu}$, the d'Alembertian $\square \equiv \nabla^{\mu}\nabla_{\mu}$ and the Riemann tensor $R^{\alpha\;\beta}_{\:\mu\:\nu}$ are defined with respect to the background metric $g_{\mu\nu}$, and $\bar{h}_{\mu\nu}\equiv h_{\mu\nu}-\frac{1}{2}g_{\mu\nu}h$ is the trace-reversed metric perturbation.
We work in the Lorenz gauge, $Z_{\mu}=0$, in which the field equations simplify and $\bar{h}_{\mu\nu}$ is relatively well-behaved,
\begin{align}\label{eq:einstein_field_eq}
\mathcal{E}_{\mu\nu}[h]\equiv\square\bar{h} _{\mu\nu}+2R^{\alpha\;\beta}_{\:\mu\:\nu}\bar{h}_{\alpha\beta}=-16\pi T_{\mu\nu} \, .
\end{align}
The principal part of this operator acts identically on each component of $\bar h_{\mu\nu}$, so that, in the frequency domain, each $m$-mode will obey a system of elliptic equations with a simple, component-independent principal part (see Sec.~\ref{sec:first_order_flux}).
These equations are equivalent to the linearized Einstein equations only when the gauge condition $Z_\mu=0$ is also satisfied.

Beyond this structural simplicity, the Lorenz gauge has several features that make it well suited to self-force calculations, particularly at second order.
First, the singular field near the particle is approximately isotropic in this gauge, and well-established procedures exist for regularizing it.
Second, the asymptotic behavior of the first-order Lorenz-gauge metric perturbation toward the horizon and future null infinity is well understood.
Third, the Lorenz gauge is one of the most widely used gauges in the self-force community and forms the basis of most second-order calculations to date~\cite{Pound_2020,Warburton:2021kwk,Wardell_2023}.
It is therefore natural to compute the first-order field in this gauge, so that it can serve directly as input to second-order calculations.

\subsection{\texorpdfstring{$m$}{m}-mode decomposition} 
We use horizon-penetrating ingoing Eddington-Finkelstein coordinates, $(v,r,z,\varphi)$, where $v\equiv t+r_{*}$, $z\equiv\cos\theta$, and $\varphi$ is Kerr's well-behaved azimuthal coordinate
\begin{align}
\varphi(\phi,r)=\phi+\frac{a}{r_{+}-r_{-}}\ln\left(\frac{r-r_{+}}{r-r_{-}}\right) \, .
\end{align}
We have also introduced the tortoise coordinate $r_*$
\begin{align}
r_{*}=r+\frac{2M}{r_{+}-r_{-}}\left[r_{+}\ln\left(\frac{r-r_{+}}{2M}\right)-r_{-}\ln\left(\frac{r-r_{-}}{2M}\right)\right] \, ,
\end{align}
with 
\begin{align}
\frac{dr_{*}}{dr}=\frac{r^{2}+a^{2}}{\Delta} \, .
\end{align}
We separate the $\varphi$ and $v$ variables by representing the metric perturbation $\bar{h}_{\mu\nu}$ as a sum over $m$-modes 
\begin{align}\label{eq:h_mode_expansion}
&\bar{h}_{\mu\nu}(v,r,z,\varphi)=\sum^{\infty}_{m=-\infty}\bar h^{(m)}_{\mu\nu}(r,z)\, e^{im\Delta\varphi} \, ,
\\
& 
\label{eq:m_inverse}
\bar h^{(m)}_{\mu\nu}(r,z) = \frac{1}{2\pi}\int_0^{2\pi}\bar{h}_{\mu\nu}(v,r,z,\varphi)\,e^{-im\Delta\varphi}\,d\Delta\varphi \, ,
\\
& \Delta\varphi = \varphi - \Omega \, v \, ,
\end{align}
where each coefficient $\bar h^{(m)}_{\mu\nu}$ involves $(r,z)$ dependence, and we only consider frequencies that appear as harmonics of the circular orbital angular frequency: $\omega = m\Omega$. Note that we next introduce hyperboloidal slicing to describe time dependence of the metric perturbation, after which $v$ will only be used near the horizon (except for indicating tensor components, where $v$ is used throughout).
It is convenient to introduce factors allowing each numerical field to approach a finite non-zero value at the boundaries
\begin{align}
\label{eq:eq_MP}
\bar h^{(m)}_{\mu\nu}(r,z)=\left[\frac{1}{r}\right]^{\kappa^{(\mu\nu)}}\left[\sin\theta\right]^{\lambda^{(\mu\nu)}_{m}}(\psi_{m})_{\mu\nu}(r,z) \, ,
\end{align}
where $\kappa^{(\mu\nu)}$ and $\lambda^{(\mu\nu)}_{m}$ are powers that are motivated by regularity at each boundary\footnote{The radial prefactors are the same as in Refs.~\cite{Osburn:2026uct, Dolan_2013}. We additionally factor out the behavior at the angular poles through $[\sin\theta]^{\lambda^{(\mu\nu)}_{m}}$}~\cite{adria} (see Appendix~\ref{sec:prefactor} for full details).
Usage of ingoing Eddington-Finkelstein coordinates was key to achieving correct behavior of $\bar h^{(m)}_{\mu\nu}$ near the horizon (see Sec.~\ref{sec:BC}).

\subsection{\texorpdfstring{$vtu$}{vtu} slicing}
We implement hyperboloidal slicing, specifically $vtu$\nobreakdash-slicing, in a way that largely follows recent related work~\cite{Vu:2026ypc}, with the addition of compactification in the $u$ region. The spacetime is foliated into slices according to $s$, which is related to $t$ through a radially-dependent height function $h$
\begin{align}
s=s(t,r):=t-h(r_{*}) \, .
\end{align}
Our height function is chosen to be a continuous function of $r_*$.\footnote{The same $vtu$-slicing method, but with discontinuous height function is used in Refs.~\cite{Miller:2023ers,Leather:2026zhl}} However, our $h$ is not smooth because we abruptly switch from $t$ to $v$ near the future horizon and abruptly switch to $u\equiv t-r_{*}$ outside the particle's orbit 
\begin{align}
h(r_{*})=\begin{cases}
-(r_{*}-r^{v}_{*}), & r_{*}\leq r^{v}_{*}\quad\quad\quad\;(v\text{\:domain})\\
0, & r^{v}_{*}\leq r_{*}\leq r^{u}_{*}\:\:\:\:(t\text{\:domain})\\
+(r_{*}-r^{u}_{*}), & r_{*}\geq r^{u}_{*}\quad\quad\quad\;(u\text{\:domain})
\end{cases} \, .
\end{align}
One advantage of $vtu$ slicing is that it eliminates the asymptotic oscillations of the field toward the horizon and future null infinity. This alleviates the need to resolve highly oscillatory behavior there, thereby reducing the resolution requirements and associated computational cost. Another advantage is that $vtu$ slicing has been found to be convenient for second-order source construction near the boundaries~\cite{Upton:2025bja} by avoiding the need to implement second-order boundary punctures~\cite{Miller:2023ers}. Considering that one of the primary motivations of this work is second-order calculations, $vtu$ slicing is an obvious choice. 

This choice of height function introduces a piecewise-constant boost function $H(r_{*})\equiv \frac{dh}{dr_{*}}$, defined as
\begin{align}
H(r_{*})=\begin{cases}
-1, & r_{*} < r^{v}_{*}\quad\quad\quad\;(v\text{\:domain})\\
0, & r^{v}_{*}\leq r_{*}< r^{u}_{*}\:\:\:\:(t\text{\:domain})\\
+1, & r^{u}_{*} \leq r_{*} \quad\quad\quad\;(u\text{\:domain})
\end{cases} \, .
\end{align}
This transformation affects our PDEs via the chain rule
\begin{align}\label{sec:vtu_radial}
\frac{\partial}{\partial r_{*}}&\rightarrow\frac{\partial}{\partial r_{*}}+im\Omega H \, ,
\\
\frac{\partial^{2}}{\partial r^{2}_{*}}&\rightarrow\frac{\partial^{2}}{\partial r^{2}_{*}}+2im\Omega H\frac{\partial}{\partial r_{*}}+im\Omega\frac{\partial H}{\partial r_{*}}-m^{2}\Omega^{2}H^{2} \, ,
\end{align}
see Ref.~\cite{Vu:2026ypc} for full details.
Since $H$ is piecewise constant, $\partial H /\partial r_{*}$ is a sum of delta functions at the transitions; rather than keeping this term, we impose the equivalent junction conditions.
In particular, this implies that the jump in the derivative is
\begin{align}\label{eq:jump_vt}
\partial_{r_*}(\psi_m)_{\mu\nu}\big|_{(r^{v}_*)^+} - \partial_{r_*}(\psi_m)_{\mu\nu}\big|_{(r^{v}_*)^-} = -im\Omega\,(\psi_m)_{\mu\nu}\big|_{r^v_*}\, ,
\end{align}
at the transition from $v$ to $t$ and that it is
\begin{align}\label{eq:jump_tu}
\partial_{r_*}(\psi_m)_{\mu\nu}\big|_{(r^{u}_*)^+} - \partial_{r_*}(\psi_m)_{\mu\nu}\big|_{(r^{u}_*)^-} = -im\Omega\,(\psi_m)_{\mu\nu}\big|_{r^u_*}\, .
\end{align}
at the transition from $t$ to $u$. Our numerical DG method (Sec.~\ref{sec:discretisation}) is well suited to account for such jumps. 

\subsection{Equations in first-order flux form}\label{sec:first_order_flux}
The elliptic PDEs that govern each $(\psi_m)_{\mu\nu}$, defined in Eq.~\eqref{eq:eq_MP}, take the first-order flux form in terms of 2D coordinates $q^i = (r,z)$ after simplification of Eq.~\eqref{eq:einstein_field_eq}
\begin{align}
\label{eq:first_order_flux_form}
\begin{split}
\triangle_{m}(\psi_{m})_{\mu\nu} &= -\partial_{i}(F^{i}_{m})_{\mu\nu}+\beta^{\lambda\rho}_{\mu\nu}(\psi_{m})_{\lambda\rho} \\
&\quad +\gamma^{i,\lambda\rho}_{\mu\nu}\,\partial_{i}(\psi_{m})_{\lambda\rho}=(S_m)_{\mu\nu}
\end{split}
\end{align}
with 
\begin{align}\label{eq:flux}
(F^{i}_{m})_{\mu\nu}=\left\{ \alpha_r\frac{\partial(\psi_{m})_{\mu\nu}}{\partial r},\alpha_z\frac{\partial(\psi_{m})_{\mu\nu}}{\partial z}\right\} 
\end{align}
\begin{align}
\alpha_r = \frac{\Delta}{r^{2}+a^{2}}, \qquad \alpha_z=\frac{1-z^{2}}{r^{2}+a^{2}}
\end{align}
Each of $\beta^{\lambda\rho}_{\mu\nu}$, $\gamma^{r, \lambda\rho}_{\mu\nu}$ and $\gamma^{z, \lambda\rho}_{\mu\nu}$ is a matrix of known functions of $r$ and $z$. Considering that the metric is a symmetric tensor, these can be represented in terms of $10\times10$ matrices~\cite{Osburn:2026uct}.
The partial derivative is $\partial_i \equiv \frac{\partial}{\partial q^i}$, and the divergence $-\partial_i F^{i}$ produces the principal part of $\triangle_{m}(\psi_{m})_{\mu\nu}$. 
At the horizon ($r=r_+$) and at the poles ($z=\pm1$), the component of the flux normal to the boundary vanishes. Note that there is one case where our coefficients in Eq.~\eqref{eq:first_order_flux_form} do not directly follow from the field equation of Eq.~\eqref{eq:einstein_field_eq}: the $m=0$ mode with vanishing frequency. This mode involves residual freedom, which we resolve by decoupling one of the metric perturbation components in a way that requires the overall mode to be stationary (see Appendix~\ref{sec:static_mode}).

\subsection{Compactification}
\label{sec:compactification}
Our recent work~\cite{Vu:2026ypc} used a coordinate map proportional to $1/r$ to resolve the wave zone in the $u$ region and to extend it to large but finite radius.
In the GR case, however, we observed that convergence was spoiled by the finite outer radius and the approximate boundary condition. 
We therefore proceed to compactify toward null infinity in this work. 
To this end, we introduce the radial grid coordinate $\sigma$, which coincides with the Boyer--Lindquist radius $r$ in the $v$ and $t$ regions and is compactified in the $u$ region,
\begin{align}\label{eq:sigma}
\sigma(r)=\begin{cases}
r, & r\leq r_{u}\quad(v\text{ and }t\text{ domains})\\
2r_{u}-\dfrac{r^{2}_{u}}{r}, & r\geq r_{u}\quad(u\text{ domain})
\end{cases} \, ,
\end{align}
so that future null infinity, $r\to\infty$, is located at the finite value $\sigma=2r_u$.
Both $\sigma$ and $d\sigma/dr$ are continuous across $r_u$ (i.e., the map is $C^1$), so no coordinate jump condition is needed.
Since $\partial_r = (d\sigma/dr)\,\partial_\sigma$, the flux form must be rewritten in terms of $\sigma$-derivatives. Dividing the transformed equation by $d\sigma/dr$ to match the generic structure of the first-order flux form gives the substitutions
\begin{align}
&\alpha_r \rightarrow \alpha_r \frac{d\sigma}{dr}, && \alpha_z \rightarrow \alpha_z \frac{dr}{d\sigma}, \\
&\beta \rightarrow \beta\,\frac{dr}{d\sigma}, && \gamma_z \rightarrow \gamma_z\,\frac{dr}{d\sigma},
\end{align}
while $\gamma_r$ is unchanged. For $r\leq r_u$, $d\sigma/dr = 1$ and these substitutions reduce to the identity. With these substitutions, Eqs.~\eqref{eq:first_order_flux_form} and \eqref{eq:flux} hold with $q^i=(\sigma,z)$.
The source $(S_m)_{\mu\nu}$ is also divided by $d\sigma/dr$. At first order, however, the source is nonzero only in the $t$ region, so this has no effect.

\subsection{Boundary conditions}
\label{sec:BC}
Our formulation of the field equations avoids coordinate singularities at the horizon and angular poles, and therefore reduces the boundary conditions to simple regularity conditions. 
These regularity conditions are particularly well suited to spectral numerical methods, because such methods expand fields in a polynomial series that inherently selects regular solutions. 
As foreshadowed in Ref.~\cite{Osburn:2026uct}, this resolves the near-horizon behavior issues encountered by the finite-difference scheme. 

In the $u$ region, the flux is
\begin{align}
(F^{i}_{m})_{\mu\nu}=\left\{\alpha_r\frac{d\sigma}{dr}\frac{\partial(\psi_{m})_{\mu\nu}}{\partial \sigma},\ \alpha_z\frac{dr}{d\sigma}\frac{\partial(\psi_{m})_{\mu\nu}}{\partial z}\right\}
\text{.}
\end{align}
At future null infinity $(\sigma=2r_u)$, the Jacobian $d\sigma/dr$ vanishes and the radial flux vanishes with it, so only regularity is required at $\mathcal{I}^{+}$. 

\section{Regularization}\label{sec:regularization}

Near the worldline, the metric perturbation diverges as $h_{\mu\nu} \sim 1/\lambda$, where we introduce $\lambda$ as a formal order-counting parameter that counts powers of distance away from the worldline.
The strength of this divergence implies that the $m$-modes of the retarded field then diverge as $h^{(m)}_{\mu\nu} \sim \log\lambda$.
In order to sidestep issues caused by divergent behavior in our computational domain, we adopt the effective-source method~\cite{Barack_2007a,Vega_2008,Wardell_2012,Warburton:2013lea,Wardell:2015ada} and introduce an analytically determined puncture $h^{\calP}_{\mu\nu}$ that directly cancels the divergent pieces of $h_{\mu\nu}$.
This leaves a residual field
\begin{equation}
    h^{\calR}_{\mu\nu} \coloneqq h_{\mu\nu} - h^{\calP}_{\mu\nu},
\end{equation}
that is regular at the worldline.
We substitute this singular-regular split into the field equations and move the puncture term to the right-hand side, thereby creating an effective source, so that
\begin{align}
    \calE_{\mu\nu}[h^{\calR}] &= -16\pi T_{\mu\nu} -\calE_{\mu\nu}[h^{\calP}] \nonumber \\
        &\eqqcolon S^{\text{eff}}_{\mu\nu}.
\end{align}
We then solve directly for the residual field without ever having to numerically evaluate diverging quantities.
We can then construct the self-force directly from the residual field. We can also add back on the puncture to get the retarded field, and use that as input to construct the source for the second-order field equations~\cite{Upton:2025bja}.

\subsection{Puncture construction}

This work makes use of a puncture that was formulated in Ref.~\cite{Leather:2026gnc}. We summarize here only the details relevant for the calculation at hand, and direct interested readers to Ref.~\cite{Leather:2026gnc} for in-depth details on its construction.

The puncture is an approximation to the Detweiler-Whiting singular field~\cite{Detweiler_2003} and features four total orders in distance,
\begin{equation}
    h^{\calP}_{\mu\nu} \sim \frac{1}{\lambda} + \lambda^0 + \lambda^1 + \lambda^2.
\end{equation}
This leaves a residual field that is $C^2$ on the worldline, whose $m$-modes $h^{\calR(m)}_{\mu\nu}$ have a $\lambda^4 \log \lambda$ non-smoothness, and whose sum over $m$-modes converges as $m^{-4}$. The corresponding $m$-mode effective source, $(S^{\text{eff}}_{m})_{\mu\nu}$, has a $\lambda^2 \log \lambda$ non-smoothness at the worldline.

The puncture is constructed in comoving coordinates, $(T,X,Y,Z)$, that are adapted to the particle's motion so that the particle sits at the origin of the coordinate system for all $T$.
In these coordinates the metric reduces to the Minkowski metric on the worldline, the four-velocity is given by $u^\alpha = \delta^\alpha_T$, and the resulting form of the puncture is significantly simplified.
Explicitly, the relations between the two coordinates are given by
\begin{subequations}
\begin{align}
    T ={}& \calE t - \Lz \phi, \\
    X ={}& \frac{r_p\,\Delta r}{\sqrt{\Delta_p}}, \\
    Y ={}& - r_p \cos\theta, \\
    Z ={}& z_c\sin\biggl(\frac{\Delta\phi}{2}\biggr),
\end{align}
\end{subequations}
where $\Delta r \equiv r-r_p$, $\Delta_p \equiv \Delta|_{r=r_p}$, $z_c \equiv 2u^t\sqrt{\Delta_p}$ and $\Delta\phi \equiv \phi - \phi_p$.
In this coordinate system, each component of the puncture then has the form
\begin{equation}
    h^{\calP}_{\mu\nu} \sim \sum_{ijkn}c_{ijkn} X^i Y^j (Z')^k R^n,
\end{equation}
where $-1\leq i+j+k+n\leq 2$, $c_{ijkn}$ is a constant that depends on the mass, spin and orbital radius.
To allow analytic integration of the puncture, 
we introduce $Z'=z_c\sin\Delta\phi = 2Z + \calO(Z^3)$. The comoving radius is defined as $R^2 \equiv X^2 +Y^2 +Z^2$.

Building on Ref.~\cite{Bourg:2024cgh}, Ref.~\cite{Leather:2026gnc} demonstrated that, after integrating over $\Delta\phi$ according to Eq.~\eqref{eq:m_inverse}, the $m$-modes of the puncture are given by
\begin{align}
    h^{\calP(m)}_{\mu\nu} \sim \sum d_{ijkn}X^iY^j \mathcal{J}_{knm}(X,Y)
\end{align}
where $\mathcal{J}_{knm}$ satisfies the recursion relation
\begin{equation}
    \mathcal{J}_{knm} = \frac{iz_c}{2}\bigl(\mathcal{J}_{k-1,n,m+1}-\mathcal{J}_{k-1,n,m-1}\bigr),
\end{equation}
with base case
\begin{equation}
    \mathcal{J}_{0nm} = \frac{(-1)^m(\varrho\calR)^{n/2}}{(1+n/2)_{m}}\,P^m_{n/2}(\bar{z}),
\end{equation}
where $\varrho^2\equiv X^2+Y^2$, $\mathcal{R}^2\equiv \varrho^2 + z_c^2$, $\bar{z}\equiv (\varrho^2+z_c^2/2)/(\varrho\mathcal{R})$, $(a)_n=\Gamma(a+n)/\Gamma(a)$ is the Pochhammer symbol and $P_{\mu}^{\nu}(x)$ is the associated Legendre function valid for $x>1$.
Expressions of this form were encountered in our previous work on the scalar case~\cite{Vu:2026ypc}, and we refer readers to Sec.~III~B of that reference for details of the approach we used to accurately evaluate these expressions numerically.

\subsection{Worldtube scheme}
As in our previous work~\cite{Vu:2026ypc}, we introduce a rectangular worldtube in $r$ and $z$ around the particle's position.
Inside this region, we solve for the residual fields $h^{\calR(m)}_{\mu\nu}$ using the effective source, and outside we solve directly for $h^{(m)}_{\mu\nu}$ using the homogeneous equations.
We then transition between the residual field and the retarded field at the boundary of the worldtube via the puncture. 
Each $m$-mode field is extracted via Eqs.~\eqref{eq:m_inverse} and~\eqref{eq:eq_MP}: $h^{\calP(m)}_{\mu\nu} \rightarrow (\psi_{m}^{\calP})_{\mu\nu}$.
This leaves us with a similar setup to our previous work, but now solving the Lorenz gauge field equations:
\begin{alignat}{2}
(S^{\text{eff}}_{m})_{\mu\nu} ={}& \mathrlap{(S_{m})_{\mu\nu} -\triangle_{m}(\psi^\calP_{m})_{\mu\nu}}
\\
\triangle_{m}(\psi^\calR_{m})_{\mu\nu} ={}& (S^{\text{eff}}_{m})_{\mu\nu} &\quad& (\text{interior}) \\
\triangle_{m}(\psi_{m})_{\mu\nu} ={}& 0 &\quad& (\text{exterior}) \\
(\psi^\calR_{m})_{\mu\nu} ={}& (\psi_{m})_{\mu\nu} - (\psi^\calP_{m})_{\mu\nu} &\quad& (\text{across the interface}). \label{eq:worldtube_interface}
\end{alignat}

The worldtube jumps to transform between $(\psi^\calR_{m})_{\mu\nu}$ and $(\psi_{m})_{\mu\nu}$ are imposed by appropriately modifying the DG fluxes across elements following Ref.~\cite{Vu:2026ypc} (see Sec.~\ref{sec:DG_res}).

\section{Numerical methods}\label{sec:solver}

Using the Lorenz gauge and working in the frequency domain, we have reduced the linearized Einstein field equations for each azimuthal $m$-mode to a system of ten coupled elliptic PDEs on the two-dimensional $(\sigma,z)$ domain, Eq.~\eqref{eq:first_order_flux_form}.
To solve this system numerically, we need a discretization that turns the PDEs into a matrix equation, $\mathbf{A} \mathbf{u}=\mathbf{b}$, together with a linear solver that applies the equivalent of the inverse matrix, schematically $\mathbf{u}=\mathbf{A}^{-1}\mathbf{b}$. Here, the state vector $\mathbf{u}$ collects all discrete degrees of freedom (the values of $(\psi_m)_{\mu\nu}$ on a grid), $\mathbf{b}$ is the discrete representation of the source, and $\mathbf{A}$ is an $N_\text{DOF}\times N_\text{DOF}$ matrix representing the discretized operator. 
Previous frequency-domain $m$-mode calculations have taken different approaches to this problem. References~\cite{Osburn_2022,Osburn:2026uct} use a second-order finite-difference scheme, which yields a sparse linear system for the field values at each grid point in $r$ and $\theta$. Reference~\cite{Macedo_2024} instead uses a multidomain spectral collocation method for a scalar toy model, in which each subdomain is discretized on a Chebyshev grid; this approach was recently extended to the gravitational perturbation case~\cite{Cownden:2026msq}.

Elliptic problems of this kind are also central to numerical relativity (NR), where the Einstein constraint equations must be solved to construct initial data for binary black hole simulations. To this end, DG methods~\cite{Vu:2021,Vu:2024} and associated preconditioned iterative solvers~\cite{Vu_2022} have been developed. These methods are well suited for self-force problems, as they achieve exponential convergence under adaptive mesh refinement despite the presence of non-smoothness at the particle, handle regularity conditions at domain boundaries through their spectral expansion in a polynomial basis, and allow for jump conditions between elements at the worldtube and at $vtu$ domain boundaries. 
Following our previous work, \ccite{Vu:2026ypc}, we therefore use the elliptic solver of the numerical relativity platform \spectre{}~\cite{spectrecode}.
In the following, we summarize these methods and focus on the changes needed for the gravitational $m$-mode problem.

\subsection{Discontinuous Galerkin discretization}
\label{sec:discretisation}
To reduce Eq.~\eqref{eq:first_order_flux_form} to a finite-dimensional linear problem, we use a DG discretization, a spectral finite-element method. The domain is divided into non-overlapping elements that couple only to their nearest neighbors through numerical fluxes on shared boundaries, which makes the scheme well suited to parallel computation~\cite{Vu_2022}. Within each element, the fields are expanded in high-order polynomials, so the discretization error of smooth fields decreases exponentially with the polynomial order.
The solutions in our problem are not smooth everywhere. Where surface discontinuities are fixed and known in advance, we place them on element boundaries and impose them by modifying the numerical flux~\cite{Vu:2026ypc}, so the fields remain smooth inside each element. Near the particle, where the residual field exhibits a point of non-smoothness, we instead place the particle at a corner between four element boundaries and split nearby elements into smaller ones ($h$ refinement) while increasing the polynomial order elsewhere ($p$ refinement). Such a combined $hp$ refinement strategy recovers exponential convergence even for non-smooth solutions (see \cref{sec:refinement}).
With the field equations already written in the first-order flux form of Eq.~\eqref{eq:first_order_flux_form}, we only need to supply the coefficients and the source, and can then use the elliptic DG scheme of \spectre{}~\cite{spectrecode,Vu:2021,Vu:2024} directly. We summarize the scheme below and refer to Refs.~\cite{Vu:2021,Vu:2024,Vu:2026ypc} for details.

\begin{figure*}
    \centering
    \includegraphics[width=\textwidth]{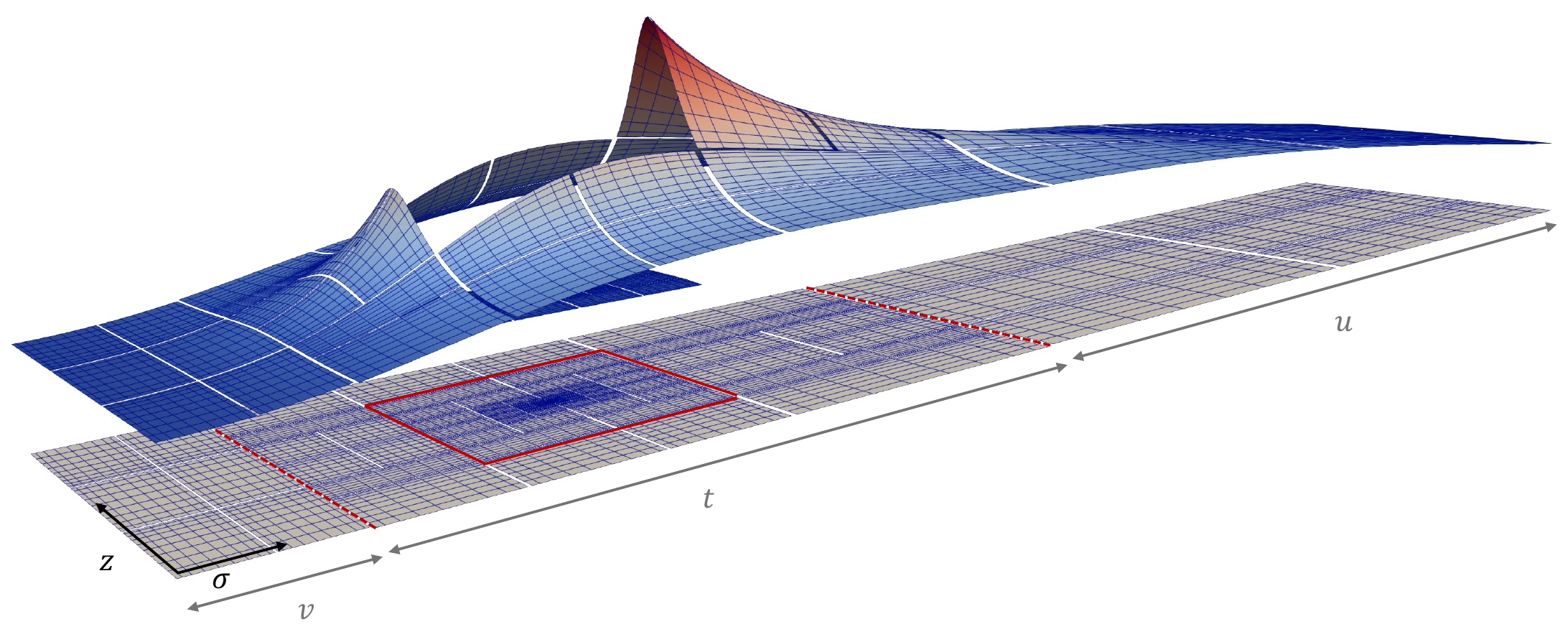}
    \caption{Computational domain and solution for $a/M=0.6$, $r_0/M=8$, $m=2$. The domain extends radially from the horizon $r_+$ to future null infinity, and in angle from $z=\cos\theta=-1$ to $1$. The bottom shows the DG mesh projected onto the $(\sigma,z)$ plane, with the red square marking the worldtube and red dashed lines marking the two transitions separating the $v$-, $t$-, and $u$-slicing regions. The surface shows the real part of
    $(\psi_{m})_{vv}=r\sin^{-2}\theta\;\bar h^{(m)}_{vv}$, except in the worldtube, where the residual field $(\psi^{\mathcal R}_m)_{vv}=(\psi_m)_{vv}-(\psi^{\mathcal P}_m)_{vv}$ 
    is plotted instead. Null slicing ($H=\pm 1$) in the $v$ and $u$ regions removes the oscillations of the field toward the horizon and future null infinity.}
    \label{fig:field3d}
\end{figure*}

\subsubsection{Domain decomposition}\label{sec:domain_decomposition}

Figure~\ref{fig:field3d} shows our decomposition of the computational domain. We divide the two-dimensional domain, with coordinates $q^i=(\sigma,z)$, into $5\times 3$ rectangular \emph{blocks}.
Radially, we have a horizon-penetrating wavezone ($v$\nobreakdash-slicing), a transition region ($t$-slicing), the worldtube region ($t$\nobreakdash-slicing) containing the particle, another transition region ($t$-slicing), and an outer wavezone ($u$\nobreakdash-slicing), which extends to future null infinity.
In the angular direction, the middle row of blocks contains the worldtube and the two outer rows extend to the poles at $z=\pm 1$. The block boundaries are input parameters that depend on the orbital configuration ($a$, $r_0$) but are the same for all $m$.
We choose the worldtube boundaries to balance two requirements: the worldtube must be large enough to avoid steep gradients in the solution near the particle, yet small enough to limit the cost of resolving small-scale features in the residual field. In addition, we center the worldtube on the particle. This keeps the particle on the corner shared by neighboring elements at every refinement level, rather than inside an element, where the limited smoothness of the field would spoil spectral convergence.
Each block is mapped to logical coordinates, $\xi^i=(\xi,\eta)\in[-1,1]^2$, by a linear map in $(\sigma,z)$.
Because the compactification is already built into $\sigma$, Eq.~\eqref{eq:sigma}, no additional nonlinear radial map is needed, unlike the $1/r$ map used in Ref.~\cite{Vu:2026ypc}.

\subsubsection{Mesh refinement}\label{sec:refinement}
\begin{figure}
    \centering
    \includegraphics[width=\columnwidth]{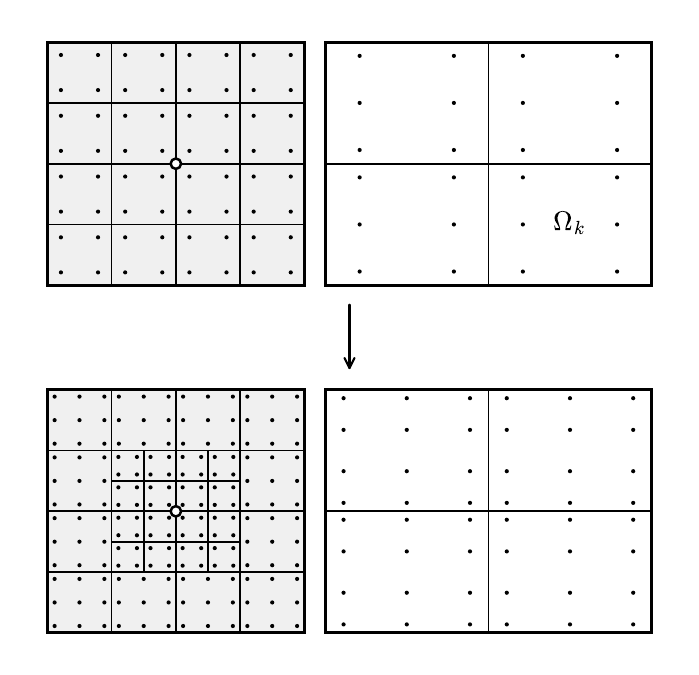}
    \caption{Schematic of two consecutive refinement steps (top to bottom). The gray block is the worldtube block containing the particle (open circle), and the white block is its radial neighbor. Thick lines are block boundaries, thin lines element boundaries, and dots Legendre--Gauss collocation points; $\Omega_k$ labels one element. Elements containing the particle are halved in both dimensions ($h$ refinement), subject to two-to-one balance, and all other elements gain one collocation point per dimension ($p$ refinement).}
    \label{fig:mesh}
\end{figure}

We split the blocks into smaller \emph{elements}, $\Omega_k$, by halving the logical cube along $\xi$, $\eta$, or both ($h$ refinement). In each refinement step, every element containing the particle is halved in both dimensions, so that the particle always remains on the corner shared by four elements. Throughout, we enforce a two-to-one balance: neighboring elements may differ by at most one refinement level in each dimension, so that, in particular, each element face borders at most two neighbors. Elements that would violate this condition are split as well.
Figure~\ref{fig:mesh} illustrates two consecutive refinement steps, with the worldtube block shaded gray and the particle marked by an open circle. Each of the four elements around the particle is split into four, so the number of elements grows only near the particle. The new elements differ by one refinement level from the elements surrounding them, consistent with the two-to-one balance, while the elements of the neighboring block are not split.

Within each element $\Omega_k$, we expand the fields in a polynomial basis whose order can be increased ($p$ refinement). To this end, we define a grid of $N_k=N_\xi \times N_\eta$ Legendre--Gauss collocation points, $\bm{\xi}_p=(\xi_{p_1}, \eta_{p_2})$, in each element, where the $\xi_{p_1}$ are the roots of the Legendre polynomial of degree $N_\xi$, and likewise for $\eta_{p_2}$, visible as grid lines in \cref{fig:field3d} and as points in \cref{fig:mesh}. Unlike Legendre--Gauss--Lobatto points, Legendre--Gauss points lie strictly inside the element, so no grid point sits on an element or block boundary.

On this grid, we expand fields in a Lagrange polynomial basis with respect to the collocation points $\bm{\xi}_p$, as detailed in \ccite{Vu:2026ypc}. In this \emph{nodal} picture, fields are represented by their value at the Legendre--Gauss collocation points, as opposed to the equivalent \emph{modal} picture, where they are instead represented by their spectral coefficients with respect to the Legendre basis that underpins the Legendre--Gauss grid.

The number of grid points, and hence the polynomial order, can be chosen independently in each element and dimension.
In each refinement step, we increase it by one in both dimensions in every element that is not split. In \cref{fig:mesh}, the elements away from the particle, including those of the neighboring block, gain one grid point per dimension from one step to the next, while the newly split elements around the particle keep their number of points. For smooth fields, the spectral expansion converges exponentially with the polynomial order. Moreover, since the fields are represented in a polynomial basis, the discrete solutions are regular by construction.

\subsubsection{DG residuals}\label{sec:DG_res}

To discretize the field equations, $\triangle_m(\psi_m)_{\mu\nu} = (S_m)_{\mu\nu}$, Eq.~\eqref{eq:first_order_flux_form}, we project the ten components of the equations onto the same polynomial basis that is used to expand fields in each element to obtain nodal DG residuals in strong form, as detailed in \ccite{Vu:2026ypc}.
The coefficients $\beta^{\lambda\rho}_{\mu\nu}$ and $\gamma^{i,\lambda\rho}_{\mu\nu}$ couple components of the field equations, acting effectively as $10\times 10$ matrices at each grid point. The coefficients $\dgFA_m^{ij}=\mathrm{diag}(\alpha_r,\alpha_z)$ act identically on all ten components of the flux, $(\dgF_{m}^{i})_{\mu\nu}\equiv\dgFA_m^{ij}\partial_j (\psi_{m})_{\mu\nu}$. Together, the coefficients $\alpha_r$, $\alpha_z$, $\beta^{\lambda\rho}_{\mu\nu}$, and $\gamma^{i,\lambda\rho}_{\mu\nu}$ define the first-order flux form that is detailed in \cite{Vu:2026ypc}.

We choose the internal penalty numerical flux of \ccite{Vu:2026ypc} with penalty parameter $C=1.5$ to couple neighboring elements through fluxes exchanged across their faces, applied to all ten components of the equations. The numerical flux also imposes jump conditions across the worldtube boundary, Eq.~\eqref{eq:worldtube_interface}, and across the interfaces between the $v$, $t$, and $u$ domains, Eqs.~\eqref{eq:jump_vt} and \eqref{eq:jump_tu}. It does so by modifying the exterior quantities, $(\psi_m^\mathrm{ext})_{\mu\nu}$ and $(n_i\dgF_m^{i,\mathrm{ext}})_{\mu\nu}$, after an element has received them from its neighbor, where $n_i$ is the unit normal to the element face. At the worldtube boundary, we replace
\begin{align}\label{eq:worldtube_jump}
    (\psi_m^\mathrm{ext})_{\mu\nu} &\to (\psi_m^\mathrm{ext})_{\mu\nu} \pm (\psi_m^\mathcal{P})_{\mu\nu} \\
    (n_i\dgF_m^{i,\mathrm{ext}})_{\mu\nu} &\to (n_i\dgF_m^{i,\mathrm{ext}})_{\mu\nu} \mp n_i\dgFA_m^{ij}\partial_j(\psi_m^\mathcal{P})_{\mu\nu}
\end{align}
where the upper signs apply to elements outside the worldtube and the lower signs to elements inside. The opposite sign in the flux arises because the received normal flux is defined with the face normal of the sending element. In the same way, at the $vtu$ domain interfaces we replace
\begin{equation}\label{eq:vtu_jump}
    (n_i\dgF_m^{i,\mathrm{ext}})_{\mu\nu} \to (n_i\dgF_m^{i,\mathrm{ext}})_{\mu\nu} - im\Omega\,\{\!\{(\psi_m)_{\mu\nu}\}\!\}
    \text{,}
\end{equation}
with $\{\!\{(\psi_m)_{\mu\nu}\}\!\}=\frac{1}{2} ((\psi_m^\mathrm{int})_{\mu\nu} + (\psi_m^\mathrm{ext})_{\mu\nu})$ the average of the field components across the interface. The sign change between the two sides of the interface is absorbed by the opposite orientation of the face normal $n_i$.

\begin{figure*}
    \centering
    \includegraphics[width=\textwidth]{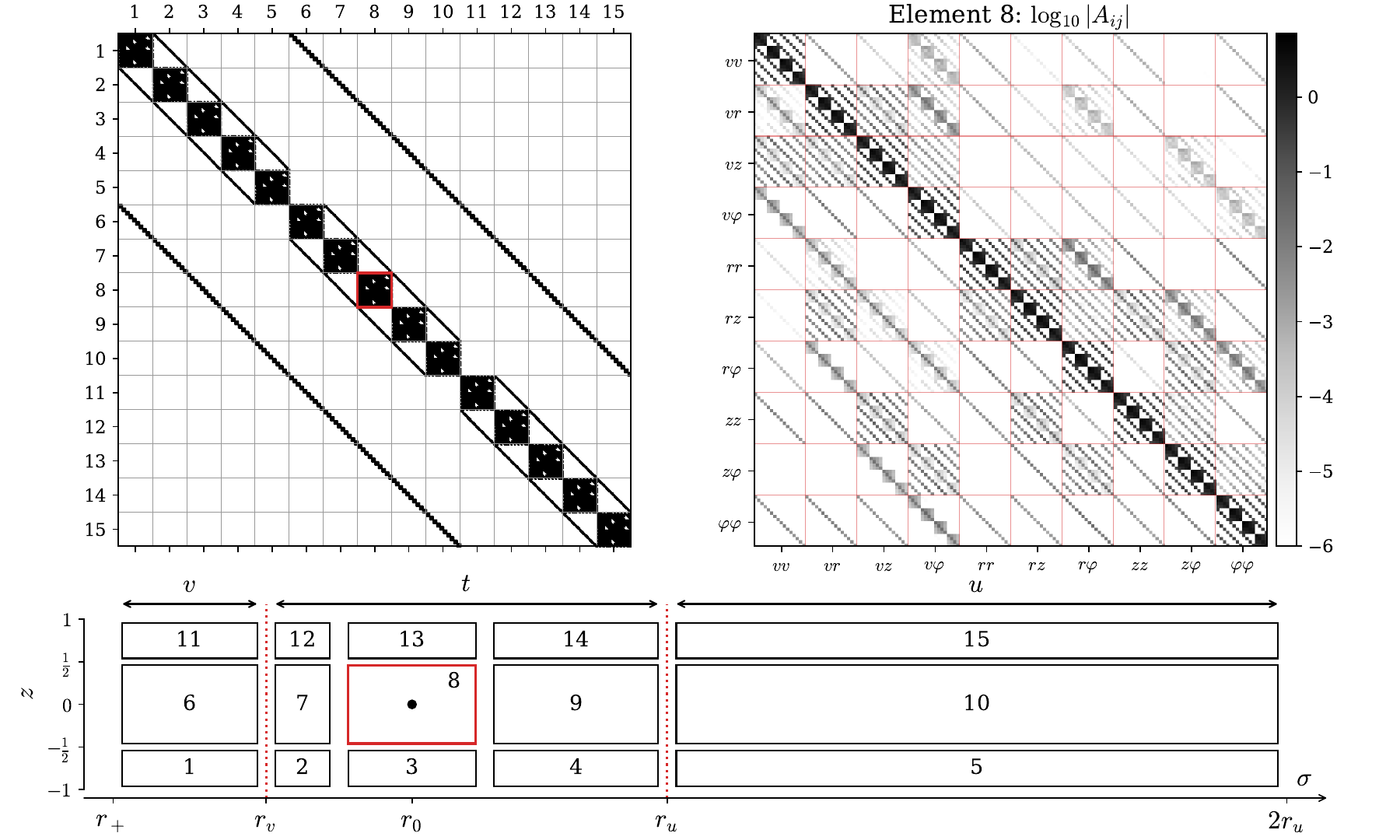}
    \caption{Matrix representation of the DG operator $\mathbf{A}$ for $a/M=0.6$, $r_0/M=10$, $m=2$, assembled on the coarsest mesh.
    Top left: nonzero entries of the full matrix, with rows and columns grouped by element.
    Top right: $\log_{10}|A_{ij}|$ of the diagonal block of element 8 (red square in the top-left panel), with rows and columns grouped by the components of $(\psi_m)_{\mu\nu}$.
    Bottom: numbering of the DG elements in the $(\sigma,z)$ domain, with the particle marked by the black dot. The red rectangle marks the worldtube boundary and the red dotted lines the $vtu$ transitions at $r_v$ and $r_u$.}
    \label{fig:dg_M_block}
\end{figure*}

\subsubsection{Matrix representation}

To turn the DG residuals into a discrete system, they are evaluated by Gauss quadrature on the Legendre--Gauss collocation points of each element~\cite{Vu:2021,Vu:2024,Vu:2026ypc}. The result is a set of $N_\text{DOF}$ coupled linear equations for the state vector $\mathbf{u}$, which is composed of the ten field components of $(\psi_m)_{\mu\nu}$ on all Legendre--Gauss collocation points across elements in the domain. These linear equations, schematically $\mathbf{A}\mathbf{u}=\mathbf{b}$, couple the field components and grid points through multiple effects:
\begin{itemize}
    \item The field equations couple components of $(\psi_m)_{\mu\nu}$ through the pointwise matrices $\beta^{\lambda\rho}_{\mu\nu}$ and $\gamma^{i,\lambda\rho}_{\mu\nu}$.
    \item Derivatives in the equations couple grid points within an element through dense spectral differentiation matrices per grid dimension.
    \item Numerical fluxes couple grid points across element faces, which include the jump conditions.
\end{itemize}
To illustrate this structure, Fig.~\ref{fig:dg_M_block} shows the operator $\mathbf{A}$ assembled explicitly on the coarsest mesh, where each of the $5\times3$ blocks consists of a single element $\Omega_k$ with $4\times4$ grid points. With ten components of $(\psi_m)_{\mu\nu}$ per grid point, the matrix has dimension $2400\times2400$.
Its sparsity pattern reflects the locality of the DG scheme: the diagonal blocks couple the degrees of freedom within one element, while the off-diagonal bands couple face-neighboring elements through the numerical flux, either in the radial direction (adjacent elements) or in the angular direction (elements offset by five). Elements that do not share a face, such as elements 5 and 6, are not coupled. The worldtube jumps, Eq.~\eqref{eq:worldtube_jump}, and the $vtu$ jumps, Eq.~\eqref{eq:vtu_jump}, enter through these same interface couplings. Within the diagonal block of the worldtube element, the diagonal sub-blocks, which carry the principal part of the operator acting on each component, dominate, while the couplings between different components are weaker by several orders of magnitude (right panel). In the solver, however, $\mathbf{A}$ is never assembled explicitly; it is only used to compute matrix-vector products. Therefore, only constituents such as the one-dimensional logical differentiation matrices are computed explicitly and cached globally in the simulation. This matrix-free scheme allows computations to scale effectively to high resolution, because no large matrices must be assembled and stored, but it requires iterative linear solvers, as discussed in the next section.
\begin{figure*}
    \centering
    \includegraphics[width=\textwidth]{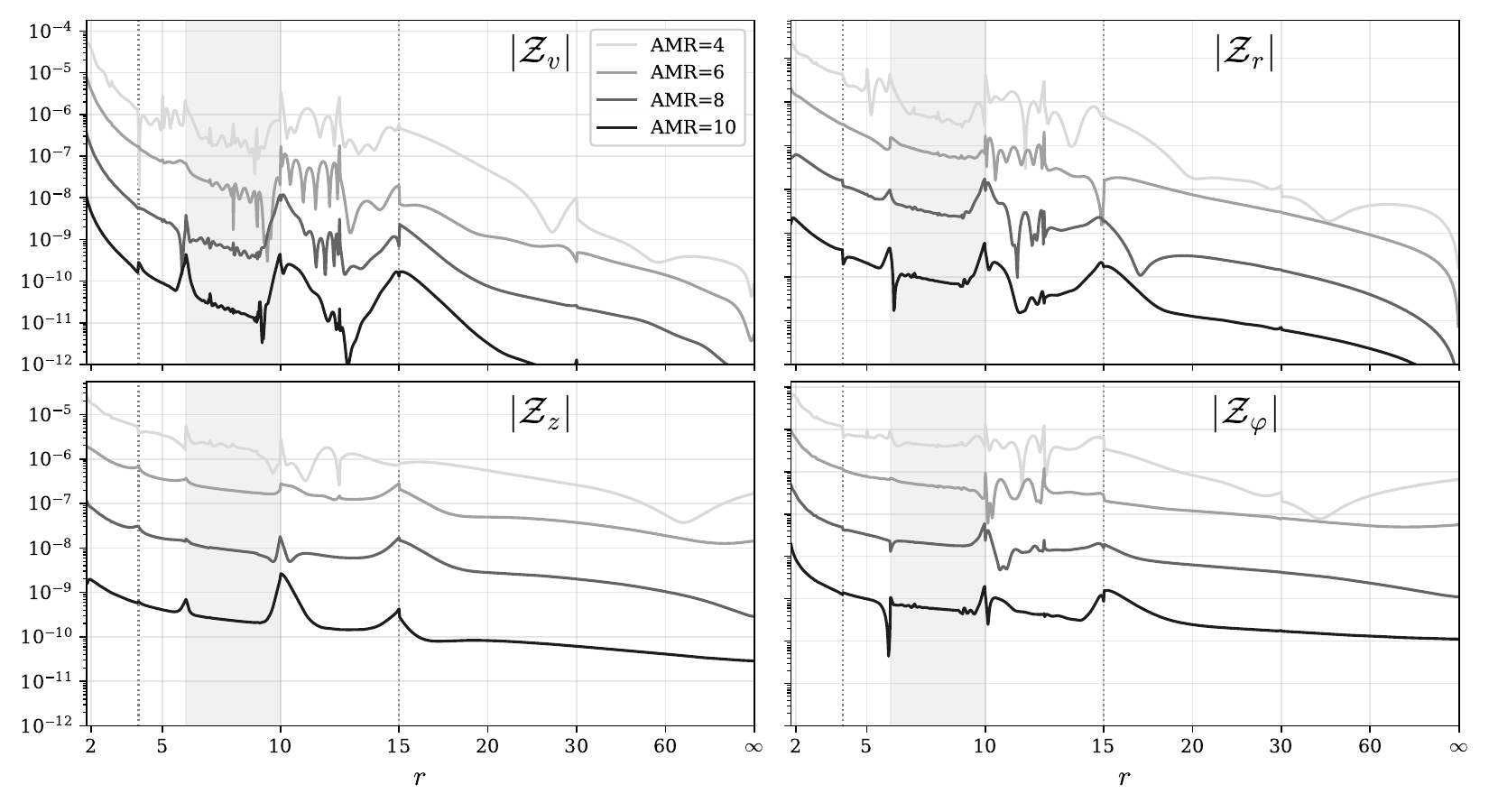}
    \caption{
    The Lorenz-gauge violation decreases exponentially with AMR level throughout the domain.
    Shown is $|\mathcal{Z}^{(m)}_{\alpha}|$, for $\alpha=v,r,z,\varphi$, for the $m=2$ mode with $a/M=0.6$ and $r_0/M=8$, along $\theta=5\pi/12$. 
    Dotted lines mark the $vtu$ transitions.
    Inside the worldtube (gray band) we show $\mathcal{Z}^{(m)}_{\alpha}[\psi^{\mathcal{R}}_m+\psi^{\mathcal{P}}_m]$, and outside it $\mathcal{Z}^{(m)}_{\alpha}[\psi_m]$.}
    \label{fig:LG_4_conditions_m2}
\end{figure*}
\subsection{Iterative linear solver}\label{sec:linsolv}

To solve the discretized elliptic problem, we must effectively invert the matrix $\mathbf{A}$ to obtain the state vector $\mathbf{u}=\mathbf{A}^{-1}\mathbf{b}$ that solves the discretized equations. Note that solving this discretized problem even to machine precision does not mean that the continuum problem is solved exactly; the error made by evaluating the fields and the equations on a finite-resolution grid remains as discretization error.

We employ a preconditioned iterative algorithm to solve the discretized linear problems fast, at high resolution, and without explicitly constructing the matrix $\mathbf{A}$. Our complex-valued GMRES algorithm with multigrid-Schwarz preconditioner is implemented in the open-source \spectre{} code~\cite{spectrecode} and detailed in \ccites{Vu:2026ypc,Vu_2022}. It constructs a multigrid hierarchy through successive AMR levels and runs one multigrid V-cycle per GMRES iteration. The multigrid hierarchy enables resolution-independent iteration counts by solving small-scale and large-scale modes in the solution simultaneously throughout the GMRES iterations.

On every multigrid level we run two passes of an additive Schwarz smoother, which constructs a subdomain problem centered on each DG element, solves them all in parallel, and combines the result in a weighted sum. The subdomain problems are built by restricting the full linear operator $\mathbf{A}$ to the grid points of each element and a few points of overlap into its face neighbors. These subdomain problems are small enough that we construct their matrix representations explicitly and invert them directly via LU decomposition. We do not use more sophisticated iterative subdomain solvers in this work, as described in \ccite{Vu:2026ypc}, but note this as a primary direction of future work to improve the efficiency of the numerical solver.
We also do not use the bottom solver described in \ccite{Vu:2026ypc} in this work, as the additive Schwarz smoother turned out to be more efficient at solving the larger ten-component gravitational problem than explicitly constructing bottom matrices.
We do, however, employ the complex shift of the Schwarz subdomain matrices described in \ccite{Vu:2026ypc} to improve preconditioning of our Helmholtz-type elliptic problems (see \cref{sec:results_flux}).

\section{Results}\label{sec:results}
In this section, we present numerical results that demonstrate the accuracy of our approach. We first verify that our solutions satisfy the Lorenz gauge condition (Sec.~\ref{sec:results_LG}). We then examine the structure of the $m$-mode metric perturbation, confirming that the solutions behave as expected: they approach constant values toward the radial and polar boundaries, show no oscillations owing to the hyperboloidal slicing, and reflect the equatorial symmetry of the problem. We also compare them directly with independent $m$-mode data (Sec.~\ref{sec:results_mmode}). Next, we turn to the redshift invariant, which we use to demonstrate the self-convergence of our scheme and to compare with reference values across a range of spins, finding a relative difference below $10^{-8}$ (Sec.~\ref{sec:results_redshift}). Finally, we compare the radiated energy flux toward infinity (Sec.~\ref{sec:results_flux}).
\begin{figure*}
    \centering
    \includegraphics[width=\textwidth]{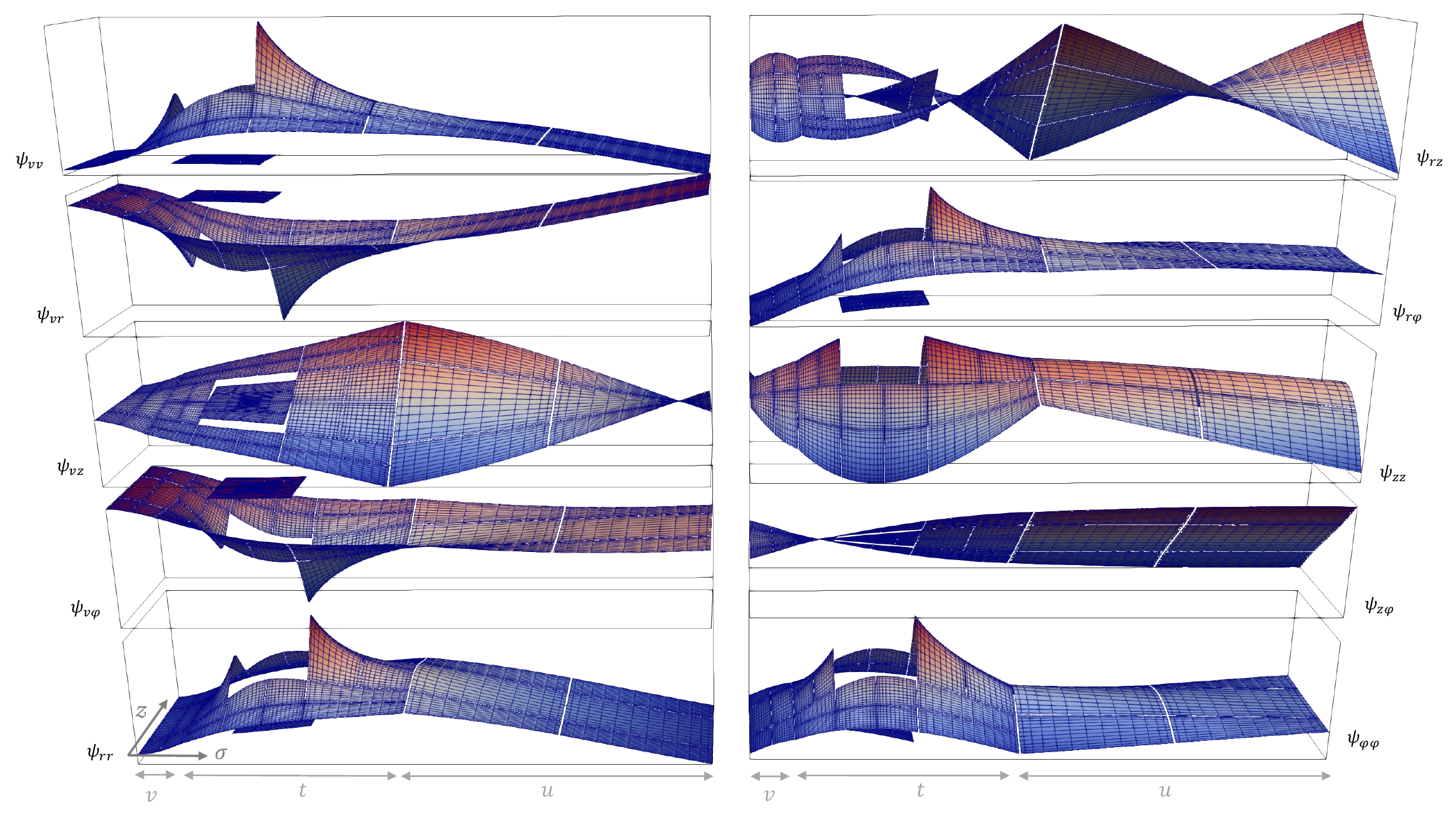}
    \caption{Real part of the ten components $(\psi_{m=2})_{\mu\nu}$ for $a/M = 0.6$ and $r_0/M = 8$. Each component is normalized
    by its own maximum. Surfaces are drawn over the $(\sigma,z)$ plane, with $\sigma$ the radial grid coordinate of Eq.~\eqref{eq:sigma}; the right edge, $\sigma=2r_u$, is future null infinity. Inside the worldtube we show the residual field $(\psi^{\mathcal{R}}_m)_{\mu\nu}=(\psi_m)_{\mu\nu}-(\psi^{\mathcal{P}}_m)_{\mu\nu}$, and outside it the retarded field $(\psi_m)_{\mu\nu}$.
    }
    \label{fig:field_m2}
\end{figure*}

\subsection{Lorenz gauge conditions}\label{sec:results_LG}
A solution of the Lorenz-gauge field equations satisfies the linearized Einstein equations only if the gauge condition $Z_\mu\equiv\nabla^\nu\bar h_{\mu\nu}=0$ is also satisfied. We do not directly impose this condition,\footnote{With our $\sim\text{sin}^{|m|}\theta$ factorization, Lorenz gauge violating solutions are irregular at the poles (according to the PDE coefficients themselves), so finding a unique regular solution does indirectly impose the Lorenz gauge conditions.} so its satisfaction is an independent check of the numerical solution. We expect the Lorenz gauge violation to decrease as the resolution increases. We evaluate $Z_\mu$ directly from the numerical solution, in the ingoing Kerr coordinates $(v,r,z,\varphi)$ in which $(\psi_m)_{\mu\nu}$ is defined. The $z$ component carries a coordinate pole at the axis, which we remove by plotting
\begin{align}\label{eq:LG_weighting}
\mathcal{Z}^{(m)}_{v}&=Z_{v}, &
\mathcal{Z}^{(m)}_{r}&=Z_{r}, \nonumber\\
\mathcal{Z}^{(m)}_{z}&=\left(1-z^2 \right)\,Z_{z}, &
\mathcal{Z}^{(m)}_{\varphi}&=Z_{\varphi}.
\end{align}
Figure~\ref{fig:LG_4_conditions_m2} shows these four components for $m=2$, $a/M=0.6$ and $r_0/M=8$ along the radial slice $\theta=5\pi/12$, which passes through the interior of the worldtube.
Outside the worldtube, $(\psi_m)_{\mu\nu}$ is the retarded field and we plot $\mathcal{Z}^{(m)}_\mu[(\psi_m)_{\mu\nu}]$. Inside, we solve for the residual field $(\psi^{\mathcal{R}}_m)_{\mu\nu}=(\psi_m)_{\mu\nu}-(\psi^{\mathcal{P}}_m)_{\mu\nu}$ (Eq.~\eqref{eq:worldtube_interface}), which does not satisfy the gauge condition by itself, since the truncated puncture $(\psi^{\mathcal{P}}_m)_{\mu\nu}$ does not. We therefore add the analytic puncture back in and plot the gauge violation of the retarded field, $\mathcal{Z}^{(m)}_\mu[(\psi^{\mathcal{R}}_m)_{\mu\nu}]+\mathcal{Z}^{(m)}_\mu[(\psi^{\mathcal{P}}_m)_{\mu\nu}]$.
Throughout the domain, from the horizon to future null infinity, the violation decreases with each AMR level, by roughly an order of magnitude per two levels.
The violation shows small kinks at the worldtube boundaries and at the $vtu$ transitions. They reflect how accurately the numerical flux enforces the jump conditions, Eqs.~\eqref{eq:worldtube_jump} and~\eqref{eq:vtu_jump}, and they decrease with resolution like the violation elsewhere. Similar features in $10M\lesssim r\lesssim15M$ arise at element boundaries inside that block and decrease with resolution as well.

\begin{figure*}
    \centering
    \includegraphics[width=\textwidth]{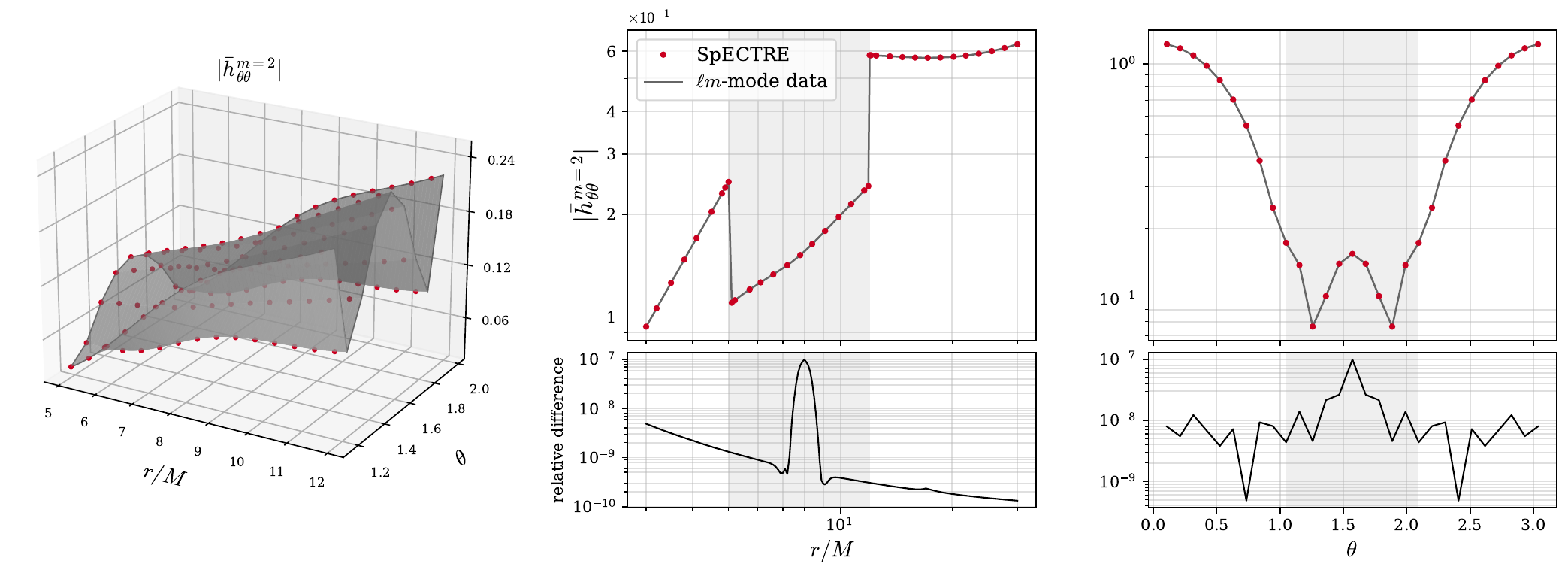}
    \caption{Agreement of our $m=2$ field with the independent Lorenz-gauge calculation of Ref.~\cite{Leather:2026gnc} to within $10^{-7}$.
    Shown is $\lvert\bar h^{(m=2)}_{\theta\theta}\rvert$ for $a/M=0.6$ and $r_0/M=8$; inside the worldtube (gray band) we compare residual fields, and outside it retarded fields.
    Left: residual field inside the worldtube over $(r,\theta)$, with the $\ell m$-mode data as a gray surface and our solution as red dots.
    Middle and right: cuts through the particle at $\theta=\pi/2$ and $r=r_0$, respectively, comparing the fields (top) and showing their relative difference (bottom).
}
    \label{fig:lm_to_m}
\end{figure*}
\subsection{\texorpdfstring{$m$}{m}-mode metric perturbation}\label{sec:results_mmode}
Figure~\ref{fig:field_m2} shows the ten components $(\psi_m)_{\mu\nu}$ of the $m=2$ mode for $a/M=0.6$ and $r_0/M=8$ over the full computational domain from the horizon to future null infinity.
Owing to the $vtu$ slicing and the compactification, the solutions show no oscillations toward the horizon or $\mathcal{I}^+$, and the prefactors in Eq.~\eqref{eq:eq_MP} ensure that $(\psi_m)_{\mu\nu}$ approaches finite values at the radial boundaries and at the poles $z=\pm1$.
In our worldtube scheme, the variable we solve for differs across the worldtube boundary: inside the worldtube we solve for the residual field $(\psi^{\mathcal{R}}_{m})_{\mu\nu}=(\psi_m)_{\mu\nu}-(\psi^{\mathcal{P}}_{m})_{\mu\nu}$, and outside it for the retarded field $(\psi_m)_{\mu\nu}$. The solutions therefore jump across the worldtube boundary by the puncture $(\psi^{\mathcal{P}}_{m})_{\mu\nu}$~\eqref{eq:worldtube_interface}.
At the $vtu$ transitions, by contrast, $(\psi_m)_{\mu\nu}$ is continuous but its radial derivative jumps, producing the kinks visible in Fig.~\ref{fig:field_m2}, as dictated by the jump condition~\eqref{eq:vtu_jump}.
Finally, the solutions reflect the symmetry of the problem under reflection through the equatorial plane: the components  
$(\psi_{m})_{vz}$, $(\psi_{m})_{rz}$ and $(\psi_{m})_{z\varphi}$, 
which carry a single $z$ index, are antisymmetric about $z=0$, while all other components are symmetric.

Beyond these qualitative features, we validate the field itself by comparing our solution with an independent Lorenz-gauge calculation~\cite{Leather:2026gnc}.
Unlike our direct solution of the coupled $(r,\theta)$ equations for each $m$-mode, this calculation reconstructs the metric perturbation from solutions of separable Teukolsky-type equations, solved on hyperboloidal slices for each spin-weighted spheroidal $(\ell,m)$ mode, and obtains each $m$-mode by summing over $\ell$.
Since both calculations use the same puncture, not only the retarded fields but also the residual fields should agree. This allows us to test the accuracy of our solution near the particle. 
The resulting data, which we refer to as the $\ell m$-mode data, are given as components of $h^{(m)}_{\mu\nu}$ in the Kinnersley tetrad, in Boyer--Lindquist coordinates.
To compare with our solution, we convert them to the trace-reversed field in ingoing Eddington--Finkelstein coordinates.

Figure~\ref{fig:lm_to_m} compares the trace-reversed component $|\bar h^{(m)}_{\theta\theta}|$ of the $m=2$ mode for $a/M=0.6$ and $r_0/M=8$ with the $\ell m$-mode data. Both calculations use the same worldtube, so inside it we compare residual fields and outside it retarded fields.
The left panel shows the $\ell m$-mode data as a gray surface over $(r,\theta)$, with our solution overlaid (red dots). The middle and right panels show cuts through the particle along $\theta=\pi/2$ and $r=r_0$, respectively. The top row compares the fields, and the bottom row shows the relative difference $|\bar h_{\rm SpECTRE}-\bar h_{\rm \ell m\text{-mode data}}|/|\bar h_{\rm SpECTRE}|$.
The relative difference stays at or below $10^{-7}$ everywhere, including inside the worldtube (gray band), and is largest at the particle.
Since the two calculations solve different equations with different numerical methods and share only the puncture, this provides a stringent check of our solution, even near the particle.
\begin{figure*}
    \centering
    \includegraphics[width=\textwidth]{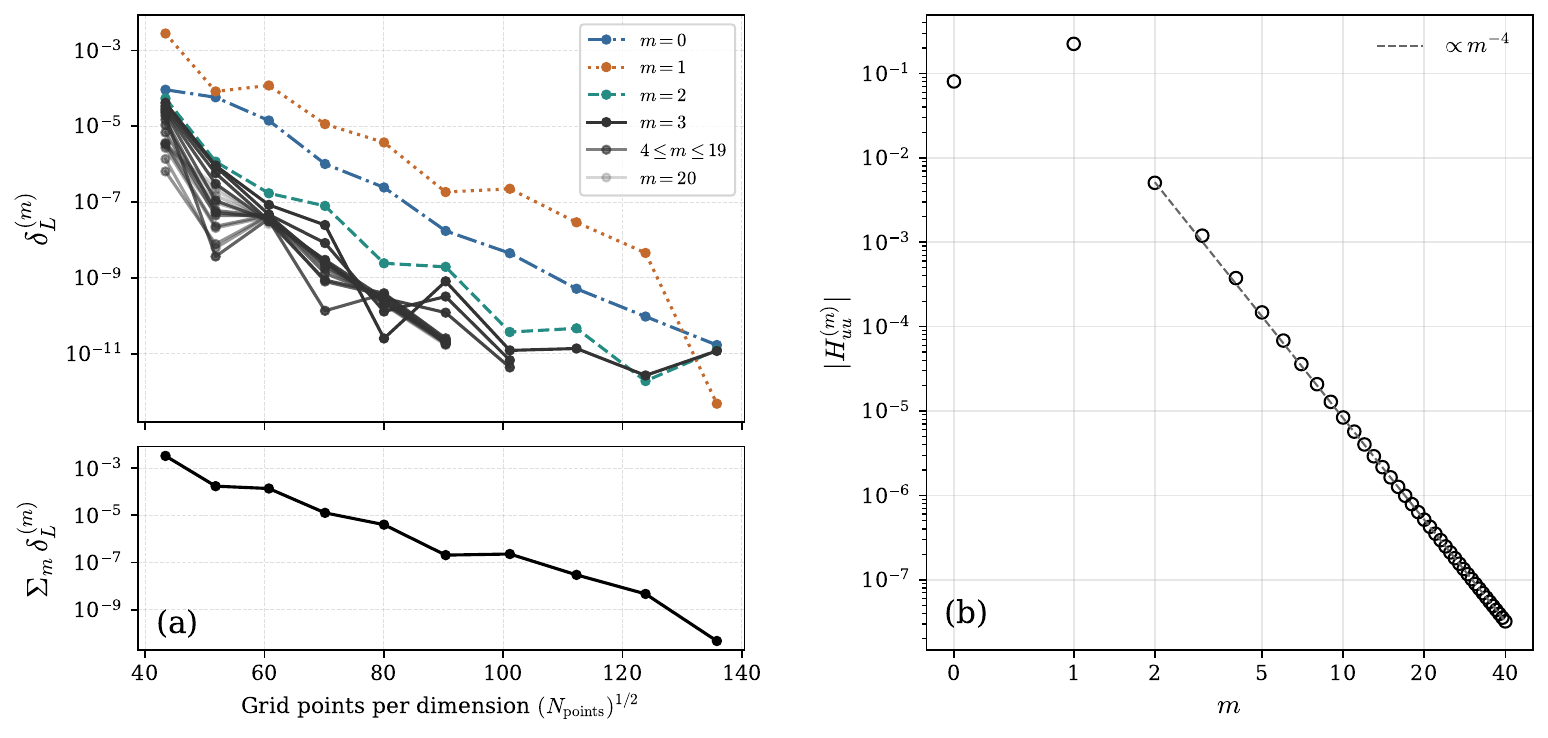}
    \caption{(a)~Exponential convergence with resolution.
    Top: relative change $\delta^{(m)}_L$ [Eq.~\eqref{eq:self_convergence}] of each $m$-mode contribution between successive AMR levels, for $0\leq m\leq 20$.
    Bottom: $\Sigma_{m\leq40}\,\delta^{(m)}_L$, an upper bound on the relative change of $\Delta z$; modes that have reached their final level contribute zero.
    (b)~Power-law falloff of the mode contributions $|H^{(m)}_{uu}|$; the dashed line indicates $m^{-4}$. Both panels are for $a/M=0.6$, $r_0/M=8$.}
    \label{fig:redshift_convergence}
\end{figure*}
\subsection{Redshift invariant}\label{sec:results_redshift}

Next, we test our solution in the vicinity of the particle, where we solve for the residual field. To probe the field at the particle and the conservative sector of the self-force, we compute Detweiler's redshift invariant~\cite{Detweiler:2008ft}. In this section, we use it to demonstrate the convergence of our solver and to compare with reference values across a range of spins.

Under the self-force, the secondary deviates from a geodesic of the background Kerr spacetime. Its motion is nonetheless geodesic in the effective metric $\tilde g_{\mu\nu}=g_{\mu\nu}+\epsilon h^{\calR}_{\mu\nu}$, where $h^{\calR}_{\mu\nu}$ is the Detweiler--Whiting regular field, which is smooth on the worldline. For circular orbits, the effective spacetime is invariant along the helical Killing vector $k^\mu=\xi^\mu_{(t)}+\Omega\,\xi^\mu_{(\phi)}$, which is tangent to the orbit. The four-velocity with respect to the proper time $\tilde\tau$ of $\tilde g_{\mu\nu}$ is therefore $\tilde u^\mu=\tilde u^t k^\mu$, and the redshift is defined as
\begin{align}
    z \defeq \frac{1}{\tilde u^t} = \frac{d\tilde\tau}{dt}.
\end{align}
Considered as a function of $\Omega$, $z$ is invariant under gauge transformations that preserve the helical symmetry~\cite{Detweiler:2008ft}; see Ref.~\cite{Pound:2014koa} for its extension to second order. It has become a standard benchmark for conservative self-force calculations, including for circular orbits in Kerr spacetime.
Expanding at fixed $\Omega$, $z=z_0+\epsilon\,\Delta z+\mathcal{O}(\epsilon^2)$, the zeroth-order piece is the geodesic value $z_0=1/u^t$, with $u^t$ given in Eq.~\eqref{eq:ut}. Normalizing $\tilde u^\mu$ with the effective metric, $\tilde g_{\mu\nu}\tilde u^\mu\tilde u^\nu=-1$, yields the first-order correction
\begin{align}\label{eq:redshift_def}
    \Delta z = -\frac{1}{2}\, z_0\, h^{\calR}_{uu},
    \qquad
    h^{\calR}_{uu} \defeq h^{\calR}_{\mu\nu}\,u^\mu u^\nu ,
\end{align}
evaluated on the worldline. Note that $h^\calR_{uu}$ is invariant for circular orbits~\cite{Akcay:2015pza}.
Inside the worldtube, we solve for the residual field $\bar h^{\mathcal{R}}_{\mu\nu}=\bar h_{\mu\nu}-\bar h^{\mathcal{P}}_{\mu\nu}$.
We undo the trace reversal, $h^{\mathcal{R}}_{\mu\nu}=\bar h^{\mathcal{R}}_{\mu\nu}-\frac{1}{2}g_{\mu\nu}\,g^{\alpha\beta}\bar h^{\mathcal{R}}_{\alpha\beta}$, and contract with $u^\mu$ to obtain the $m$-mode contributions at the particle, $H^{(m)}_{uu}$, such that $h^R_{uu}=\sum_m H^{(m)}_{uu}$.

Computing the redshift in Lorenz gauge requires one further step. 
A stationary Lorenz-gauge perturbation cannot simultaneously be regular at the horizon, have $h_{tt} \to 0$ at large radius (so that $t$ remains the proper time of distant static observers), and leave the black hole inside the orbit with its background parameters.
This was shown in Schwarzschild spacetime in Refs.~\cite{Dolan_2013, Berndtson:2007gsc}, and one finds the same in Kerr spacetime.
The solution found by Berndtson in Ref.~\cite{Berndtson:2007gsc} is regular at the horizon and preserves intuitive features of the $t$ coordinate, but it shifts the mass of the larger black hole.
We adopt Berndtson-type solutions by imposing regularity at the horizon and, through the radial prefactors of Appendix~\ref{sec:prefactor}, the appropriate $1/r$ decay at large radius. In Kerr spacetime both the mass and the angular momentum $J = aM$ are shifted.
The primary black hole has mass and spin
\begin{align}
    M_\bullet = M + \epsilon\,\delta M, \qquad J_\bullet = J + \epsilon\,\delta J .
\end{align}
A comparison of gauge-invariant quantities between different calculations must be
made at fixed $\Omega$ and for the same black-hole parameters. 
Evaluating Eq.~\eqref{eq:redshift_def} with our $h^\calR_{uu}$ gives $\Delta z^{\rm Bernd}$. Because our solution is of Berndtson type, $z_0(\Omega; M, J) + \epsilon\,\Delta z^{\rm Bernd}$ is the redshift of an orbit around a black hole with parameters $(M_\bullet, J_\bullet)$. Re-expanding about the background parameters $(M, J)$ gives the first-order redshift
\begin{align}
\Delta z = \Delta z^{\rm Bernd}
- \frac{\partial z_0}{\partial M}\bigg|_{\Omega,J} \delta M
- \frac{\partial z_0}{\partial J}\bigg|_{\Omega,M} \delta J .
\label{eq:redshift_correction}
\end{align}
Appendix~\ref{sec:mass_correction} describes how we obtain $\delta M$ and $\delta J$ from the $m = 0$ mode and gives the derivatives of $z_0$.

\begin{figure}
    \centering
    \includegraphics[width=\columnwidth]{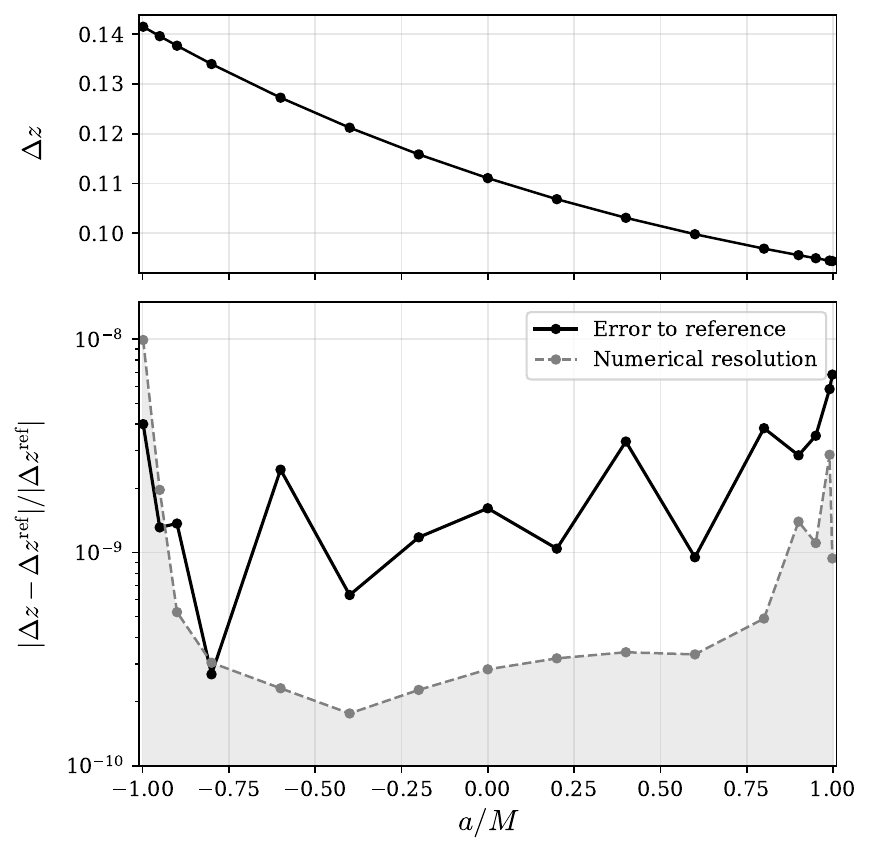}
    \caption{The redshift agrees with the reference values~\cite{Nasipak:2025tby,Nasipak_pybhpt_2026} to better than $10^{-8}$ for all spins, including the near-extremal $a/M=\pm0.998$, for $r_0/M=8$.
    Top: first-order redshift correction $\Delta z$ as a function of spin.
    Bottom: relative difference between our results, including the fitted tail $m>40$, and the reference.
    The gray region shows our estimated numerical errors.}
    \label{fig:redshift}
\end{figure}

Before comparing with reference values, we assess the convergence of our solution.
We monitor the change in each mode contribution between successive AMR levels $L$,
normalized by the total $\Delta z$,
\begin{align}\label{eq:self_convergence}
    \delta^{(m)}_L \defeq \frac{z_0}{2}\,\frac{\big|H^{(m)}_{uu,L}-H^{(m)}_{uu,L-1}\big|}{|\Delta z|}.
\end{align}
The top of Fig.~\ref{fig:redshift_convergence}(a) shows $\delta^{(m)}_L$ for $a/M=0.6$ and $r_0/M=8$ as a function of the grid points per dimension.
For the low modes, $m\leq 3$, $\delta^{(m)}_L$ decreases approximately exponentially in $N_\mathrm{points}^{1/2}$ and reaches $\sim10^{-11}$ at the finest level, $L=10$.
The mode $m=1$ converges most slowly and dominates $\sum_m\delta^{(m)}_L$ at nearly every level.
The higher modes converge faster, reaching $\delta^{(m)}_L\lesssim10^{-10}$ by $N^{1/2}_{\rm points}\approx90$. 
We therefore stop the AMR for these modes early, at $L=7$ for $4\leq m\leq6$ and at $L=6$ for $m\geq7$, which substantially reduces the cost.
The bottom of Fig.~\ref{fig:redshift_convergence}(a) shows $\sum_{m=0}^{40}\delta^{(m)}_L$, where modes that have reached their final level $L^{(m)}_{\max}$ contribute $\delta^{(m)}_L=0$ thereafter.
It decreases to $\sim5\times10^{-11}$ at $L=10$.
Summing the final changes $\delta^{(m)}_{L}$ of all modes, each taken at its own finest level, gives $\sim4\times10^{-10}$, which we adopt as our numerical error in Fig.~\ref{fig:redshift}.

Figure~\ref{fig:redshift_convergence}(b) shows $|H^{(m)}_{uu}|$  for the same configuration, up to $m=40$.
For $m\gtrsim10$, the modes follow a power law, $H^{(m)}_{uu}\propto m^{-4}$. This tail does not come from the regular field itself, which is smooth and whose $m$-modes would already have converged by these $m$. Instead, it arises from truncating the puncture expansion at finite order.
The error from truncating the mode sum at $m_{\max}$ therefore decreases only as $m_{\max}^{-3}$: at $m_{\max}=40$ it amounts to a relative error of $\approx2\times10^{-6}$ in $\Delta z$, far above our numerical resolution.
We therefore estimate the contribution of the modes $m>m_{\max}$ by fitting
\begin{align}
H^{(m)}_{uu} \simeq A\,m^{-4} + B\,m^{-6} + C\,m^{-8}
\end{align}
over $20\leq m\leq 40$. The powers are fixed, and the coefficients $A$, $B$, and $C$ are determined by linear least squares. We then sum the fitted model over $m>m_{\max}$ analytically using the Hurwitz zeta function.

The top of Fig.~\ref{fig:redshift} shows the resulting $\Delta z$ as a function of spin ($a/M\in[-0.998,0.998]$). The bottom shows its relative difference from the reference values computed with
\pybhpt{}~\cite{Nasipak_pybhpt_2026}, which obtains the redshift from the metric perturbation reconstructed from the Weyl scalars $\psi_0$ and $\psi_4$, truncating the sum over $m$ once successive terms fall below a prescribed threshold (see Ref.~\cite{Nasipak:2025tby} for details).
With the large-$m$ tail included, the relative difference is $(0.3\text{--}7)\times10^{-9}$ for all spins, well within the uncertainty of the reference values ($3\times10^{-8}$--$3\times10^{-7}$). 
For $a/M \lesssim -0.65$, the orbit at $r_0=8M$ lies inside the innermost stable circular orbit (ISCO), and the agreement is equally good in this region.
The gray region indicates our numerical error, estimated from the change in each $H^{(m)}_{uu}$ between the two finest AMR levels, relative to $\Delta z$. 
The numerical error grows toward high spins, spreading over many high-$m$ modes rather than dominated by any single one. The convergence of these modes at high spin is sensitive to the domain layout, as discussed at the end of Sec.~\ref{sec:results_flux}. Nevertheless, our redshift agrees with the reference to better than $10^{-8}$ for all spins, including the near-extremal $a/M=\pm0.998$.
Table~\ref{tab:redshift_comparison} lists the values for selected spins.
\begin{table}[t]
    \centering
    \begin{tabular}{cccc}
        \hline\hline
        $a/M$ & This work & Reference & Rel.\ difference \\
        \hline
        $-0.998$ & $0.141484828(1)$ & $0.14148483(4)$ & $4.0\times10^{-9}$ \\
        $-0.8$ & $0.13399014563(4)$ & $0.13399015(2)$ & $2.7\times10^{-10}$ \\
        $-0.6$ & $0.12723224130(3)$ & $0.12723224(2)$ & $2.4\times10^{-9}$ \\
        $+0.0$ & $0.11107483971(3)$ & $0.111074840(6)$ & $1.6\times10^{-9}$ \\
        $+0.6$ & $0.09980690304(3)$ & $0.099806903(3)$ & $1.1\times10^{-9}$ \\
        $+0.8$ & $0.09691323996(5)$ & $0.096913240(3)$ & $3.8\times10^{-9}$ \\
        $+0.998$ & $0.09441078312(9)$ & $0.094410782(6)$ & $6.8\times10^{-9}$ \\
        \hline\hline
    \end{tabular}
    \caption{First-order redshift correction $\Delta z$ at $r_0/M=8$ for selected spins. Our values include the modes $m\leq40$ and the fitted tail $m>40$.
    The reference values are from Refs.~\cite{Nasipak:2025tby,Nasipak_pybhpt_2026}.
    Numbers in parentheses give the uncertainty in the last digit.
    The relative difference is $\lesssim10^{-8}$ for all spins, below the uncertainty of the reference values.
    }
    \label{tab:redshift_comparison}
\end{table}

\subsection{Comparison of energy flux}\label{sec:results_flux}

\begin{figure}
    \centering
    \includegraphics[width=\columnwidth]{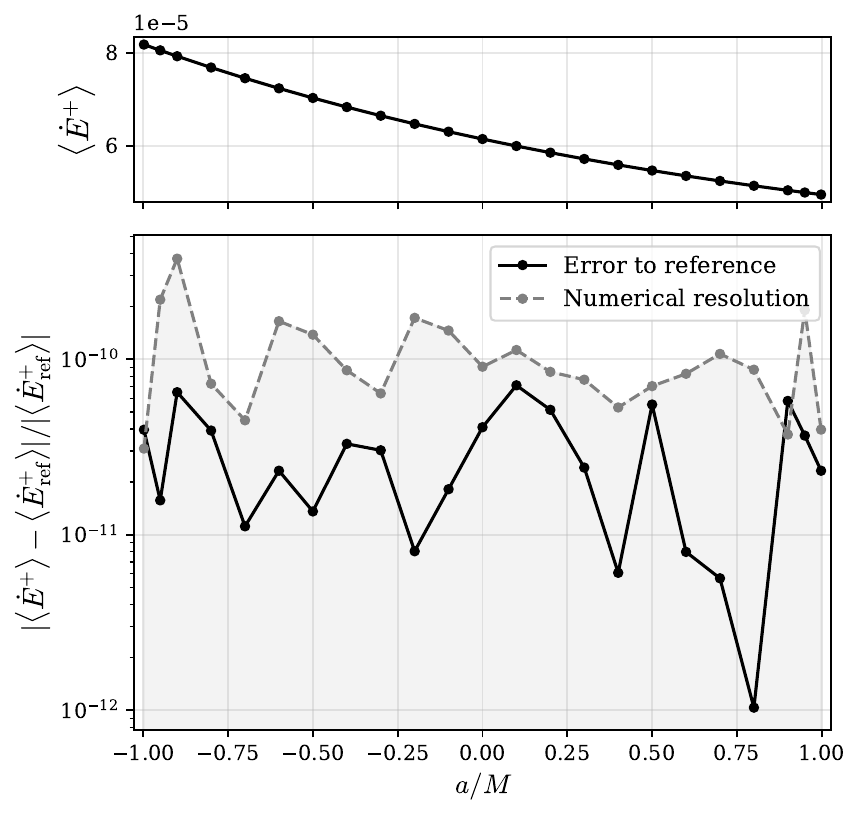} 
    \caption{The energy flux agrees with the reference to better than $10^{-10}$ for $a/M\in[-0.998,0.998]$.
    Top: outgoing energy flux $\langle\dot E\rangle$ at $r_0/M=10$, truncated at $m=20$. Bottom: relative difference compared to the \pybhpt{}~\cite{Nasipak_pybhpt_2026} (solid), and estimated numerical resolution error, given by the relative difference between the two highest AMR levels (dashed). }
    \label{fig:flux_spin}
\end{figure}

The energy flux radiated to future null infinity drives the dissipative evolution of the orbit, and is therefore a central ingredient of self-force applications at both first- and second-order. The energy flux follows from the Isaacson effective stress-energy tensor~\cite{PhysRev.166.1272}, which is quadratic in the first-order perturbation $h_{\mu\nu}$. Integrating it over a large sphere and averaging over an orbital period gives
\begin{align}
\langle\dot{E}^{+}\rangle&=\sum^{\infty}_{m=1}2\langle\dot{E}^{+}_{m}\rangle\\
\langle\dot{E}^{+}_{m}\rangle&=\frac{m^{2}\Omega^{2}}{16}\int^{\pi}_{0}\Big(|C^{(m)}_{\theta\theta}|^{2}+2|\csc\theta\,C^{(m)}_{\theta\varphi}|^{2}\nonumber\\
&\qquad\qquad+|\csc^{2}\theta\,C^{(m)}_{\varphi\varphi}|^{2}\Big)\sin\theta\,d\theta \, ,
\end{align}
where shear components are defined as
\begin{align}
C^{(m)}_{\theta\theta}=&\frac{1}{2}\lim_{r\rightarrow\infty}\left(\frac{\sin^{2}\theta}{r}\bar{h}_{zz}^{(m)}-\frac{1}{r\sin^{2}\theta}\bar{h}_{\varphi\varphi}^{(m)}\right)\, ,
\\\csc\theta\:C^{(m)}_{\theta\varphi}&=-\lim_{r\rightarrow\infty}\frac{1}{r}\bar{h}_{z\varphi}^{(m)} \, ,
\\\csc^{2}\theta\;C^{(m)}_{\varphi\varphi}&=\frac{1}{2}\lim_{r\rightarrow\infty}\left(\frac{1}{r\sin^{2}\theta}\bar{h}^{(m)}_{\varphi\varphi}-\frac{\sin^{2}\theta}{r}\bar{h}_{zz}^{(m)}\right) \, . \qquad
\end{align}
We compare the flux at infinity with reference values computed with \pybhpt{}~\cite{Nasipak_pybhpt_2026}, which solves the $s=-2$ Teukolsky equation in spin-weighted spheroidal $(\ell,m)$ modes. For each $m$, the reference sums the $\ell$-mode contributions until two consecutive contributions fall below $10^{-12}$ relative to the running sum. For both our data and the reference, we truncate the sum over $m$ at $m=20$.
Figure~\ref{fig:flux_spin} shows the total flux and its relative difference from the reference for orbits at $r_0/M=10$, for spins $-0.998\leq a/M\leq 0.998$. We also show an estimate of the numerical error, given by the relative difference between the two highest AMR levels.
Across all spins, the relative difference from the reference ranges from $\sim10^{-12}$ to $\sim7\times10^{-11}$. Overall, it is comparable to or below our resolution estimate, so our results agree with the reference to within our numerical error.

\begin{figure}
    \centering
    \includegraphics[width=\columnwidth]{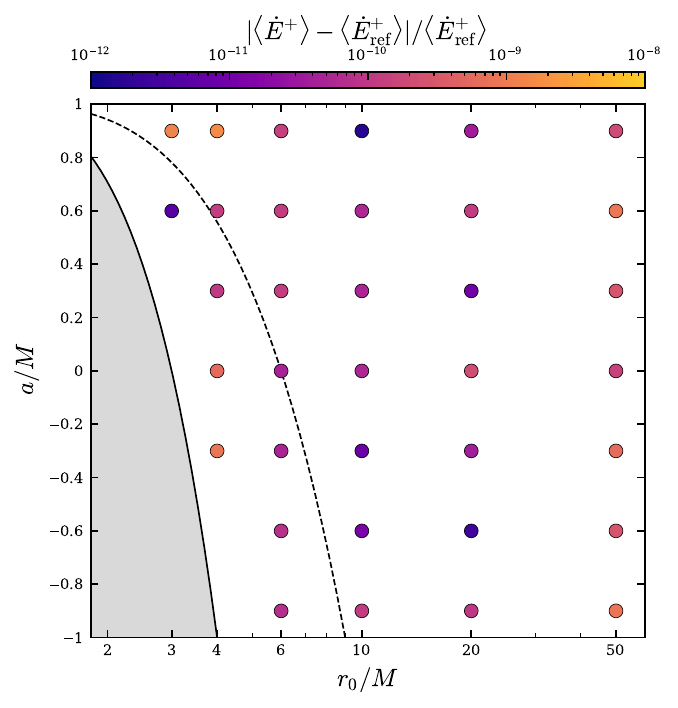} 
    \caption{Accuracy of the energy flux across the $(r_0,a)$ parameter space. Shown is the relative difference from the reference, with both truncated at $m=20$. The gray region lies inside the light ring, and the dashed line marks the ISCO; configurations inside or close to the light ring are omitted.}
    \label{fig:flux_across_params}
\end{figure}

To assess the accuracy of our method beyond a single orbital radius, Fig.~\ref{fig:flux_across_params} shows the relative difference between the outgoing energy flux and the reference across the $(r_0,a)$ parameter space, for $a/M\in\{0,\pm0.3,\pm0.6,\pm0.9\}$ and orbital radii up to $r_0/M=50$. For all configurations considered, the relative difference is below $2\times10^{-9}$.
While extending the calculation across this parameter space, we found that the convergence of our solutions is sensitive to the layout of the computational domain, in particular to the size of the worldtube and the locations of the $vtu$ slicing transitions.
The relative difference shows no systematic dependence on the spin, and varies irregularly between neighboring configurations, from $\sim10^{-11}$ to $\sim10^{-9}$. It is smallest at intermediate radii, $r_0/M\simeq10$--$20$, and somewhat larger close to the light ring and at $r_0/M=50$.

More importantly, we found that a small complex shift in the preconditioner is crucial for higher $m$-modes in the strong-field regime. Without it, numerical noise can excite spurious oscillations that render standard preconditioners ineffective, a known difficulty for Helmholtz-type problems~\cite{Erlangga2006,vanGijzen2007}. In our previous scalar-field work~\cite{Vu:2026ypc}, we added a small complex shift of order $m^2\Omega^2$ to the subdomain matrices to damp such oscillations in the Schwarz smoother, but found it to be unnecessary once hyperboloidal slicing was introduced, since the solution then loses its global wave-like behavior. In the gravitational case considered here, in contrast, we find that the complex shift remains crucial for higher $m$-modes in the strong field, even with hyperboloidal slicing. Moreover, a single prescription for the shift does not work across all modes. A shift proportional to $m^2\Omega^2$ works well for low and intermediate $m$, but at higher $m$ it grows too large and stalls the solver, so a substantially smaller value is required.
A systematic understanding of these sensitivities, and a generalization of the domain layout and complex shift to the full parameter space, is left for future work.
We stress, however, that even without such systematic optimization, the relative difference from the reference is below $2\times10^{-9}$ for all configurations considered here. 
Together with the agreement of the redshift, which tests the field at the particle (Sec.~\ref{sec:results_redshift}), this shows that our method provides metric perturbations accurate enough to serve as input for second-order self-force calculations.

\subsection{Computational cost}\label{sec:computational_cost}
\begin{figure}
    \centering
    \includegraphics[width=\columnwidth]{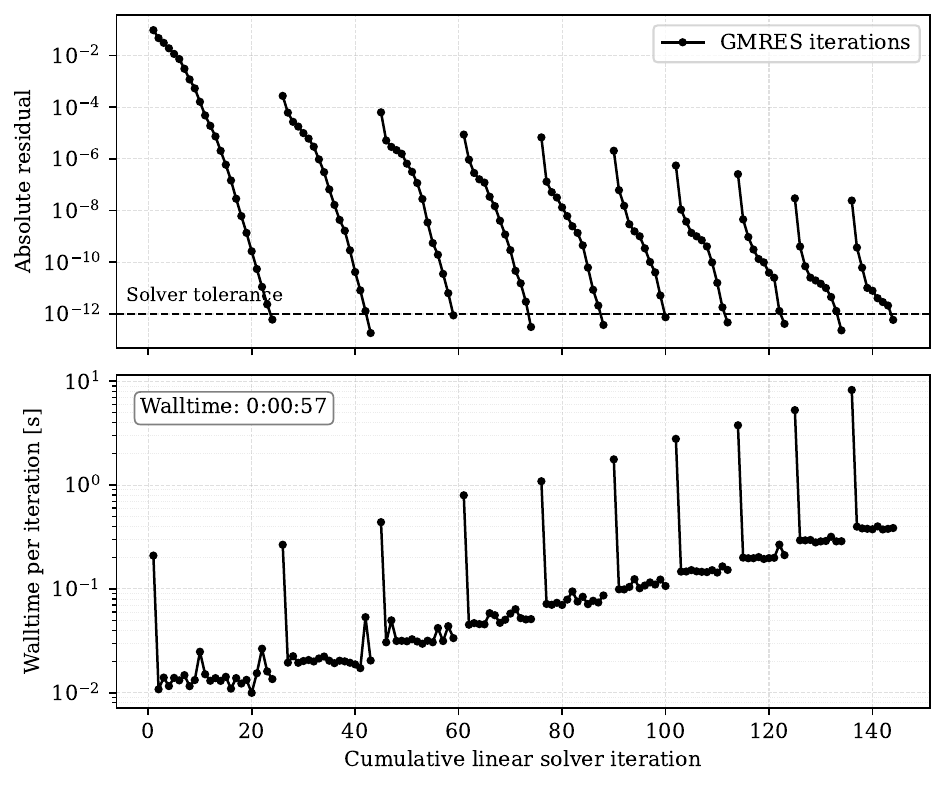}
    \caption{Convergence of the preconditioned linear solver for the $m=2$ mode with $a/M=0.6$, $r_0/M=8$. Top: GMRES residual over the cumulative number of iterations, with each segment corresponding to one AMR level. The dashed line marks the solver tolerance. Bottom: walltime per iteration. The first iteration on each AMR level is the most expensive, since it builds and caches the subdomain matrices.}
    \label{fig:walltime}
\end{figure}

Finally, we assess the performance of our preconditioned iterative linear solver. Figure~\ref{fig:walltime} shows the GMRES residual over consecutive AMR levels for the $m=2$ mode (same run as Fig.~\ref{fig:redshift_convergence}(a)), until the relative error between the last two AMR iterations reaches $\sim10^{-11}$.
Compared with our scalar case~\cite{Vu:2026ypc}, where the solver converged in fewer than 10 iterations per AMR level, the GR problem requires more iterations. This reflects the increased complexity of the system, which now consists of ten coupled elliptic PDEs with ten degrees of freedom per grid point (see Fig.~\ref{fig:dg_M_block} for the structure of the resulting DG operator).
Importantly, the number of GMRES iterations per AMR level does not grow as the resolution increases. The cost per iteration, however, grows with each AMR level, as shown in the bottom panel. The first iteration on each new level is the most expensive, since it constructs and caches the Schwarz subdomain matrices, and it accounts for a large fraction of the walltime at high resolution. These subdomain solves can be parallelized over elements in our additive Schwarz algorithm. In total, this mode converges in about 1 minute of walltime, using two nodes, each with two Intel Xeon Platinum 8360Y CPUs (72 cores per node).

\section{Conclusions and future directions}\label{sec:conclusion}
We have computed first-order Kerr metric perturbations directly in Lorenz gauge for a point mass following a circular-equatorial geodesic. Decomposing the field into azimuthal $m$-modes in the frequency domain reduces the linearized Einstein equations to a system of ten coupled elliptic PDEs in $(r,z)$ for each mode. This extends our previous scalar-field implementation in \spectre{}~\cite{Vu:2026ypc} to gravitational perturbations. 

Since higher-order perturbations are our ultimate goal, several ingredients of our formulation were chosen with second-order self-force calculations in mind. We use horizon-penetrating ingoing Eddington--Finkelstein coordinates, and factor out the leading behavior of each component toward the poles and toward infinity, so that the numerical variables $(\psi_m)_{\mu\nu}$ approach constant values at all boundaries. We adopt a continuous $vtu$ hyperboloidal slicing, which eliminates the asymptotic oscillations of the field toward the horizon and future null infinity, and for which second-order metric perturbations and sources are well-behaved. We compactify the $u$ region onto a finite domain that extends to future null infinity.
As a result, the boundary conditions at the horizon, at the poles, and at future null infinity all reduce to regularity conditions, which the spectral expansion imposes automatically. This resolves the near-horizon issues encountered with the finite-difference scheme of Ref.~\cite{Osburn:2026uct}. Issues related to residual freedom within the $m=0$ mode (and associated vanishing frequency) were resolved by decoupling one of the metric perturbation components in a way that requires the overall mode to be stationary.

To solve the resulting equations, we adapted the elliptic solver of the numerical relativity code \spectre{}, originally developed to construct initial data for binary black hole simulations. The equations are discretized with a spectral DG method, in which the jump conditions across the worldtube boundary and across the $vtu$ transitions are imposed through the numerical DG flux, so that the non-smooth features of the solution lie on element boundaries. We refine the grid with an $hp$-AMR strategy that splits the elements adjacent to the particle and increases the polynomial order everywhere else to achieve exponential convergence, and we solve the discretized system with a complex-valued GMRES algorithm with a multigrid-Schwarz preconditioner.

We validated our solutions in several independent ways.
Since the Lorenz-gauge field equations are equivalent to the linearized Einstein equations only when the gauge condition $Z_\mu=0$ holds, we monitored the gauge violation of each $m$-mode and found that it decreases exponentially with resolution throughout the entire domain.
We also compared our $m$-mode field with an independent Lorenz-gauge calculation based on metric reconstruction from Teukolsky-type solutions in a spheroidal $(\ell,m)$ basis~\cite{Leather:2026gnc} and found a relative difference at or below $10^{-7}$ throughout the domain. This provides a stringent cross-check between two independent methods developed toward second-order calculations in Kerr spacetime.

Finally, we compared gauge-invariant quantities with reference values, starting with Detweiler's redshift invariant. We achieve agreement with the reference values~\cite{Nasipak:2025tby} to relative errors below $10^{-8}$ over the full spin range, including spins up to $|a|/M=0.998$ (the Thorne limit), by regularizing the power-law tail of our numerical $m$-mode results.
In the dissipative sector, we validated the energy flux at future null infinity with reference values computed with \pybhpt{}~\cite{Nasipak_pybhpt_2026} and find relative errors below $10^{-10}$ over the full spin range, comparable to or below our estimated numerical error. 
Across a range of spins and orbital radii, from near the light ring to $r_0/M=50$, the relative errors in the energy flux remain below $2\times10^{-9}$ for all configurations considered.

There remain several directions in which our framework can be further improved. At first order, enhanced preconditioners for the Helmholtz-type problem can improve the convergence of our iterative solver.
The main goal of this program, however, is second-order Kerr metric perturbations.
Construction of appropriate second-order sources from first-order data, together with corresponding second-order punctures, is in progress. The $m$-mode approach is well suited to this construction: since the $(r,\theta)$ dependence is resolved directly on a two-dimensional grid, the discretization near the particle is approximately isotropic, which avoids the slow convergence of an angular mode sum there (although, simultaneous work has mitigated this slow angular convergence~\cite{Leather:2026gnc}). By combining a DG discretization, AMR, and a preconditioned iterative solver, we obtain first-order solutions accurate enough to serve directly as input to this construction.
Once the source is constructed from the first-order data, solving for the second-order field requires no new machinery: since the same linear differential operator governs the metric perturbation at every perturbative order (with nonlinearities confined to the sources), the same solver applies to the second-order $m$-mode field equations. Equipped with an appropriate second-order source, this approach can produce second-order Kerr self-force results in collaboration with other Kerr solvers~\cite{Leather:2026zhl,Cownden:2026msq}.

Another natural extension of this work is to consider more generic orbits. Forthcoming work will take a step towards this by considering eccentric orbits for a scalar field toy model in Kerr. Preliminary results indicate that our methods largely carry over to the eccentric case, although there are some hindrances due to the Gibbs phenomenon associated with the Fourier transform of a moving non-smooth source. Advancing to the generic case, where orbits are both eccentric and inclined, is another important challenge for future work.

\begin{acknowledgments}
We thank Brad Cownden, Rodrigo Panosso Macedo, Adam Pound and Ayush Roy for helpful discussions, as well as the \spectre{} development team for the collaborative work. We thank Zachary Nasipak for providing comparison redshift data.
NN is grateful to Peter James Nee and Clemens Dittmer for their guidance and support during the early stages of working with \spectre{}.
NLV gratefully acknowledges support from the Swiss National Science Foundation (SNSF) Ambizione Grant
No.~PZ00-2\_232961.
TO gratefully acknowledges support from the National Science Foundation (NSF) under Grant No.\ PHY-2309020 at SUNY Geneseo and from the Fulbright U.S. Scholar Program, which is sponsored by the U.S. Department of State and the Fulbright Commission in Ireland.
This work was supported in part by the Sherman Fairchild Foundation, by NSF Grants PHY-2407742,
PHY-2308615, PHY-2607486, OAC-2513338, and NASA award 80NSSC26K0340 at Cornell University.
SDU acknowledges the support of a Royal Society University Research Fellowship and the ERC Consolidator/UKRI Frontier Research Grant GWModels (selected by the ERC and funded by UKRI [grant number EP/Y008251/1]).
BL gratefully acknowledges funding from the European Union’s Horizon Europe research and innovation programme under the Marie Sklodowska-Curie grant agreement No. 101209791.
This work is supported by ERC grant EMRIWaveforms (DOI: \href{https://doi.org/10.3030/101200625}{10.3030/101200625}).
Any opinions, findings, conclusions or recommendations expressed in this article are solely the responsibility of the authors and do not necessarily reflect the views of the NSF, NASA, the Fulbright Program, the U.S. Government, or the Fulbright Commission in Ireland.
Computations were performed on the Urania HPC systems at the Max Planck Computing and
Data Facility. The figures in this article were produced with \matplotlib{}~\cite{matplotlib1, matplotlib2} and \paraview{}~\cite{paraview}.
This work makes use of the Black Hole Perturbation Toolkit~\cite{BHPToolkit}, Wolfram Mathematica~\cite{Mathematica} and xAct~\cite{xAct,Martin-Garcia:2008ysv,xCoba}.

\textit{AI usage:}
The formulation and conceptualisation of the problem, derivations, calculations and initial drafts of this manuscript were prepared entirely by the authors without the assistance of AI tools. Claude (Opus 5.5) was used to assist with code development, and in the late stages of the project to improve figure presentation, to assist with drafting and refining the final manuscript, and to check equations.
All AI products were independently verified by the authors, who take full responsibility for the accuracy and content of this paper.
\end{acknowledgments}

\appendix
\section{Radial and angular prefactors}\label{sec:prefactor}
As mentioned in Sec.~\ref{sec:formulation}, the prefactors in
Eq.~\eqref{eq:eq_MP} were introduced so that the functions $(\psi_m)_{\mu\nu}$ we solve for
approach a constant amplitude near the poles $z=\pm1$ and in the far-field region
$r\to\infty$.
Here we derive these prefactors and demonstrate the resulting improvement in
accuracy, following Ref.~\cite{adria}. Consider the $m$-mode expansion of the metric perturbation,
\begin{equation}
\bar{h}_{\mu\nu}(v,r,\theta,\varphi)
=\sum_{m=-\infty}^{\infty}
\bar{h}_{\mu\nu}^{(m)}(v,r,\theta)
\,e^{im\varphi}
\text{.}
\end{equation}
In Cartesian-like coordinates, the mode coefficients $\bar{h}_{\alpha\beta}^{(m)}(v,r,\theta)$ take the form
\begin{equation}\label{eq:cartesian_scaling}
\bar{h}_{\alpha\beta}^{(m)}(v,r,\theta)
=
\frac{1}{r}\,
\sin^{|m|}\theta\,
(\zeta_{m})_{\mu\nu}
\,,
\end{equation}
where
\begin{equation}
\alpha,\beta\in\{\hat{v},\hat{x},\hat{y},\hat{z}\}
\quad\text{with}\quad
\left\{\begin{array}{ll}
\hat{v}=&v\\
\hat{x}=&r\cos\varphi\sin\theta\\
\hat{y}=&r\sin\varphi\sin\theta\\
\hat{z}=&r\cos\theta
\end{array}\right.,
\end{equation}
and $(\zeta_{m})_{\mu\nu}$ are functions of $v$, $r$ and $\theta$, which approach non-zero constants at the poles, $\theta\in\{0,\pi\}$, and at the radial boundaries.
Equation~\eqref{eq:cartesian_scaling} follows from four properties: 
(1) the coordinate system is regular everywhere outside the inner-horizon,
(2) the negative modes are the complex conjugate of the positive modes, i.e., $\bar{h}_{\alpha\beta}^{(-m)}
=
\bar{h}_{\alpha\beta}^{(m)*}
$ 
since
$
\bar{h}_{\alpha\beta}\in\mathbb{R}
$,
(3) the solutions of the linearized Einstein equations decay as $1/r$ at large $r$, 
(4) the components of the tensor field must scale as $\sin^{|m|}\theta$ in order to have smoothness of the functions $\bar{h}_{\alpha\beta}^{(m)}(v,r,\theta)\
e^{im\varphi}$ near the poles.
Transforming Eq.~\eqref{eq:cartesian_scaling} as a tensor to the Eddington--Finkelstein coordinates $(v,r,z,\varphi)$, with $z=\cos\theta$, gives
\begin{equation}\label{eq:h_full}
\bar{h}_{\mu\nu}=\sum_{m=-\infty}^{\infty}\frac{\sin^{|m|}(\theta)}{r}\sum_{\alpha\beta}\left[\frac{\partial\hat{x}^\alpha}{\partial x^\mu}\frac{\partial\hat{x}^\beta}{\partial x^\nu}\zeta^{(\alpha\beta)}_{m}\right]
e^{im\varphi}
\text{,}
\end{equation}
where $\frac{\partial\hat{x}^\alpha}{\partial x^\mu}
=
\Upsilon_{(\alpha,\mu)}$ and
\begin{equation}
\resizebox{0.45\textwidth}{!}{$
\Upsilon
=
\left[\begin{array}{cccc}
    1 & 0 & 0 & 0\\
    0 & \sqrt{1-z^2}\cos\varphi & -\frac{rz\cos\varphi}{\sqrt{1-z^2}} & -r\sqrt{1-z^2}\sin\varphi\\
    0 & \sqrt{1-z^2}\sin\varphi & -\frac{rz\sin\varphi}{\sqrt{1-z^2}} & r\sqrt{1-z^2}\cos\varphi\\
    0 & z & r & 0
\end{array}\right].
$}
\end{equation}
We define the exponents $\kappa^{(\mu\nu)}$ and $\lambda^{(\mu\nu)}_{m}$ through
\begin{equation}\label{eq:lambda_kappa_def}
\bar{h}_{\mu\nu}^{(m)}
=
\left[\frac{1}{r}\right]^{\kappa^{(\mu\nu)}}
\left[\sin\theta\right]^{\lambda^{(\mu\nu)}_{m}}
(\psi_{m})_{\mu\nu}(r,z)\;,
\end{equation}
where the functions $(\psi_{m})_{\mu\nu}(r,z)$ are regular and generically nonzero at $z=\pm 1$ and tending to a finite limit as $r\to\infty$. 
Inserting Eq.~\eqref{eq:h_full} into Eq.~\eqref{eq:lambda_kappa_def} yields the
exponents listed in Table~\ref{tab:lambda}.
\begin{table}[h!]
\begin{minipage}{1\linewidth}
\begin{tabular}{l|c}
\toprule
\toprule
$\;\,\kappa^{(\mu\nu)}$ & $\forall m\in\mathbb{Z}$\\
\hline
$\;vv$ & $\phantom{-}1$\\ 
$\;vr$ & $\phantom{-}1$\\
$\;vz$ & $\phantom{-}0$\\
$\;v\varphi$ & $\phantom{-}0$\\ 
$\;rr$ & $\phantom{-}1$\\
$\;rz$ & $\phantom{-}0$\\
$\;r\varphi$ & $\phantom{-}0$\\
$\;zz$ & $-1$\\
$\;z\varphi$ & $-1$\\ 
$\;\varphi\varphi$ & $-1$\\
\bottomrule
\bottomrule
\end{tabular}
\hfill
\begin{tabular}{l|ccc}
\toprule
\toprule
$\;\,\lambda^{(\mu\nu)}_{m}$ & $m=0$ & $|m|=1$ & $|m|\geq2$\\
\hline
$\;vv$ & $0$ & $1$ & $|m|$ \\ 
$\;vr$ & $0$ & $1$ & $|m|$ \\
$\;vz$ & $0$ & $-1$& $|m|-2$  \\
$\;v\varphi$ & $2$ & $1$ & $|m|$ \\ 
$\;rr$ & $0$ & $1$ & $|m|$ \\
$\;rz$ & $0$ & $-1$ & $|m|-2$ \\
$\;r\varphi$ & $2$ & $1$ & $|m|$ \\
$\;zz$ & $-2$ & $-1$ & $|m|-4$ \\
$\;z\varphi$ & $2$ & $1$ & $|m|-2$ \\ 
$\;\varphi\varphi$ & $2$ & $3$ & $|m|$ \\
\bottomrule
\bottomrule
\end{tabular}
\end{minipage}
\caption[]{Values of the exponents $\kappa^{(\mu\nu)}$ and $\lambda^{(\mu\nu)}_{m}$.
The function $\bar h^{(m)}_{\mu\nu}$ in the far-field region scales in terms of $r$ as 
$\bar h^{(m)}_{\mu\nu}\sim\left[\frac{1}{r}\right]^{\kappa^{(\mu\nu)}}$.
The function $\bar h^{(m)}_{\mu\nu}$ near the poles $z=\pm1$ scales in terms of $\theta$ as $\bar h^{(m)}_{\mu\nu}\sim\left[\sin\theta\right]^{\lambda^{(\mu\nu)}_{m}}$.}
\label{tab:lambda}
\end{table}

\begin{figure}[h!]
    \centering
    \includegraphics[width=\columnwidth]{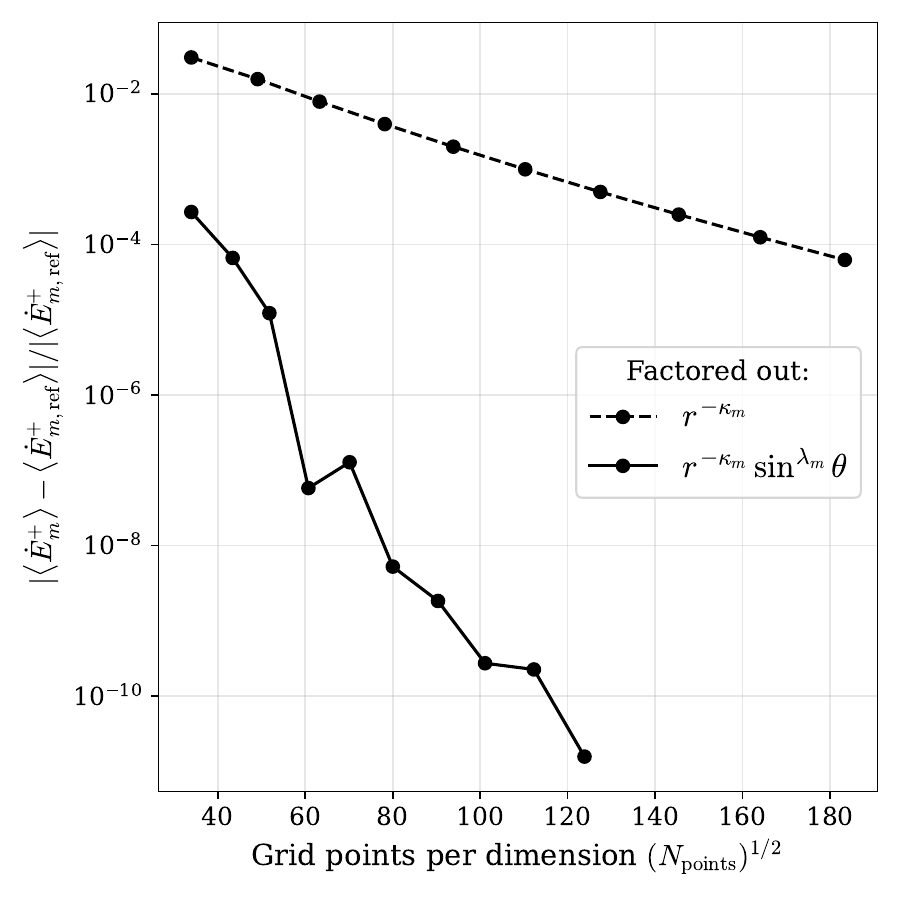}
    \caption{Factoring out the angular behavior at the poles greatly accelerates convergence.
    Shown is the relative difference of the $m=2$ outgoing energy flux, for $a/M=0.6$ and $r_0/M=10$.
    Both runs use the same number of AMR iterations; the dashed run (radial prefactor only) requires additional $p$-refinement at the poles and therefore extends to more grid points.
    For readability, the component indices $(\mu\nu)$ of $\kappa$ and $\lambda$ are omitted in the legend.}
    \label{fig:v2_v3_comp}
\end{figure}

Ref.~\cite{Dolan_2013} factored out the far-field $r$-dependence of $\bar{h}_{\mu\nu}$.
Here we additionally factor out the behavior near the poles, 
which substantially improves the accuracy of the numerical solution.
Figure~\ref{fig:v2_v3_comp} compares the two choices.
With only the radial prefactor $r^{-\kappa}$ (dashed line), additional
$p$-refinement of the elements adjacent to the poles was required, so that at each
AMR level this run uses more grid points.
Even so, its error decreases much more slowly with resolution than when the angular
prefactor $\sin^{\lambda_m}\theta$ is also factored out (solid line). For example, at
$(N_\mathrm{points})^{1/2}\approx120$, the relative error is $\sim5\times10^{-4}$
for the former and $\sim10^{-11}$ for the latter.

\vspace{3mm}

\section{Zero frequency $m=0$ mode}\label{sec:static_mode}
For $m\neq0$, we solve the elliptic system in first-order flux form, Eq.~\eqref{eq:first_order_flux_form}, with coefficients $\beta$ and $\gamma^i$ derived from the Lorenz-gauge linearized Einstein field equations, Eq.~\eqref{eq:einstein_field_eq}. 
The zero-frequency $m=0$ mode, however, has a residual freedom that must be fixed by imposing additional conditions.
The Kerr metric in Boyer--Lindquist coordinates is invariant under $(t,\phi)\rightarrow(-t,-\phi)$, and for $m=0$, so is the source of a circular orbit. The physical perturbation therefore shares this symmetry, which requires
\begin{align}
\bar h^{\rm BL}_{tr}=\bar h^{\rm BL}_{t\theta}=\bar h^{\rm BL}_{r\phi}
=\bar h^{\rm BL}_{\theta\phi}=0.
\end{align}
We refer to these as the stationary conditions; as we show next, imposing them explicitly is essential to fix the residual freedom.
Throughout this appendix, $m=0$, and we drop the mode label; we write $\psi_{\mu\nu}\equiv(\psi_{m=0})_{\mu\nu}$ for the components.

\begin{figure}
    \centering
    \includegraphics[width=\columnwidth]{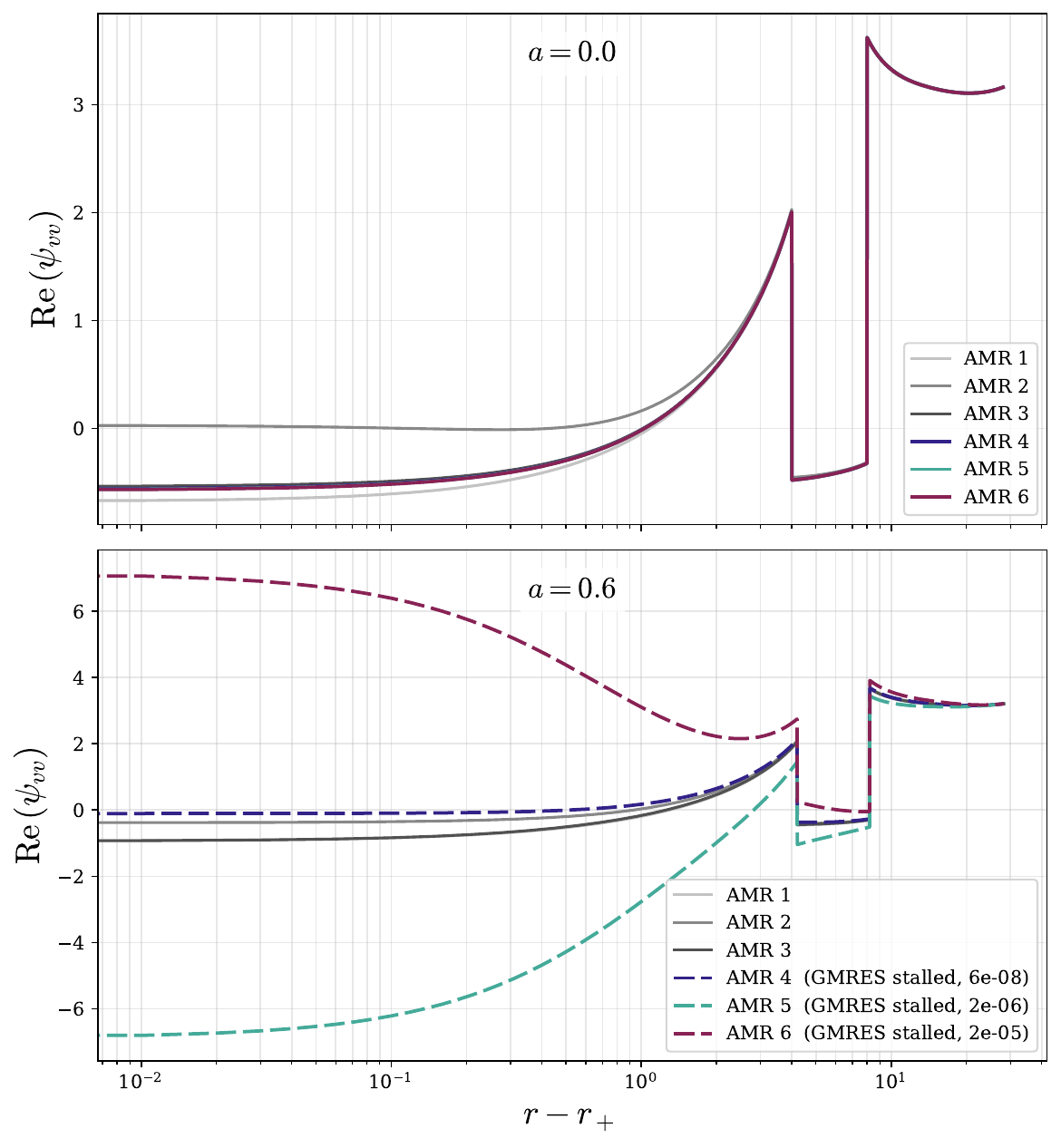}
    \caption{
    Without imposing the stationary conditions, the $m=0$ mode either converges to an incorrect solution or fails to converge.
    Shown is $\mathrm{Re}(\psi_{vv})$ along $\theta=\pi/2$ at each AMR level, for $r_0/M=8$.
    Top: for $a/M=0$, the solution converges, but to an incorrect solution (Fig.~\ref{fig:m0_nonspin_static_condition}).
    Bottom: for $a/M=0.6$, GMRES stalls at higher AMR levels (dashed lines; the legend gives the final residual), and the field near the horizon changes strongly between levels.}
    \label{fig:m0_field_incorrect_solution}
\end{figure}

Without these conditions, solving the $m=0$ mode with the original $\beta$ and $\gamma^i$ does not yield a consistent solution, as illustrated in Fig.~\ref{fig:m0_field_incorrect_solution}, which shows the real part of $\psi_{vv}$ at $\theta=\pi/2$ over successive AMR levels. For $a/M=0$ (top panel), the solution converges with refinement, but to a solution that violates the stationary conditions as we show below. For $a/M=0.6$ (bottom panel), it does not converge at all: the field near the horizon oscillates strongly between AMR levels, and from AMR level~4 onwards (dashed lines) GMRES stalls before reaching the residual tolerance. We show $\psi_{vv}$ here, but the other components behave similarly.

\begin{figure}
    \centering
    \includegraphics[width=\columnwidth]{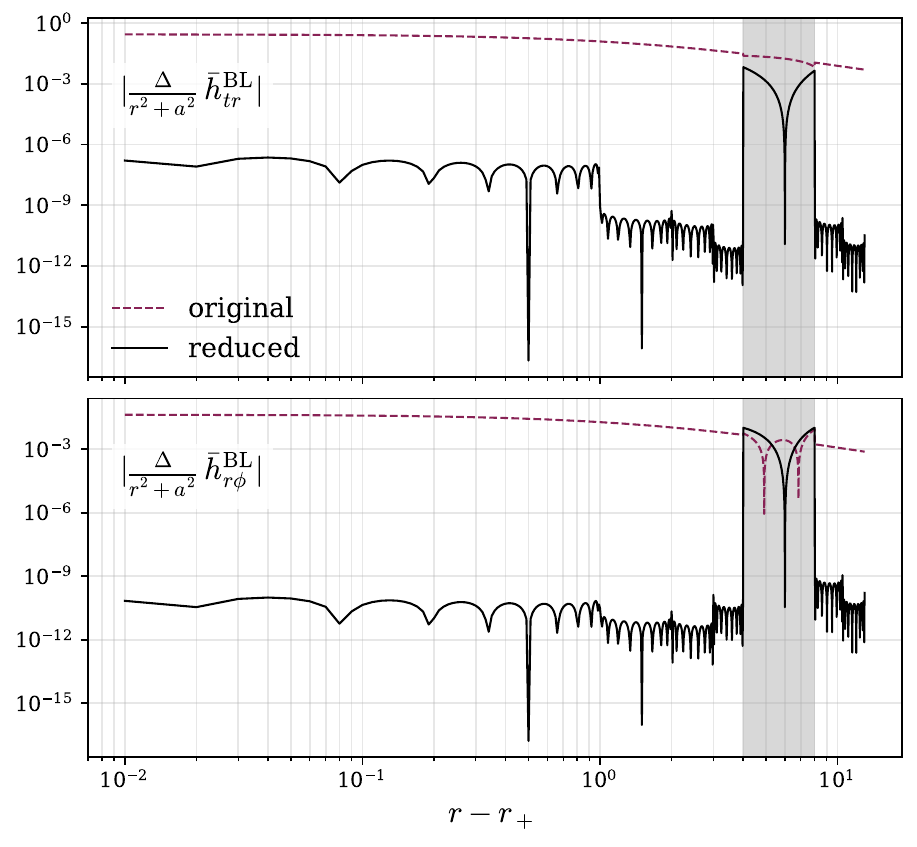}
    \caption{With the reduced matrices, the solution satisfies the stationary conditions outside the worldtube; with the original matrices, it violates them.
    Shown are the two stationary conditions $\bar h^{\rm BL}_{tr}=0$ and $\bar h^{\rm BL}_{r\phi}=0$ along $\theta=\pi/2$ for $a/M=0$ and $r_0/M=8$; the original solution (dashed purple) is the final AMR level of Fig.~\ref{fig:m0_field_incorrect_solution}.
    To remove the $1/\Delta$ factor arising from the transformation between Eddington--Finkelstein and Boyer--Lindquist coordinates, both components are multiplied by $dr/dr_*=\Delta/(r^2+a^2)$.
    The worldtube region is shaded gray.}
    \label{fig:m0_nonspin_static_condition}
\end{figure}

To fix this, we impose the stationary conditions directly on the system by eliminating four components of $\psi_{\mu\nu}$ in favor of the remaining six. 
To this end, we rewrite the first-order flux form, Eq.~\eqref{eq:first_order_flux_form},
\begin{align}\label{eq:m0_flux_form}
&F^i_{\mu\nu}=\left\{ \frac{\Delta}{r^{2}+a^{2}}\,\partial_{r}\psi_{\mu\nu},\
\frac{1-z^{2}}{r^{2}+a^{2}}\,\partial_{z}\psi_{\mu\nu}\right\}, \quad i=r,z,\\
-&\partial_{i}F^i_{\mu\nu}+\gamma^{i,\lambda\rho}_{\mu\nu}\,\partial_{i}\psi_{\lambda\rho}
+\beta_{\mu\nu}^{\lambda\rho}\,\psi_{\lambda\rho}=S_{\mu\nu}.
\end{align}
The elimination modifies $\beta$ and $\gamma^i$; the corresponding change to the source is discussed below. Since we solve for $\psi_{\mu\nu}$, we write the conditions in Eddington--Finkelstein coordinates, including the radial and angular prefactors of Appendix~\ref{sec:prefactor}:
\begin{align}
\psi_{vz}&=\psi_{z\varphi}=0,\label{eq:m0_vtheat_thetavarphi}\\
\psi_{vr}&=-\frac{r^2+a^2}{\Delta}\,\psi_{vv}-\frac{ar(1-z^2)}{\Delta}\,\psi_{v\varphi},\label{eq:m0_vr}\\
\psi_{v\varphi}&=-\frac{\Delta}{r^2+a^2}\,\psi_{r\varphi}-\frac{ar}{r^2+a^2}\,\psi_{\varphi\varphi}.\label{eq:m0_vphi}
\end{align} 
Substituting Eq.~\eqref{eq:m0_vphi} into Eq.~\eqref{eq:m0_vr} gives
\begin{align}
\psi_{vr}&=-\frac{B}{\Delta}+\frac{ar(1-z^2)}{r^2+a^2}\,\psi_{r\varphi},\\
B&\equiv(r^2+a^2)\,\psi_{vv}-\frac{a^2r^2(1-z^2)}{r^2+a^2}\,\psi_{\varphi\varphi}.\label{eq:m0_B}
\end{align}
Expressing $\psi_{vr}$ in terms of $\psi_{vv}$ therefore introduces $1/\Delta$ into the reduced coefficients, which diverges at the horizon. 
To avoid this, we introduce the new variable
\begin{align}
\psi_B\equiv\frac{B}{\Delta},\label{eq:m0_psiB}
\end{align}
and solve for it instead of $\psi_{vv}$.
Inverting Eqs.~\eqref{eq:m0_B} and~\eqref{eq:m0_psiB} expresses $\psi_{vv}$ in terms of $\psi_B$, and the $vr$ relation above becomes free of $1/\Delta$:
\begin{align}
\psi_{vv}&=\frac{1}{r^2+a^2}\left(\Delta\,\psi_B+\frac{a^2r^2(1-z^2)}{r^2+a^2}\,\psi_{\varphi\varphi}\right),\label{eq:m0_vv_psiB}\\
\psi_{vr}&=-\psi_B+\frac{ar(1-z^2)}{r^2+a^2}\,\psi_{r\varphi}.\label{eq:m0_vr_psiB}
\end{align}
We use Eqs.~\eqref{eq:m0_vtheat_thetavarphi}, \eqref{eq:m0_vphi}, and~\eqref{eq:m0_vr_psiB} to eliminate the four components $(\psi_{vr},\psi_{vz},\psi_{v\varphi},\psi_{z\varphi})$ and solve for $(\psi_B,\psi_{rr},\psi_{rz},\psi_{r\varphi},\psi_{zz},\psi_{\varphi\varphi})$. 
The equation for $\psi_B$ is obtained by applying the same combination, Eq.~\eqref{eq:m0_B}, to the $\psi_{vv}$ and $\psi_{\varphi\varphi}$ equations and their sources and dividing by $\Delta$.
With this choice, the principal part of the $\psi_B$ equation is again $-\partial_i F^i[\psi_B]$ with $F^i_{\mu\nu}$ as in Eq.~\eqref{eq:m0_flux_form}, and $1/\Delta$ does not appear in $\beta$ or $\gamma^i$. For the same reason we eliminate $\psi_{v\varphi}$ rather than $\psi_{r\varphi}$: solving for $\psi_{r\varphi}$ instead would introduce $1/\Delta$.

The resulting matrices are then re-embedded as $10\times10$ matrices, in which the four eliminated rows are algebraic and carry no flux. The effective source was constructed for the original operator, and the puncture does not satisfy the stationary conditions, so the effective source must be re-derived. Since the flux divergence is unchanged for the kept rows and absent for the eliminated rows, the reduced and original operators differ only in the non-flux part,
\begin{align}
k[\psi_{\mu\nu}]=\beta_{\mu\nu}^{\lambda\rho}\,\psi_{\lambda\rho}+\gamma^{i,\lambda\rho}_{\mu\nu}\,\partial_i\psi_{\lambda\rho}
\text{.}
\end{align}
The effective source consistent with the reduced operator is therefore
\begin{align}
S^{\rm eff}_{\mu\nu}&\to S^{\rm eff}_{\mu\nu}+k^{\rm orig}[\psi^\mathcal{P}_{\mu\nu}]-k^{\rm red}[\psi^\mathcal{P}_{\mu\nu}]
\quad\text{(kept rows)},\\
S^{\rm eff}_{\mu\nu}&\to -k^{\rm red}[\psi^\mathcal{P}_{\mu\nu}]
\quad\text{(eliminated rows)}.
\end{align}
Because only $k$ changes, this correction requires only the puncture and its first derivatives. Here $k^{\rm orig}$ is the original operator transformed in the same way as the reduced one, i.e., with $\psi_{vv}$ expressed through $\psi_B$ and the $vv$ equation replaced by the combination of Eq.~\eqref{eq:m0_B}, and $\psi^\mathcal{P}_{\mu\nu}$ is the puncture with its $\psi_{vv}$ component converted to ${\psi}^\mathcal{P}_B$ via Eq.~\eqref{eq:m0_psiB}. Since the correction is applied only inside the worldtube, where $\Delta\neq0$, this conversion is regular.

Figure~\ref{fig:m0_nonspin_static_condition} shows the stationary conditions $\bar h^{\rm BL}_{tr}$ and $\bar h^{\rm BL}_{r\phi}$ along $\theta=\pi/2$ for $a/M=0$ and $r_0/M=8$. Since our solution is computed in Eddington--Finkelstein coordinates, the transformation to Boyer--Lindquist coordinates introduces factors of $1/\Delta$ that diverge at the horizon; we therefore rescale both components by $dr/dr_*=\Delta/(r^2+a^2)$.
The solution obtained with the original $\beta$ and $\gamma^i$ (dashed purple) converges with refinement (as shown in the top panel of Fig.~\ref{fig:m0_field_incorrect_solution}), but it violates both conditions, most strongly near the horizon. 
The solution obtained with the reduced matrices (black) satisfies both conditions outside the worldtube. We show $a/M=0$ here, but the same holds for all spins we consider, and we validate it through its contribution to the redshift in Sec.~\ref{sec:results_redshift}, including the near-extremal spins $a/M=\pm0.998$.

\section{Mass and angular-momentum shifts of the Berndtson-type solution}\label{sec:mass_correction}
As discussed in Sec.~\ref{sec:results_redshift}, our
regular Lorenz-gauge solution describes a black hole with
mass $M_\bullet = M + \epsilon\,\delta M$ and angular momentum
$J_\bullet = J + \epsilon\,\delta J$. Here we explain what the shifts are, how we obtain them from the $m = 0$ mode, and the derivatives of $z_0$ in Eq.~\eqref{eq:redshift_correction}.

Two solutions for the same source that are regular at the horizon and infinity have the same Weyl scalars $\psi_0$ and $\psi_4$, which are gauge invariant.
By Wald's theorem \cite{Wald:1973wwa}, their difference is then, up to a gauge transformation, a change of Kerr parameters,
\begin{equation}
    \delta M \frac{\partial g_{\mu\nu}}{\partial M}\bigg|_{J} + \delta J \frac{\partial g_{\mu\nu}}{\partial J}\bigg|_{M}
    \label{eq:kerr_parameter_change}
\end{equation}
The shifts $\delta M$ and $\delta J$ are the coefficients of this difference between our solution and a solution for a black hole described by the background parameters.
In general this difference also contains a gauge part, so $\delta M$ and $\delta J$ cannot be read off from individual components of ${\bar h}_{\mu\nu}$.
We therefore use surface integrals that measure the mass and angular momentum carried by the perturbation inside a sphere.

For a Killing vector $X^\alpha$ of the background, these integrals are built from the 2-form~\cite{ABBOTT198276, Dolan_2013},
\begin{align}\label{eq:komar_two_form}
F_{\alpha\beta}=-\frac{1}{8\pi}\left( X^\lambda \bar{h}_{\lambda[\alpha;\beta]}
+X^\lambda{}_{;[\alpha}\, \bar{h}_{\beta]\lambda} + X_{[\alpha} Z_{\beta]} \right),
\end{align}
where $Y_{[\alpha\beta]}\equiv\frac12(Y_{\alpha\beta}-Y_{\beta\alpha})$ and
$Z_\alpha \equiv \bar h_{\alpha\beta}{}^{;\beta}$, which vanishes in Lorenz gauge.
On a sphere $\partial\Sigma_{r}$ of constant $v$ and $r$,
\begin{align}\label{eq:komar_integral}
\mathcal{F}(X,\partial\Sigma_r)=\frac{1}{2}\oint_{\partial\Sigma_r}F^{\alpha\beta}\, d\Sigma_{\alpha\beta} 
= \oint_{\partial\Sigma_r} F^{vr}\sqrt{-g} d\theta d\varphi.
\end{align}
This integral has three properties that we use~\cite{Dolan_2013}.
It vanishes for any pure-gauge perturbation whose gauge vector is smooth on the sphere.
It takes the same value on every sphere between the horizon and the worldline, since $F^{\alpha\beta}_{;\beta} = T^{\alpha\beta}X_{\beta}$ vanishes there, and it changes by $\mu X^{\alpha}u_{\alpha}$ across the worldline.
Finally, for the Kerr perturbation of Eq.~\eqref{eq:kerr_parameter_change} it gives ${\cal F}(\partial_{v}) = -\delta M$ and ${\cal F}(\partial_{\varphi}) = \delta J$.
A solution for black hole with the background parameters therefore has ${\cal F} = 0$ inside the worldline, and for our solution
\begin{align}\label{eq:dM_dJ}
\delta M = -\mathcal{F}(\partial_v,\partial\Sigma_r), \qquad
\delta J = \mathcal{F}(\partial_{\varphi},\partial\Sigma_r),
\end{align}
on any sphere between the horizon and the worldtube.
Outside the worldline, the same integrals give the shifts plus the energy $\mu u_{t}$ and angular momentum $\mu u_{\phi}$ of the secondary.
Here $\partial_{v}$ and $\partial_{\varphi}$ coincide with the Boyer--Lindquist
Killing vectors $\partial_t$ and $\partial_\phi$, since $v$ and $\varphi$ differ from
$t$ and $\phi$ only by functions of $r$.
The $m\neq0$ parts of $F^{\alpha\beta}$ are proportional to $e^{im\varphi}$ and integrate to zero over the sphere, so only the $m=0$ mode contributes. We evaluate Eq.~\eqref{eq:dM_dJ} on the sphere at
the $v$-to-$t$ transition radius $r_v$, which lies between the horizon and the worldtube.

Finally, at fixed $\Omega$ a circular geodesic extremizes $z_{0}^{2} = -g_{\mu\nu} k^{\mu}k^{\nu}$ with respect to the orbital radius, so the derivatives in Eq.~\eqref{eq:redshift_correction}:
\begin{align}
    \frac{\partial z_{0}}{\partial M}\bigg|_{\Omega, J} &= \frac{1}{z_{0}}\left[ \frac{a^{2}\Omega^{2}}{M} - \frac{1 - a^{2}\Omega^{2}}{r_{0}} \right],\\
    \frac{\partial z_{0}}{\partial J}\bigg|_{\Omega, M} &= \frac{1}{z_{0}}\left[ \frac{2\Omega(1 - a\Omega)}{r_{0}} - \frac{a\Omega^{2}}{M} \right].
\end{align}
The $\delta J$ term contributes even for $a = 0$.
For $a = 0$, applying the $\delta M$ term to Berndtson's solution gives the same $\Delta z$ as the correction of Ref.~\cite{Sago:2008id} to the asymptotically nonflat Lorenz-gauge solution.

\bibliography{main}

@article{Ansorg:2004ds,
  author        = {Ansorg, Marcus and Bruegmann, Bernd and Tichy, Wolfgang},
  title         = {A Single-domain spectral method for black hole puncture data},
  eprint        = {gr-qc/0404056},
  archiveprefix = {arXiv},
  reportnumber  = {CGPG-03-12-3},
  doi           = {10.1103/PhysRevD.70.064011},
  journal       = {Phys. Rev. D},
  volume        = {70},
  pages         = {064011},
  year          = {2004}
}

@article{Assumpcao2021-es,
  author = "Assumpcao, Thiago and Werneck, Leonardo R. and Jacques, Terrence Pierre and Etienne, Zachariah B.",
  title = "{Fast hyperbolic relaxation elliptic solver for numerical relativity: Conformally flat, binary puncture initial data}",
  eprint = "2111.02424",
  archivePrefix = "arXiv",
  primaryClass = "gr-qc",
  doi = "10.1103/PhysRevD.105.104037",
  journal = "Phys. Rev. D",
  volume = "105",
  number = "10",
  pages = "104037",
  year = "2022",
}

@dataset{BAMcatalog,
  author    = {Hamilton, Eleanor and
               Fauchon-Jones, Edward and
               Hannam, Mark and
               Hoy, Charlie and
               Kalaghatgi, Chinmay and
               London, Lionel and
               Thompson, Jonathan and
               Yeeles, Dave and
               Ghosh, Shrobana and
               Khan, Sebastian and
               Kolitsidou, Panagiota and
               Vano-Vinuales, Alex},
  title     = {Precessing binary-black-hole numerical relativity 
               catalogue (complete data release)},
  month     = feb,
  year      = 2023,
  publisher = {Zenodo},
  version   = 1,
  doi       = {10.5281/zenodo.7677297},
  url       = {https://doi.org/10.5281/zenodo.7677297},
  note      = {DOI: \href{https://doi.org/10.5281/zenodo.7677297}{10.5281/zenodo.7677297}}
}

@article{Barack_2007a,
  author        = {Barack, Leor and Golbourn, Darren A.},
  title         = {{Scalar-field perturbations from a particle orbiting a black hole using numerical evolution in 2+1 dimensions}},
  eprint        = {0705.3620},
  archiveprefix = {arXiv},
  primaryclass  = {gr-qc},
  doi           = {10.1103/PhysRevD.76.044020},
  journal       = {Phys. Rev. D},
  volume        = {76},
  pages         = {044020},
  year          = {2007}
}

@article{Barack_2007b,
  author        = {Barack, Leor and Golbourn, Darren A. and Sago, Norichika},
  title         = {{m-Mode Regularization Scheme for the Self Force in Kerr Spacetime}},
  eprint        = {0709.4588},
  archiveprefix = {arXiv},
  primaryclass  = {gr-qc},
  doi           = {10.1103/PhysRevD.76.124036},
  journal       = {Phys. Rev. D},
  volume        = {76},
  pages         = {124036},
  year          = {2007}
}

@article{Barack:2018yvs,
  author        = {Barack, Leor and Pound, Adam},
  title         = {{Self-force and radiation reaction in general relativity}},
  eprint        = {1805.10385},
  archiveprefix = {arXiv},
  primaryclass  = {gr-qc},
  doi           = {10.1088/1361-6633/aae552},
  journal       = {Rept. Prog. Phys.},
  volume        = {82},
  number        = {1},
  pages         = {016904},
  year          = {2019}
}

@article{Bourg:2024cgh,
  author        = {Bourg, Patrick and Pound, Adam and Upton, Samuel D. and Panosso Macedo, Rodrigo},
  title         = {{Simple, efficient method of calculating the Detweiler-Whiting singular field to very high order}},
  eprint        = {2404.10082},
  archiveprefix = {arXiv},
  primaryclass  = {gr-qc},
  doi           = {10.1103/PhysRevD.110.084007},
  journal       = {Phys. Rev. D},
  volume        = {110},
  number        = {8},
  pages         = {084007},
  year          = {2024}
}

@article{Burke_2024,
  author        = {Burke, Ollie and Piovano, Gabriel Andres and Warburton, Niels and Lynch, Philip and Speri, Lorenzo and Kavanagh, Chris and Wardell, Barry and Pound, Adam and Durkan, Leanne and Miller, Jeremy},
  title         = {{Assessing the importance of first postadiabatic terms for small-mass-ratio binaries}},
  eprint        = {2310.08927},
  archiveprefix = {arXiv},
  primaryclass  = {gr-qc},
  doi           = {10.1103/PhysRevD.109.124048},
  journal       = {Phys. Rev. D},
  volume        = {109},
  number        = {12},
  pages         = {124048},
  year          = {2024}
}

@article{Detweiler_2003,
  author        = {Detweiler, Steven L. and Whiting, Bernard F.},
  title         = {{Self-force via a Green's function decomposition}},
  eprint        = {gr-qc/0202086},
  archiveprefix = {arXiv},
  doi           = {10.1103/PhysRevD.67.024025},
  journal       = {Phys. Rev. D},
  volume        = {67},
  pages         = {024025},
  year          = {2003}
}

@article{Dolan_2011a,
  author        = {Dolan, Sam R. and Barack, Leor},
  title         = {{Self force via m-mode regularization and 2+1D evolution: Foundations and a scalar-field implementation on Schwarzschild}},
  eprint        = {1010.5255},
  archiveprefix = {arXiv},
  primaryclass  = {gr-qc},
  doi           = {10.1103/PhysRevD.83.024019},
  journal       = {Phys. Rev. D},
  volume        = {83},
  pages         = {024019},
  year          = {2011}
}

@article{Dolan_2011b,
  author        = {Dolan, Sam R. and Barack, Leor and Wardell, Barry},
  title         = {{Self force via $m$-mode regularization and $2+1$D evolution: II. Scalar-field implementation on Kerr spacetime}},
  eprint        = {1107.0012},
  archiveprefix = {arXiv},
  primaryclass  = {gr-qc},
  doi           = {10.1103/PhysRevD.84.084001},
  journal       = {Phys. Rev. D},
  volume        = {84},
  pages         = {084001},
  year          = {2011}
}

@article{Dolan_2013,
  author        = {Dolan, Sam R. and Barack, Leor},
  title         = {{Self-force via $m$-mode regularization and 2+1D evolution: III. Gravitational field on Schwarzschild spacetime}},
  eprint        = {1211.4586},
  archiveprefix = {arXiv},
  primaryclass  = {gr-qc},
  doi           = {10.1103/PhysRevD.87.084066},
  journal       = {Phys. Rev. D},
  volume        = {87},
  pages         = {084066},
  year          = {2013}
}

@article{Dolan_2022,
  author        = {Dolan, Sam R. and Kavanagh, Chris and Wardell, Barry},
  title         = {{Gravitational Perturbations of Rotating Black Holes in Lorenz Gauge}},
  eprint        = {2108.06344},
  archiveprefix = {arXiv},
  primaryclass  = {gr-qc},
  doi           = {10.1103/PhysRevLett.128.151101},
  journal       = {Phys. Rev. Lett.},
  volume        = {128},
  number        = {15},
  pages         = {151101},
  year          = {2022}
}

@article{Dolan_2024,
  author        = {Dolan, Sam R. and Durkan, Leanne and Kavanagh, Chris and Wardell, Barry},
  title         = {{Metric perturbations of Kerr spacetime in Lorenz gauge: circular equatorial orbits}},
  eprint        = {2306.16459},
  archiveprefix = {arXiv},
  primaryclass  = {gr-qc},
  doi           = {10.1088/1361-6382/ad52e3},
  journal       = {Class. Quant. Grav.},
  volume        = {41},
  number        = {15},
  pages         = {155011},
  year          = {2024}
}

@article{Erlangga2006,
  author  = {Erlangga, Y. A. and Oosterlee, C. W. and Vuik, C.},
  title   = "{A Novel Multigrid Based Preconditioner For Heterogeneous Helmholtz Problems}",
  journal = {SIAM J. Sci. Comput.},
  volume  = {27},
  number  = {4},
  pages   = {1471-1492},
  year    = {2006},
  doi     = {10.1137/040615195},
  url     = {https://doi.org/10.1137/040615195}
}

@article{Ferguson:2023vta,
  author        = {Ferguson, Deborah and others},
  title         = {{Second MAYA catalog of binary black hole numerical relativity waveforms}},
  eprint        = {2309.00262},
  archiveprefix = {arXiv},
  primaryclass  = {gr-qc},
  reportnumber  = {UTWI-32-2023},
  doi           = {10.1103/gk7x-9cds},
  journal       = {Phys. Rev. D},
  volume        = {112},
  number        = {4},
  pages         = {044043},
  year          = {2025}
}

@article{Healy:2022wdn,
  author        = {Healy, James and Lousto, Carlos O.},
  title         = {{Fourth RIT binary black hole simulations catalog: Extension to eccentric orbits}},
  eprint        = {2202.00018},
  archiveprefix = {arXiv},
  primaryclass  = {gr-qc},
  doi           = {10.1103/PhysRevD.105.124010},
  journal       = {Phys. Rev. D},
  volume        = {105},
  number        = {12},
  pages         = {124010},
  year          = {2022}
}

@article{Jani:2016wkt,
  author        = {Jani, Karan and Healy, James and Clark, James A. and London,
                   Lionel and Laguna, Pablo and Shoemaker, Deirdre},
  title         = {{Georgia Tech Catalog of Gravitational Waveforms}},
  eprint        = {1605.03204},
  archiveprefix = {arXiv},
  primaryclass  = {gr-qc},
  doi           = {10.1088/0264-9381/33/20/204001},
  journal       = {Class. Quant. Grav.},
  volume        = {33},
  number        = {20},
  pages         = {204001},
  year          = {2016}
}

@software{Lorene,
  title        = {{LORENE}},
  author       = {Gourgoulhon, E. and Grandclément, P. and Marck, J.-A. and Novak, J. and Taniguchi, K.},
  url          = {http://www.lorene.obspm.fr},
  howpublished = {\url{http://www.lorene.obspm.fr}},
  year         = {2016}
}

@article{Macedo_2024,
  author        = {Panosso Macedo, Rodrigo and Bourg, Patrick and Pound, Adam and Upton, Samuel D.},
  title         = {{Multidomain spectral method for self-force calculations}},
  eprint        = {2404.10083},
  archiveprefix = {arXiv},
  primaryclass  = {gr-qc},
  doi           = {10.1103/PhysRevD.110.084008},
  journal       = {Phys. Rev. D},
  volume        = {110},
  number        = {8},
  pages         = {084008},
  year          = {2024}
}

@article{matplotlib1,
  author = "Hunter, John D.",
  title = "{Matplotlib: A 2D Graphics Environment}",
  doi = "10.1109/MCSE.2007.55",
  journal = "Comput. Sci. Eng.",
  volume = "9",
  number = "3",
  pages = "90--95",
  year = "2007",
}

@software{matplotlib2,
  author       = {Thomas A Caswell and others},
  title        = {\texttt{matplotlib v3.3.3}},
  month        = nov,
  year         = 2020,
  publisher    = {Zenodo},
  doi          = {10.5281/zenodo.4268928},
  url          = {https://github.com/matplotlib/matplotlib},
  howpublished = {\href{https://doi.org/10.5281/zenodo.4268928}{10.5281/zenodo.4268928}}
}

@article{Miller_2016,
  author        = {Miller, Jeremy and Wardell, Barry and Pound, Adam},
  title         = {{Second-order perturbation theory: the problem of infinite mode coupling}},
  eprint        = {1608.06783},
  archiveprefix = {arXiv},
  primaryclass  = {gr-qc},
  doi           = {10.1103/PhysRevD.94.104018},
  journal       = {Phys. Rev. D},
  volume        = {94},
  number        = {10},
  pages         = {104018},
  year          = {2016}
}

@article{Mino_1997,
  author        = {Mino, Yasushi and Sasaki, Misao and Tanaka, Takahiro},
  title         = {{Gravitational radiation reaction to a particle motion}},
  eprint        = {gr-qc/9606018},
  archiveprefix = {arXiv},
  reportnumber  = {OU-TAP-38, KUNS-1394},
  doi           = {10.1103/PhysRevD.55.3457},
  journal       = {Phys. Rev. D},
  volume        = {55},
  pages         = {3457--3476},
  year          = {1997}
}

@article{Osburn_2022,
  author        = {Osburn, Thomas and Nishimura, Nami},
  title         = {{New self-force method via elliptic partial differential equations for Kerr inspiral models}},
  eprint        = {2206.07031},
  archiveprefix = {arXiv},
  primaryclass  = {gr-qc},
  doi           = {10.1103/PhysRevD.106.044056},
  journal       = {Phys. Rev. D},
  volume        = {106},
  number        = {4},
  pages         = {044056},
  year          = {2022}
}

@article{Papenfort:2021hod,
  author        = {Papenfort, L. Jens and Tootle, Samuel D. and Grandcl\'ement, Philippe and Most, Elias R. and Rezzolla, Luciano},
  title         = {New public code for initial data of unequal-mass, spinning compact-object binaries},
  eprint        = {2103.09911},
  archiveprefix = {arXiv},
  primaryclass  = {gr-qc},
  doi           = {10.1103/PhysRevD.104.024057},
  journal       = {Phys. Rev. D},
  volume        = {104},
  number        = {2},
  pages         = {024057},
  year          = {2021}
}

@inproceedings{paraview,
  title     = {\texttt{ParaView}: An End-User Tool for Large-Data Visualization},
  author    = {J. Ahrens and Berk Geveci and C. Law},
  booktitle = {Visualization Handbook},
  year      = {2005},
  isbn      = {978-0123875822},
  doi       = {10.1016/B978-012387582-2/50038-1},
  url       = {https://doi.org/10.1016/B978-012387582-2/50038-1},
  publisher = {Butterworth-Heinemann, Burlington}
}

@article{Pfeiffer2003-mt,
  author        = {Pfeiffer, Harald P. and Kidder, Lawrence E. and Scheel, Mark A. and Teukolsky, Saul A.},
  title         = {{A multidomain spectral method for solving elliptic equations}},
  eprint        = {gr-qc/0202096},
  archiveprefix = {arXiv},
  doi           = {10.1016/S0010-4655(02)00847-0},
  journal       = {Comput. Phys. Commun.},
  volume        = {152},
  pages         = {253--273},
  year          = {2003}
}

@article{Poisson:2011nh,
  author        = {Poisson, Eric and Pound, Adam and Vega, Ian},
  title         = {{The motion of point particles in curved spacetime}},
  eprint        = {1102.0529},
  archiveprefix = {arXiv},
  primaryclass  = {gr-qc},
  doi           = {10.12942/lrr-2011-7},
  journal       = {Living Rev. Rel.},
  volume        = {14},
  pages         = {7},
  year          = {2011}
}

@article{Pound_2005,
  author        = {Pound, Adam and Poisson, Eric and Nickel, Bernhard G.},
  title         = {{Limitations of the adiabatic approximation to the gravitational self-force}},
  eprint        = {gr-qc/0509122},
  archiveprefix = {arXiv},
  doi           = {10.1103/PhysRevD.72.124001},
  journal       = {Phys. Rev. D},
  volume        = {72},
  pages         = {124001},
  year          = {2005}
}

@article{Pound_2020,
  author        = {Pound, Adam and Wardell, Barry and Warburton, Niels and Miller, Jeremy},
  title         = {{Second-Order Self-Force Calculation of Gravitational Binding Energy in Compact Binaries}},
  eprint        = {1908.07419},
  archiveprefix = {arXiv},
  primaryclass  = {gr-qc},
  doi           = {10.1103/PhysRevLett.124.021101},
  journal       = {Phys. Rev. Lett.},
  volume        = {124},
  number        = {2},
  pages         = {021101},
  year          = {2020}
}

@incollection{Pound_2021,
  author        = {Adam Pound and Barry Wardell},
  booktitle     = {Handbook of Gravitational Wave Astronomy},
  publisher     = {Springer},
  title         = {Black hole perturbation theory and gravitational self-force},
  year          = {2022},
  address       = {Singapore},
  editor        = {Cosimo Bambi and Stavros Katsanevas and Konstantinos D. Kokkotas},
  month         = mar,
  archiveprefix = {arXiv},
  doi           = {10.1007/978-981-15-4702-7_38-1},
  eprint        = {2101.04592},
  primaryclass  = {gr-qc}
}

@article{Quinn_1997,
  author        = {Quinn, Theodore C. and Wald, Robert M.},
  title         = {{Axiomatic approach to electromagnetic and gravitational radiation reaction of particles in curved spacetime}},
  eprint        = {gr-qc/9610053},
  archiveprefix = {arXiv},
  doi           = {10.1103/PhysRevD.56.3381},
  journal       = {Phys. Rev. D},
  volume        = {56},
  pages         = {3381--3394},
  year          = {1997}
}

@article{Rashti:2021ihv,
  author        = {Rashti, Alireza and Fabbri, Francesco Maria and Br\"ugmann, Bernd and Chaurasia, Swami Vivekanandji and Dietrich, Tim and Ujevic, Maximiliano and Tichy, Wolfgang},
  title         = {New pseudospectral code for the construction of initial data},
  eprint        = {2109.14511},
  archiveprefix = {arXiv},
  primaryclass  = {gr-qc},
  doi           = {10.1103/PhysRevD.105.104027},
  journal       = {Phys. Rev. D},
  volume        = {105},
  number        = {10},
  pages         = {104027},
  year          = {2022}
}

@article{Sago:2008id,
    author = "Sago, Norichika and Barack, Leor and Detweiler, Steven L.",
    title = "{Two approaches for the gravitational self force in black hole spacetime: Comparison of numerical results}",
    eprint = "0810.2530",
    archivePrefix = "arXiv",
    primaryClass = "gr-qc",
    doi = "10.1103/PhysRevD.78.124024",
    journal = "Phys. Rev. D",
    volume = "78",
    pages = "124024",
    year = "2008"
}

@article{Scheel:2025jct,
  author        = {Scheel, Mark A. and others},
  title         = {{The SXS collaboration's third catalog of binary black hole simulations}},
  eprint        = {2505.13378},
  archiveprefix = {arXiv},
  primaryclass  = {gr-qc},
  doi           = {10.1088/1361-6382/adfd34},
  journal       = {Class. Quant. Grav.},
  volume        = {42},
  number        = {19},
  pages         = {195017},
  year          = {2025}
}

@software{spectrecode,
    author = "Deppe, Nils and Throwe, William and Kidder, Lawrence E. and Vu,
Nils L. and Nelli, Kyle C. and Scheel, Mark A. and Armaza, Crist\'obal and
Bonilla, Marceline S. and Carpenter, Alexander and Foucart, Francois and
H\'ebert, Fran\c{c}ois and Kim, Yoonsoo and Kumar, Prayush and Lara, Guillermo
and Lovelace, Geoffrey and Macedo, Alexandra and Moxon, Jordan and Murphy,
Michael and O'Shea, Eamonn and Pfeiffer, Harald P. and Teukolsky, Saul A. and
Wittek, Nikolas A. and others",
    title = "\texttt{SpECTRE v2026.06.30}",
    version = "2026.06.30",
    publisher = "Zenodo",
    doi = "10.5281/zenodo.21065180",
    url = "https://spectre-code.org",
    howpublished =
"\href{https://doi.org/10.5281/zenodo.21065180}{10.5281/zenodo.21065180}",
    license = "MIT",
    year = "2026",
    month = "6"
}

@article{Spiers_2024,
  author        = {Spiers, Andrew},
  title         = {{Analytically separating the source of the Teukolsky equation}},
  eprint        = {2402.00604},
  archiveprefix = {arXiv},
  primaryclass  = {gr-qc},
  doi           = {10.1103/PhysRevD.109.104059},
  journal       = {Phys. Rev. D},
  volume        = {109},
  number        = {10},
  pages         = {104059},
  year          = {2024}
}

@article{Thornburg_2017,
  author        = {Thornburg, Jonathan and Wardell, Barry},
  title         = {{Scalar self-force for highly eccentric equatorial orbits in Kerr spacetime}},
  eprint        = {1610.09319},
  archiveprefix = {arXiv},
  primaryclass  = {gr-qc},
  doi           = {10.1103/PhysRevD.95.084043},
  journal       = {Phys. Rev. D},
  volume        = {95},
  number        = {8},
  pages         = {084043},
  year          = {2017}
}

@article{Uryu2012-bo,
  author        = {Uryū, Kōji and Tsokaros, Antonios},
  title         = {New code for equilibriums and quasiequilibrium initial data of compact objects},
  eprint        = {1108.3065},
  archiveprefix = {arXiv},
  primaryclass  = {gr-qc},
  journal       = {Phys. Rev. D},
  volume        = {85},
  pages         = {064014},
  year          = {2012},
  url           = {https://doi.org/10.1103/physrevd.85.064014},
  doi           = {10.1103/physrevd.85.064014}
}

@article{vanGijzen2007,
  author  = {van Gijzen, M. B. and Erlangga, Y. A. and Vuik, C.},
  title   = "{Spectral Analysis of the Discrete Helmholtz Operator Preconditioned with a Shifted Laplacian}",
  journal = {SIAM J. Sci. Comput.},
  volume  = {29},
  number  = {5},
  pages   = {1942-1958},
  year    = {2007},
  doi     = {10.1137/060661491},
  url     = {https://doi.org/10.1137/060661491}
}

@article{Vega_2008,
  author        = {Vega, Ian and Detweiler, Steven L.},
  title         = {{Regularization of fields for self-force problems in curved spacetime: Foundations and a time-domain application}},
  eprint        = {0712.4405},
  archiveprefix = {arXiv},
  primaryclass  = {gr-qc},
  doi           = {10.1103/PhysRevD.77.084008},
  journal       = {Phys. Rev. D},
  volume        = {77},
  pages         = {084008},
  year          = {2008}
}

@article{Wald:1973wwa,
    author = "Wald, Robert M.",
    title = "{On perturbations of a Kerr black hole}",
    doi = "10.1063/1.1666203",
    journal = "J. Math. Phys.",
    volume = "14",
    number = "10",
    pages = "1453--1461",
    year = "1973"
}

@article{Vu_2022,
  author        = {Vu, Nils L. and others},
  title         = {{A scalable elliptic solver with task-based parallelism for the SpECTRE numerical relativity code}},
  eprint        = {2111.06767},
  archiveprefix = {arXiv},
  primaryclass  = {gr-qc},
  doi           = {10.1103/PhysRevD.105.084027},
  journal       = {Phys. Rev. D},
  volume        = {105},
  number        = {8},
  pages         = {084027},
  year          = {2022}
}

@article{Vu:2021,
  author = "Fischer, Nils L. and Pfeiffer, Harald P.",
  title = "{Unified discontinuous Galerkin scheme for a large class of elliptic equations}",
  eprint = "2108.05826",
  archivePrefix = "arXiv",
  primaryClass = "math.NA",
  doi = "10.1103/PhysRevD.105.024034",
  journal = "Phys. Rev. D",
  volume = "105",
  number = "2",
  pages = "024034",
  year = "2022",
}

@article{Vu:2024,
  author = "Vu, Nils L.",
  title = "{Discontinuous Galerkin scheme for elliptic equations on extremely stretched grids}",
  eprint = "2405.06120",
  archivePrefix = "arXiv",
  primaryClass = "gr-qc",
  doi = "10.1103/PhysRevD.110.084062",
  journal = "Phys. Rev. D",
  volume = "110",
  number = "8",
  pages = "084062",
  year = "2024",
}

@article{Wardell_2012,
  author        = {Wardell, Barry and Vega, Ian and Thornburg, Jonathan and Diener, Peter},
  title         = {{Generic effective source for scalar self-force calculations}},
  eprint        = {1112.6355},
  archiveprefix = {arXiv},
  primaryclass  = {gr-qc},
  doi           = {10.1103/PhysRevD.85.104044},
  journal       = {Phys. Rev. D},
  volume        = {85},
  pages         = {104044},
  year          = {2012}
}

@article{Wardell_2024,
  author        = {Wardell, Barry and Kavanagh, Chris and Dolan, Sam R.},
  title         = {{Sourced metric perturbations of Kerr spacetime in Lorenz gauge}},
  eprint        = {2406.12510},
  archiveprefix = {arXiv},
  primaryclass  = {gr-qc},
  doi           = {10.1088/1361-6382/ae0918},
  journal       = {Class. Quant. Grav.},
  volume        = {42},
  number        = {20},
  pages         = {205007},
  year          = {2025}
}

@article{Vu:2026ypc,
    author = "Vu, Nils L. and Nishimura, Nami and Osburn, Thomas and Thompson, Jonathan E. and Kidder, Lawrence E. and Upton, Samuel D. and Wardell, Barry",
    title = "{Self-force calculations with numerical relativity methods}",
    eprint = "2606.04998",
    archivePrefix = "arXiv",
    primaryClass = "gr-qc",
    doi = "10.1103/lbyk-26gh",
    journal = "Phys. Rev. D",
    volume = "114",
    number = "6",
    pages = "064077",
    year = "2026"
}

@article{Upton:2021oxf,
    author = "Upton, Samuel D. and Pound, Adam",
    title = "{Second-order gravitational self-force in a highly regular gauge}",
    eprint = "2101.11409",
    archivePrefix = "arXiv",
    primaryClass = "gr-qc",
    doi = "10.1103/PhysRevD.103.124016",
    journal = "Phys. Rev. D",
    volume = "103",
    number = "12",
    pages = "124016",
    year = "2021"
}

@article{Spiers:2023cip,
    author = "Spiers, Andrew and Pound, Adam and Moxon, Jordan",
    title = "{Second-order Teukolsky formalism in Kerr spacetime: Formulation and nonlinear source}",
    eprint = "2305.19332",
    archivePrefix = "arXiv",
    primaryClass = "gr-qc",
    doi = "10.1103/PhysRevD.108.064002",
    journal = "Phys. Rev. D",
    volume = "108",
    number = "6",
    pages = "064002",
    year = "2023"
}

@article{Detweiler:2011tt,
    author = "Detweiler, Steven",
    title = "{Gravitational radiation reaction and second order perturbation theory}",
    eprint = "1107.2098",
    archivePrefix = "arXiv",
    primaryClass = "gr-qc",
    doi = "10.1103/PhysRevD.85.044048",
    journal = "Phys. Rev. D",
    volume = "85",
    pages = "044048",
    year = "2012"
}

@article{Pound:2012nt,
    author = "Pound, Adam",
    title = "{Second-order gravitational self-force}",
    eprint = "1201.5089",
    archivePrefix = "arXiv",
    primaryClass = "gr-qc",
    doi = "10.1103/PhysRevLett.109.051101",
    journal = "Phys. Rev. Lett.",
    volume = "109",
    pages = "051101",
    year = "2012"
}

@article{Pound:2012dk,
    author = "Pound, Adam",
    title = "{Nonlinear gravitational self-force. I. Field outside a small body}",
    eprint = "1206.6538",
    archivePrefix = "arXiv",
    primaryClass = "gr-qc",
    doi = "10.1103/PhysRevD.86.084019",
    journal = "Phys. Rev. D",
    volume = "86",
    pages = "084019",
    year = "2012"
}

@article{Pound:2017psq,
    author = "Pound, Adam",
    title = "{Nonlinear gravitational self-force: second-order equation of motion}",
    eprint = "1703.02836",
    archivePrefix = "arXiv",
    primaryClass = "gr-qc",
    doi = "10.1103/PhysRevD.95.104056",
    journal = "Phys. Rev. D",
    volume = "95",
    number = "10",
    pages = "104056",
    year = "2017"
}

@article{Gralla:2012db,
    author = "Gralla, Samuel E.",
    title = "{Second Order Gravitational Self Force}",
    eprint = "1203.3189",
    archivePrefix = "arXiv",
    primaryClass = "gr-qc",
    reportNumber = "UMD-94-11",
    doi = "10.1103/PhysRevD.85.124011",
    journal = "Phys. Rev. D",
    volume = "85",
    pages = "124011",
    year = "2012"
}

@article{PhysRevLett.29.1114,
  title = {Rotating Black Holes: Separable Wave Equations for Gravitational and Electromagnetic Perturbations},
  author = {Teukolsky, Saul A.},
  journal = {Phys. Rev. Lett.},
  volume = {29},
  issue = {16},
  pages = {1114--1118},
  numpages = {0},
  year = {1972},
  month = {Oct},
  publisher = {American Physical Society},
  doi = {10.1103/PhysRevLett.29.1114},
  url = {https://link.aps.org/doi/10.1103/PhysRevLett.29.1114}
}

@article{1973ApJ...185..635T,
  author = "Teukolsky, Saul A.",
  doi = "10.1086/152444",
  journal = "Astrophys. J.",
  volume = "185",
  pages = "635--647",
  year = "1973",
  title = "{Perturbations of a Rotating Black Hole. I. Fundamental Equations for Gravitational, Electromagnetic, and Neutrino-Field Perturbations}",
}

@article{Shah:2012gu,
    author = "Shah, Abhay G. and Friedman, John L. and Keidl, Tobias S.",
    title = "{EMRI corrections to the angular velocity and redshift factor of a mass in circular orbit about a Kerr black hole}",
    eprint = "1207.5595",
    archivePrefix = "arXiv",
    primaryClass = "gr-qc",
    doi = "10.1103/PhysRevD.86.084059",
    journal = "Phys. Rev. D",
    volume = "86",
    pages = "084059",
    year = "2012"
}

@article{Pound:2013faa,
    author = "Pound, Adam and Merlin, Cesar and Barack, Leor",
    title = "{Gravitational self-force from radiation-gauge metric perturbations}",
    eprint = "1310.1513",
    archivePrefix = "arXiv",
    primaryClass = "gr-qc",
    doi = "10.1103/PhysRevD.89.024009",
    journal = "Phys. Rev. D",
    volume = "89",
    number = "2",
    pages = "024009",
    year = "2014"
}

@article{Pound:2014koa,
    author = "Pound, Adam",
    title = "{Conservative effect of the second-order gravitational self-force on quasicircular orbits in Schwarzschild spacetime}",
    eprint = "1404.1543",
    archivePrefix = "arXiv",
    primaryClass = "gr-qc",
    doi = "10.1103/PhysRevD.90.084039",
    journal = "Phys. Rev. D",
    volume = "90",
    number = "8",
    pages = "084039",
    year = "2014"
}

@article{vandeMeent:2015lxa,
    author = "van de Meent, Maarten and Shah, Abhay G.",
    title = "{Metric perturbations produced by eccentric equatorial orbits around a Kerr black hole}",
    eprint = "1506.04755",
    archivePrefix = "arXiv",
    primaryClass = "gr-qc",
    doi = "10.1103/PhysRevD.92.064025",
    journal = "Phys. Rev. D",
    volume = "92",
    number = "6",
    pages = "064025",
    year = "2015"
}

@article{Bini:2016dvs,
    author = "Bini, Donato and Damour, Thibault and Geralico, Andrea",
    title = "{High post-Newtonian order gravitational self-force analytical results for eccentric equatorial orbits around a Kerr black hole}",
    eprint = "1602.08282",
    archivePrefix = "arXiv",
    primaryClass = "gr-qc",
    doi = "10.1103/PhysRevD.93.124058",
    journal = "Phys. Rev. D",
    volume = "93",
    number = "12",
    pages = "124058",
    year = "2016"
}

@article{Merlin:2016boc,
    author = "Merlin, Cesar and Ori, Amos and Barack, Leor and Pound, Adam and van de Meent, Maarten",
    title = "{Completion of metric reconstruction for a particle orbiting a Kerr black hole}",
    eprint = "1609.01227",
    archivePrefix = "arXiv",
    primaryClass = "gr-qc",
    doi = "10.1103/PhysRevD.94.104066",
    journal = "Phys. Rev. D",
    volume = "94",
    number = "10",
    pages = "104066",
    year = "2016"
}

@article{Kavanagh:2016idg,
    author = "Kavanagh, Chris and Ottewill, Adrian C. and Wardell, Barry",
    title = "{Analytical high-order post-Newtonian expansions for spinning extreme mass ratio binaries}",
    eprint = "1601.03394",
    archivePrefix = "arXiv",
    primaryClass = "gr-qc",
    doi = "10.1103/PhysRevD.93.124038",
    journal = "Phys. Rev. D",
    volume = "93",
    number = "12",
    pages = "124038",
    year = "2016"
}

@article{Bini:2018ylh,
    author = "Bini, Donato and Damour, Thibault and Geralico, Andrea and Kavanagh, Chris and van de Meent, Maarten",
    title = "{Gravitational self-force corrections to gyroscope precession along circular orbits in the Kerr spacetime}",
    eprint = "1809.02516",
    archivePrefix = "arXiv",
    primaryClass = "gr-qc",
    doi = "10.1103/PhysRevD.98.104062",
    journal = "Phys. Rev. D",
    volume = "98",
    number = "10",
    pages = "104062",
    year = "2018"
}

@article{Toomani:2021jlo,
    author = "Toomani, Vahid and Zimmerman, Peter and Spiers, Andrew and Hollands, Stefan and Pound, Adam and Green, Stephen R.",
    title = "{New metric reconstruction scheme for gravitational self-force calculations}",
    eprint = "2108.04273",
    archivePrefix = "arXiv",
    primaryClass = "gr-qc",
    doi = "10.1088/1361-6382/ac37a5",
    journal = "Class. Quant. Grav.",
    volume = "39",
    number = "1",
    pages = "015019",
    year = "2022"
}

@article{Bourg:2024vre,
    author = "Bourg, Patrick and Leather, Benjamin and Casals, Marc and Pound, Adam and Wardell, Barry",
    title = "{Implementation of a Green-Hollands-Zimmerman-Teukolsky puncture scheme for gravitational self-force calculations}",
    eprint = "2403.12634",
    archivePrefix = "arXiv",
    primaryClass = "gr-qc",
    doi = "10.1103/PhysRevD.110.044007",
    journal = "Phys. Rev. D",
    volume = "110",
    number = "4",
    pages = "044007",
    year = "2024"
}

@article{Nasipak:2025tby,
    author = "Nasipak, Zachary",
    title = "{Metric reconstruction and the Hamiltonian for eccentric, precessing binaries in the small-mass-ratio limit}",
    eprint = "2507.07746",
    archivePrefix = "arXiv",
    primaryClass = "gr-qc",
    doi = "10.1103/7zkx-vbg9",
    journal = "Phys. Rev. D",
    volume = "113",
    number = "12",
    pages = "124051",
    year = "2026"
}

@article{Warburton:2013lea,
    author = "Warburton, Niels and Wardell, Barry",
    title = "{Applying the effective-source approach to frequency-domain self-force calculations}",
    eprint = "1311.3104",
    archivePrefix = "arXiv",
    primaryClass = "gr-qc",
    doi = "10.1103/PhysRevD.89.044046",
    journal = "Phys. Rev. D",
    volume = "89",
    number = "4",
    pages = "044046",
    year = "2014"
}

@article{Wardell:2015ada,
    author = "Wardell, Barry and Warburton, Niels",
    title = "{Applying the effective-source approach to frequency-domain self-force calculations: Lorenz-gauge gravitational perturbations}",
    eprint = "1505.07841",
    archivePrefix = "arXiv",
    primaryClass = "gr-qc",
    doi = "10.1103/PhysRevD.92.084019",
    journal = "Phys. Rev. D",
    volume = "92",
    number = "8",
    pages = "084019",
    year = "2015"
}

@article{Barack:2005nr,
    author = "Barack, Leor and Lousto, Carlos O.",
    title = "{Perturbations of Schwarzschild black holes in the Lorenz gauge: Formulation and numerical implementation}",
    eprint = "gr-qc/0510019",
    archivePrefix = "arXiv",
    doi = "10.1103/PhysRevD.72.104026",
    journal = "Phys. Rev. D",
    volume = "72",
    pages = "104026",
    year = "2005"
}

@article{Akcay:2010dx,
    author = "Akcay, Sarp",
    title = "{A Fast Frequency-Domain Algorithm for Gravitational Self-Force: I. Circular Orbits in Schwarzschild Spacetime}",
    eprint = "1012.5860",
    archivePrefix = "arXiv",
    primaryClass = "gr-qc",
    doi = "10.1103/PhysRevD.83.124026",
    journal = "Phys. Rev. D",
    volume = "83",
    pages = "124026",
    year = "2011"
}

@article{Akcay_2013,
  author = "Akcay, Sarp and Warburton, Niels and Barack, Leor",
  title = "{Frequency-domain algorithm for the Lorenz-gauge gravitational self-force}",
  eprint = "1308.5223",
  archivePrefix = "arXiv",
  primaryClass = "gr-qc",
  doi = "10.1103/PhysRevD.88.104009",
  journal = "Phys. Rev. D",
  volume = "88",
  number = "10",
  pages = "104009",
  year = "2013",
}

@article{Osburn:2014hoa,
    author = "Osburn, Thomas and Forseth, Erik and Evans, Charles R. and Hopper, Seth",
    title = "{Lorenz gauge gravitational self-force calculations of eccentric binaries using a frequency domain procedure}",
    eprint = "1409.4419",
    archivePrefix = "arXiv",
    primaryClass = "gr-qc",
    doi = "10.1103/PhysRevD.90.104031",
    journal = "Phys. Rev. D",
    volume = "90",
    number = "10",
    pages = "104031",
    year = "2014"
}

@article{Hopper:2015jxa,
    author = "Hopper, Seth and Forseth, Erik and Osburn, Thomas and Evans, Charles R.",
    title = "{Fast spectral source integration in black hole perturbation calculations}",
    eprint = "1506.04742",
    archivePrefix = "arXiv",
    primaryClass = "gr-qc",
    doi = "10.1103/PhysRevD.92.044048",
    journal = "Phys. Rev. D",
    volume = "92",
    pages = "044048",
    year = "2015"
}

@article{Thornburg:2026eeb,
    author = "Thornburg, Jonathan",
    title = "{Time-domain evolution of Lorenz-gauge metric perturbations: taming the $\ell=m=1$ gauge instability}",
    eprint = "2606.27016",
    archivePrefix = "arXiv",
    primaryClass = "gr-qc",
    month = "6",
    year = "2026",
    journal = ""
}

@article{Osburn:2026uct,
    author = "Osburn, Thomas and Wardell, Barry and Battaglia, Erin",
    title = "{Schwarzschild perturbations in Lorenz gauge via elliptic differential equations}",
    eprint = "2608.07934",
    archivePrefix = "arXiv",
    primaryClass = "gr-qc",
    month = "8",
    year = "2026",
    journal = ""
}

@article{Barack:2007tm,
    author = "Barack, Leor and Sago, Norichika",
    title = "{Gravitational self force on a particle in circular orbit around a Schwarzschild black hole}",
    eprint = "gr-qc/0701069",
    archivePrefix = "arXiv",
    doi = "10.1103/PhysRevD.75.064021",
    journal = "Phys. Rev. D",
    volume = "75",
    pages = "064021",
    year = "2007"
}

@article{Barack:2010tm,
    author = "Barack, Leor and Sago, Norichika",
    title = "{Gravitational self-force on a particle in eccentric orbit around a Schwarzschild black hole}",
    eprint = "1002.2386",
    archivePrefix = "arXiv",
    primaryClass = "gr-qc",
    doi = "10.1103/PhysRevD.81.084021",
    journal = "Phys. Rev. D",
    volume = "81",
    pages = "084021",
    year = "2010"
}

@article{Bonetto:2021exn,
    author = "Bonetto, Riccardo and Pound, Adam and Sam, Zeyd",
    title = "{Deformed Schwarzschild horizons in second-order perturbation theory: Mass, geometry, and teleology}",
    eprint = "2109.09514",
    archivePrefix = "arXiv",
    primaryClass = "gr-qc",
    doi = "10.1103/PhysRevD.105.024048",
    journal = "Phys. Rev. D",
    volume = "105",
    number = "2",
    pages = "024048",
    year = "2022"
}

@article{Mathews:2025txc,
    author = "Mathews, Josh and Wardell, Barry and Pound, Adam and Warburton, Niels",
    title = "{Postadiabatic self-force waveforms: Slowly spinning primary and precessing secondary}",
    eprint = "2510.16113",
    archivePrefix = "arXiv",
    primaryClass = "gr-qc",
    doi = "10.1103/ph3p-mscl",
    journal = "Phys. Rev. D",
    volume = "113",
    number = "6",
    pages = "064034",
    year = "2026"
}

@article{Reynolds:2020jwt,
    author = "Reynolds, Christopher S.",
    title = "{Observational Constraints on Black Hole Spin}",
    eprint = "2011.08948",
    archivePrefix = "arXiv",
    primaryClass = "astro-ph.HE",
    doi = "10.1146/annurev-astro-112420-035022",
    journal = "Ann. Rev. Astron. Astrophys.",
    volume = "59",
    pages = "117--154",
    year = "2021"
}

@article{vandeMeent:2017bcc,
    author = "van de Meent, Maarten",
    title = "{Gravitational self-force on generic bound geodesics in Kerr spacetime}",
    eprint = "1711.09607",
    archivePrefix = "arXiv",
    primaryClass = "gr-qc",
    doi = "10.1103/PhysRevD.97.104033",
    journal = "Phys. Rev. D",
    volume = "97",
    number = "10",
    pages = "104033",
    year = "2018"
}

@article{Leather:2026zhl,
    author = "Leather, Benjamin and Spiers, Andrew and Pound, Adam and Upton, Samuel D. and Wardell, Barry and Durkan, Leanne and Warburton, Niels",
    title = "{Second-order Teukolsky calculations for nonspinning, quasicircular binaries}",
    eprint = "2609.08995",
    archivePrefix = "arXiv",
    primaryClass = "gr-qc",
    month = "9",
    year = "2026",
    journal = ""
}

@article{Detweiler:2008ft,
    author = "Detweiler, Steven L.",
    title = "{A Consequence of the gravitational self-force for circular orbits of the Schwarzschild geometry}",
    eprint = "0804.3529",
    archivePrefix = "arXiv",
    primaryClass = "gr-qc",
    doi = "10.1103/PhysRevD.77.124026",
    journal = "Phys. Rev. D",
    volume = "77",
    pages = "124026",
    year = "2008"
}

@article{Akcay:2015pza,
    author = "Akcay, Sarp and Le Tiec, Alexandre and Barack, Leor and Sago, Norichika and Warburton, Niels",
    title = "{Comparison Between Self-Force and Post-Newtonian Dynamics: Beyond Circular Orbits}",
    eprint = "1503.01374",
    archivePrefix = "arXiv",
    primaryClass = "gr-qc",
    doi = "10.1103/PhysRevD.91.124014",
    journal = "Phys. Rev. D",
    volume = "91",
    number = "12",
    pages = "124014",
    year = "2015"
}

@article{Mathews:2021rod,
    author = "Mathews, Josh and Pound, Adam and Wardell, Barry",
    title = "{Self-force calculations with a spinning secondary}",
    eprint = "2112.13069",
    archivePrefix = "arXiv",
    primaryClass = "gr-qc",
    doi = "10.1103/PhysRevD.105.084031",
    journal = "Phys. Rev. D",
    volume = "105",
    number = "8",
    pages = "084031",
    year = "2022"
}

@article{Abbott_2019,
  author = "Abbott, B. P. and others",
  collaboration = "LIGO Scientific, Virgo",
  title = "{GWTC-1: A Gravitational-Wave Transient Catalog of Compact Binary Mergers Observed by LIGO and Virgo during the First and Second Observing Runs}",
  eprint = "1811.12907",
  archivePrefix = "arXiv",
  primaryClass = "astro-ph.HE",
  reportNumber = "LIGO-P1800307",
  doi = "10.1103/PhysRevX.9.031040",
  journal = "Phys. Rev. X",
  volume = "9",
  number = "3",
  pages = "031040",
  year = "2019",
}

@article{Abbott_2021,
  author = "Abbott, R. and others",
  collaboration = "LIGO Scientific, Virgo",
  title = "{GWTC-2: Compact Binary Coalescences Observed by LIGO and Virgo During the First Half of the Third Observing Run}",
  eprint = "2010.14527",
  archivePrefix = "arXiv",
  primaryClass = "gr-qc",
  reportNumber = "P2000061",
  doi = "10.1103/PhysRevX.11.021053",
  journal = "Phys. Rev. X",
  volume = "11",
  pages = "021053",
  year = "2021",
}

@article{Abbott_2022,
  author = "Abbott, R. and others",
  collaboration = "KAGRA, VIRGO, LIGO Scientific",
  title = "{GWTC-3: Compact Binary Coalescences Observed by LIGO and Virgo during the Second Part of the Third Observing Run}",
  eprint = "2111.03606",
  archivePrefix = "arXiv",
  primaryClass = "gr-qc",
  reportNumber = "LIGO-P2000318",
  doi = "10.1103/PhysRevX.13.041039",
  journal = "Phys. Rev. X",
  volume = "13",
  number = "4",
  pages = "041039",
  year = "2023",
}

@article{ET:2025xjr,
  author = "Abac, Adrian and others",
  collaboration = "ET",
  title = "{The Science of the Einstein Telescope}",
  eprint = "2503.12263",
  archivePrefix = "arXiv",
  primaryClass = "gr-qc",
  reportNumber = "ET-0036C-25",
  doi = "10.1088/1475-7516/2026/03/081",
  journal = "JCAP",
  volume = "03",
  pages = "081",
  year = "2026",
}

@unpublished{Evans:2021gyd,
  author        = {Evans, Matthew and others},
  title         = {{A Horizon Study for Cosmic Explorer: Science, Observatories, and Community}},
  eprint        = {2109.09882},
  archiveprefix = {arXiv},
  primaryclass  = {astro-ph.IM},
  reportnumber  = {CE-P2100003-v7, Cosmic Explorer technical report CE-P2100003-v6},
  month         = {9},
  year          = {2021}
}

@unpublished{LISA,
  author        = {Baker, John and others},
  title         = {{The Laser Interferometer Space Antenna: Unveiling the Millihertz Gravitational Wave Sky}},
  eprint        = {1907.06482},
  archiveprefix = {arXiv},
  primaryclass  = {astro-ph.IM},
  reportnumber  = {FERMILAB-PUB-19-436-A},
  month         = {7},
  year          = {2019}
}

@unpublished{LISA:2024hlh,
  author        = {Colpi, Monica and others},
  collaboration = {LISA},
  title         = {{LISA Definition Study Report}},
  eprint        = {2402.07571},
  archiveprefix = {arXiv},
  primaryclass  = {astro-ph.CO},
  month         = {2},
  year          = {2024}
}

@article{LISAWaveformWG:2023arg,
  author        = {Afshordi, Niayesh and others},
  collaboration = {LISA Consortium Waveform Working Group},
  title         = {{Waveform modelling for the Laser Interferometer Space Antenna}},
  eprint        = {2311.01300},
  archiveprefix = {arXiv},
  primaryclass  = {gr-qc},
  doi           = {10.1007/s41114-025-00056-1},
  journal       = {Living Rev. Rel.},
  volume        = {28},
  number        = {1},
  pages         = {9},
  year          = {2025}
}

@article{Wardell_2023,
  author        = {Wardell, Barry and Pound, Adam and Warburton, Niels and Miller, Jeremy and Durkan, Leanne and Le Tiec, Alexandre},
  title         = {{Gravitational Waveforms for Compact Binaries from Second-Order Self-Force Theory}},
  eprint        = {2112.12265},
  archiveprefix = {arXiv},
  primaryclass  = {gr-qc},
  doi           = {10.1103/PhysRevLett.130.241402},
  journal       = {Phys. Rev. Lett.},
  volume        = {130},
  number        = {24},
  pages         = {241402},
  year          = {2023}
}

@article{Warburton:2021kwk,
    author = "Warburton, Niels and Pound, Adam and Wardell, Barry and Miller, Jeremy and Durkan, Leanne",
    title = "{Gravitational-Wave Energy Flux for Compact Binaries through Second Order in the Mass Ratio}",
    eprint = "2107.01298",
    archivePrefix = "arXiv",
    primaryClass = "gr-qc",
    doi = "10.1103/PhysRevLett.127.151102",
    journal = "Phys. Rev. Lett.",
    volume = "127",
    number = "15",
    pages = "151102",
    year = "2021"
}

@article{Miller:2023ers,
  author        = {Miller, Jeremy and Leather, Benjamin and Pound, Adam and Warburton, Niels},
  title         = {{Worldtube puncture scheme for first- and second-order self-force calculations in the Fourier domain}},
  eprint        = {2401.00455},
  archiveprefix = {arXiv},
  primaryclass  = {gr-qc},
  doi           = {10.1103/PhysRevD.109.104010},
  journal       = {Phys. Rev. D},
  volume        = {109},
  number        = {10},
  pages         = {104010},
  year          = {2024}
}

@article{Upton:2025bja,
    author = "Upton, Samuel D. and Wardell, Barry and Pound, Adam and Warburton, Niels and Barack, Leor",
    title = "{Effective source for second-order self-force calculations: Quasicircular orbits in Schwarzschild spacetime}",
    eprint = "2508.00087",
    archivePrefix = "arXiv",
    primaryClass = "gr-qc",
    doi = "10.1103/f9j2-k64r",
    journal = "Phys. Rev. D",
    volume = "113",
    number = "6",
    pages = "064013",
    year = "2026"
}

@misc{BHPToolkit,
  title = {{Black Hole Perturbation Toolkit}},
  howpublished = {(\href{https://bhptoolkit.org/}{bhptoolkit.org})},
}

@misc{xAct,
  author       = {Jos{\'{e}} M. Mart{\'{i}}n-Garc{\'{i}}a},
  howpublished = {\url{https://xact.es/}},
  title        = {{xAct}: Efficient tensor computer algebra for the {Wolfram Language}},
  year  = "2026"
}

@misc{xCoba,
  author       = {David Yllanes and Jos{\'{e}} M. Mart{\'{i}}n-Garc{\'{i}}a},
  howpublished = {\url{https://xact.es/xCoba/}},
  title        = {{xCoba}: General component tensor computer algebra},
  year  = "2026"
}

@article{Martin-Garcia:2008ysv,
    author = "Mart{\'\i}n-Garc{\'\i}a, Jos{\'e} M.",
    title = "{xPerm: fast index canonicalization for tensor computer algebra}",
    eprint = "0803.0862",
    archivePrefix = "arXiv",
    primaryClass = "cs.SC",
    doi = "10.1016/j.cpc.2008.05.009",
    journal = "Comput. Phys. Commun.",
    volume = "179",
    number = "8",
    pages = "597--603",
    year = "2008"
}

@misc{Mathematica,
  author = "{Wolfram Research{,} Inc.}",
  title = {Mathematica, {V}ersion 15.0},
  url = {https://www.wolfram.com/mathematica},
  note = "{Champaign, IL}",
  year = "2026"
}

@article{Leather:2026gnc,
    author = "Leather, Benjamin and others",
    title = "{Kerr metric perturbations for second-order self-force theory}",
    eprint = "2611.03888",
    archivePrefix = "arXiv",
    primaryClass = "gr-qc",
    year = "2026",
    journal = ""
}

@software{Nasipak_pybhpt_2026,
author = {Nasipak, Zachary},
license = {GPL-3.0},
month = sep,
title = {{pybhpt}},
url = {https://github.com/znasipak/pybhpt},
version = {v1.0.0},
year = {2026}
}

@article{Pound:2009sm,
    author = "Pound, Adam",
    title = "{Self-consistent gravitational self-force}",
    eprint = "0907.5197",
    archivePrefix = "arXiv",
    primaryClass = "gr-qc",
    doi = "10.1103/PhysRevD.81.024023",
    journal = "Phys. Rev. D",
    volume = "81",
    pages = "024023",
    year = "2010"
}

@article{Rosenthal:2006iy,
    author = "Rosenthal, Eran",
    title = "{Second-order gravitational self-force}",
    eprint = "gr-qc/0609069",
    archivePrefix = "arXiv",
    doi = "10.1103/PhysRevD.74.084018",
    journal = "Phys. Rev. D",
    volume = "74",
    pages = "084018",
    year = "2006"
}

@phdthesis{Berndtson:2007gsc,
	archiveprefix = {arXiv},
	author = {Berndtson, Mark V.},
	eprint = {0904.0033},
	primaryclass = {gr-qc},
	school = {University of Colorado},
	title = {{Harmonic gauge perturbations of the Schwarzschild metric}},
	year = {2007}}

@article{Cownden:2026msq,
    author = "Cownden, Brad and Panosso Macedo, Rodrigo and Pantelidou, Christiana and Upton, Samuel D. and Wardell, Barry",
    title = "{Multi-domain spectral method for Kerr spacetimes: a hyperboloidal Teukolsky $m$-mode solver for self-force computations}",
    eprint = "2611.03870",
    archivePrefix = "arXiv",
    primaryClass = "gr-qc",
    year = "2026",
    journal = ""
}

@mastersthesis{adria,
  author      = {Gispert Fern\'andez, Adri\`a},
  title       = {Kerr Black Hole Perturbation Theory},
  school      = {University College Dublin},
  year        = {2026},
  type        = {Master's thesis},
  month       = {August}
}

@article{ABBOTT198276,
    author = "Abbott, L. F. and Deser, Stanley",
    title = "{Stability of Gravity with a Cosmological Constant}",
    reportNumber = "CERN-TH-3136",
    doi = "10.1016/0550-3213(82)90049-9",
    journal = "Nucl. Phys. B",
    volume = "195",
    pages = "76--96",
    year = "1982"
}

@article{PhysRev.166.1272,
  title = {Gravitational Radiation in the Limit of High Frequency. II. Nonlinear Terms and the Effective Stress Tensor},
  author = {Isaacson, Richard A.},
  journal = {Phys. Rev.},
  volume = {166},
  issue = {5},
  pages = {1272--1280},
  numpages = {0},
  year = {1968},
  month = {Feb},
  publisher = {American Physical Society},
  doi = {10.1103/PhysRev.166.1272},
  url = {https://link.aps.org/doi/10.1103/PhysRev.166.1272}
}

\end{document}